\documentstyle[twoside,epsf]{article}

\input{qicsty}

\newcommand{\ket}[1]{\left|\mathrm{#1}\right\rangle}
\newcommand{\bra}[1]{\left\langle\mathrm{#1}\right|}
\newcommand{\0}{\left|0\right\rangle}
\newcommand{\1}{\left|1\right\rangle}

\newcommand{\onlinecite}[1]{[\cite{#1}]}

\newcommand{\fig}[3]{
    \begin{figure}[htbp]
    \vspace{10pt}
    \centerline{\psfig{file=figures/#1.eps,width=#2\columnwidth}}
    \vspace{10pt}
    \fcaption{#3}
    \label{#1}
    \end{figure}
}

\begin{document}
\setlength{\textheight}{7.7truein}    
\runninghead{Photonic Entanglement for Fundamental Tests and
Quantum Communication $\ldots$}
            {Wolfgang Tittel and Gregor Weihs $\ldots$}

\normalsize\textlineskip \thispagestyle{empty}
\setcounter{page}{1}
\copyrightheading{} 
\vspace*{0.88truein}

    \fpage{1} \centerline{\bf Photonic Entanglement for Fundamental Tests and
Quantum Communication}
    \vspace*{0.37truein}
    \centerline{\footnotesize Wolfgang Tittel}
    \vspace*{0.015truein}
    \centerline{\footnotesize\it Group of Applied Physics, University of Geneva (GAP)}
    \baselineskip=10pt
    \centerline{\footnotesize\it 20, Rue de l'Ecole-de-M\'edecine, 1211 Geneve 4, Switzerland}
    \vspace*{10pt}
    \centerline{\footnotesize Gregor Weihs}
    \vspace*{0.015truein}
    \centerline{\footnotesize\it Institut of Experimental Physics, University of Vienna (UNIVIE)}
    \baselineskip=10pt
    \centerline{\footnotesize\it Boltzmanngasse 5, 1090 Wien, Austria}
    \vspace*{0.225truein}
    \publisher{(\today)}{(\today)}

\vspace*{0.21truein} \abstracts{Entanglement is at the heart of
fundamental tests of quantum mechanics like tests of
Bell-inequalities and, as discovered lately, of quantum
computation and communication. Their technological advance made
entangled photons play an outstanding role in entanglement
physics. We give a generalized concept of qubit entanglement and
review the state of the art of photonic experiments.}{}{}

\vspace*{10pt}  
\vspace*{1pt}\textlineskip  

\setcounter{footnote}{0}
\renewcommand{\thefootnote}{\alph{footnote}}
\newpage
\tableofcontents
\newpage

\section{Introduction} \noindent
Due to its importance for understanding the properties of the
quantum world and its role in applications in the new domain of
quantum computation and communication, entanglement got more and
more attention within the physics community throughout the last 70
years, and, lately, even in the general public.\footnote{searching
the internet for ``entanglement", we found 45.000 pages!} The
interest in entanglement, a term invented by Erwin Schr\"{o}dinger in
1935\cite{Schrodinger35a},\footnote{Entanglement is translated
from the original German word ``Verschr\"{a}nkung".} was triggered by
a paper by Einstein, Podolsky and Rosen (EPR) that was published
also in 1935.\cite{Einstein35a} In this famous paper, often
referred to as EPR paradox, the authors analyze the predictions
for a two-particle state where neither particle can be considered
in a state independent from the other. In contrast, both
subsystems, even if at arbitrarily large distance, form a single
entangled system. Based on the assumption of locality, i.e. that
the choice of the type of measurement performed on one particle
can not influence the properties of the other particle, they
argued that the description of reality as given by the
wavefunction is not complete.

The question whether or not this is true, or, in other words,
whether or not entanglement (and hence non-locality) exists
became a very important fundamental issue. It was Bell's
discovery of the so-called Bell inequalities in
1964\cite{Bell64a} and their extension to experimental conditions
by Clauser {\it{et al.}} in 1969 and
1974\cite{Clauser69a,Clauser74a} that transferred the former
purely philosophical debate to the realms of laboratory
experiments. Beginning with the first test of Bell inequalities
in 1972,\cite{Freedman72a} an increasing number of more and more
refined experiments has been
performed.\cite{Aspect99a,Grangier01a} Although the type of
particles entangled is of no importance to demonstrate the
existence of non-locality, by far most experiments relied on
entangled photons. Nowadays, although not all experimental
loopholes have been closed simultaneously in a single experiment
(but all of them have already been closed), it is commonly
believed that quantum non-locality is indeed real. Nevertheless,
there is still interest in performing more Bell-type tests. A
first motivation is to examine the boundary between the quantum
and the classical world,\cite{Zurek91a,Ghirardi80a} a second one
are experiments extending the traditional set-up for Bell-type
tests to relativistic configurations and investigating so-called
relativistic non-locality.\cite{Suarez97b} While massive
particles --- prone to decoherence --- are used for experiments of
the first kind,\cite{Hagley97a} it was again photons that served
for experiments of the second kind.\cite{Zbinden01a}

Apart from these fundamental motivations, the recent discovery
that processing and exchange of information based on quantum
systems enable new forms of computation and communication, more
powerful than its classical analogs, engendered further interest
in entangled systems (see i.e.
Refs.~\onlinecite{PhysicsWorld98a,Bouwmeester00a,Lo98a}). Best
known applications in the domain of quantum communication ---
hence in the domain where photons are most likely best suited for
--- are quantum cryptography (for a recent review see
Ref.~\onlinecite{Gisin01a}) and quantum
teleportation.\cite{Zeilinger00a}

In this article, we review experiments based on {\it photonic
entanglement}, addressing both fundamental as well as applied
aspects. However, although there has been considerable progress in
experiments based on continuous quantum variables as well (see
e.g. Refs.~\onlinecite{Ou92a,Furusawa98a,Silberhorn01a}), we will
focus only on entanglement between discrete two-level quantum
systems (now called quantum bits or qubits). This enables us to
pursue a rather formal approach, summarizing all experiments under
the aspect of experiments with entangled qubits.

The article is structured along the following lines: In
Section~\ref{superpositionprinciple} we introduce the ``quantum
toolbox" in form of sources and analyzers for qubits and entangled
qubits, respectively, and we explain various experimental
approaches. The fact that all different realizations of a qubit
are formally equivalent (a qubit is a qubit) then renders the task
of presenting the variety of experiments quite simple: With
different arrangements of these few basic building blocks various
issues can be addressed experimentally. This concerns tests of
non-locality (Section~\ref{fundamentals}) as well as applications
of entanglement in the domain of quantum communication (section
\ref{quantumcommunication}). As we will see, it is sometimes
enough to change only minor things like analyzer settings in order
to ``continuously" pass from one to the other side --- like from
tests of Bell inequalities to quantum tomography and quantum
cryptography. Finally, a short conclusion is given in Section
\ref{conclusion}.

\section{The superposition principle: photonic qubits and entanglement}
\label{superpositionprinciple} \noindent

One of the most important features of quantum theory is certainly
the superposition principle. While we find this property already
in classical wave theories, e.g. in Young's famous double-slit
experiment in classical optics, quantum theory allows for instance
to describe objects that have traditionally been considered well
localized in space as being in a superposition of different
positions, well separated in space. The superposition-principle
can be applied to any physical property, and it is at the origin
of the ``quantum-paradoxes" as well as of quantum information
theory.

\subsection{The qubit}\label{qubits}

\subsubsection{A theoretical approach}

The most important entity of classical information theory is the
bit. A bit can {\it{either}} have the value ``0" or ``1" with both
values separated by a large energy gap so that the unwanted
spontaneous transition from one to the other value is extremely
unlikely.\footnote{The bit error rate in standard
telecommunication is $10^{-9}-10^{-12}$.}

The quantum mechanical analog of the bit is the quantum bit or
qubit. It is a two-state quantum system with the basic states
$\ket{0}$ and $\ket{1}$ forming an orthogonal basis in the qubit
space. In contrast to the classical bit, it is  possible to create
qubits in a coherent {\it{superposition}} of $\ket{0}$ and
$\ket{1}$, the general state being
\begin{equation}
\ket{\psi}_{\mbox{\scriptsize qubit}}=\alpha\ket{0}+\beta
e^{i\phi}\ket{1}\hspace{1cm}(\alpha^2+\beta^2=1).
\label{generalqubitstate}
\end{equation}
Qubits can be represented graphically on the
qubit-sphere\footnote{Depending on the physical property
represented also known as Bloch- or Poincar\'e-sphere.} pictured
in Fig.~\ref{Qbsphere}. The states $\ket{0}$ and $\ket{1}$ are
localized on the poles of the sphere, any superposition of
$\ket{0}$ and $\ket{1}$ with equal coefficients $\alpha$ and
$\beta$ are represented on the equator, and qubits with different
coefficients lie on a circle with polar angle
$\tan(\varphi)=\beta/\alpha$. Note that any two states represented
on opposite sides of the sphere form a orthonormal basis in the
two-dimensional Hilbert space describing the qubit.

In contrast to classical bits, the outcome of a measurement of a
qubit is not always deterministic. For the general qubit state
given in Eq.~\ref{generalqubitstate}, one finds the value ``0"
with probability $\alpha^2$ and the value ``1" with probability
$\beta^2$.  Note that this could still be achieved with a
classical bit in a mixture between ``0" and ``1". However, the
unique feature of a quantum bit is that the basic states $\ket{0}$
and $\ket{1}$ are superposed coherently. Let us consider the state
\begin{equation}
\ket{\psi'}=\frac{1}{\sqrt{2}}\big[\ket{0}+\ket{1}\big].
\end{equation}
\noindent Measuring this state in a rotated basis with
eigenvectors $\ket{0'}=\ket{0}+\ket{1}$ and
$\ket{1'}=\ket{0}-\ket{1}$, we always find the result ``$0'$".
This contrasts with an incoherent mixture between $\ket{0}$ and
$\ket{1}$ that stays a mixture in any basis and leads to either
result with equal probabilities.

The transition from a coherent superposition to an incoherent one
can be caused by decoherence. In contrast to the first case that
is represented on the shell of the qubit sphere, an incoherent
superposition can be found closer to the origin of the sphere with
a completely incoherent one represented in the origin.

\fig{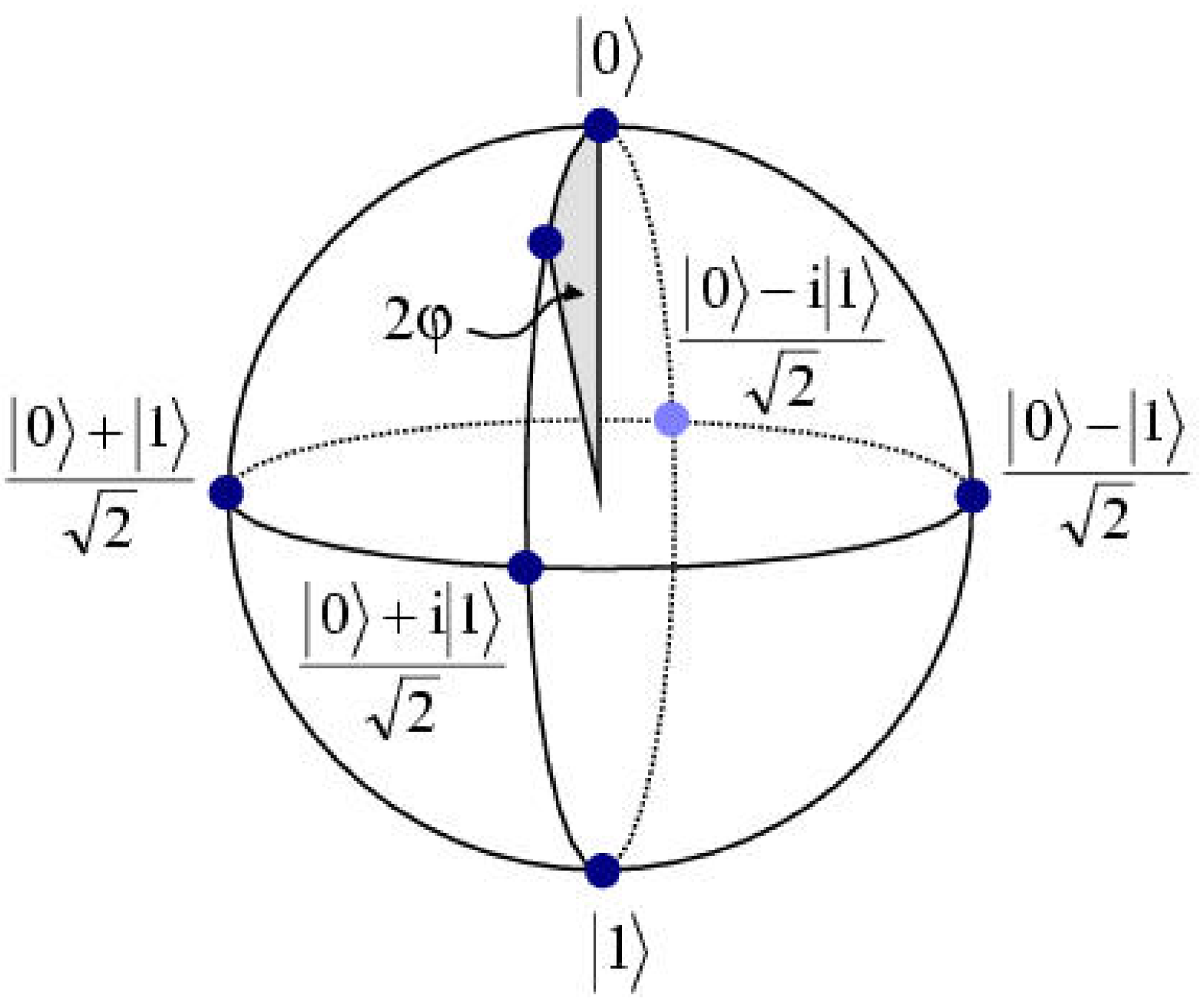}{0.55}{The general qubit sphere. Coherent
superpositions of $\0$ and $\1$ lie on the shell of the sphere,
incoherent ones closer to the origin. All states represented on
opposite sides on the shell of the sphere form a orthonormal basis
in the two-dimensional qubit space.}

\vspace{1cm}
\subsubsection{Preparing and measuring a qubit}
Although, from a theoretical point of view, a qubit is just a
qubit independent of its implementation, one must identify the
abstract qubit space with a physical property when planning an
experiment. There are various ways in which qubits can be realized
using single photons.\footnote{Note that the generation of a
{\it{single}} photon is far from being obvious. In quantum
cryptography, single photons are often mimicked by faint laser
pulses with a mean photon number of 0.1.} Every degree of freedom
that is available can in principle be exploited. The available
properties are the photons' polarization, spatial mode, emission
time, or their frequency.

In addition, depending on the specific goal, there are initial
considerations concerning the wavelength of the photons used: If
the goal is to demonstrate the existence of a certain quantum
effect, it is a good idea to work at a wavelength where high
efficiency and low noise single photon detectors (based on silicon
avalanche photo diodes (APD)) are commercially available, hence at
around 700--800 nm. If the wavelength has to be compatible with
optical fibers as often requested for quantum cryptography or
other long distance applications of quantum communication, the
absorption of the fibers require to work in the second or third
telecommunication window (at 1310 and 1550 nm, respectively).
Here, only home made detectors based on Germanium or InGaAs APDs
are available. Obviously, the same reflections hold for the
creation of entangled qubits (Section~\ref{pairsources}) as well.
For a more detailed discussion of technological issues, see Gisin
{\it et al.}\cite{Gisin01a}

\paragraph{Polarization qubits}

The most well known realization of a qubit is probably the one
using orthogonal states of polarization. In this case, the
qubit-sphere is identical with the well-known Poincar\'e sphere.
We identify left $\ket{l}$ and right $\ket{r}$ circular polarized
photons as our basis states $\0$ and $\1$; they are represented
on the poles of the sphere. Linear polarization of any
orientation as an equally weighted coherent superposition of
$\ket{l}$ and $\ket{r}$ can be found on the equator, and
elliptically polarized light elsewhere. Completely depolarized
light as an incoherent superposition of right and left circular
polarized photons is represented by a point located at the origin.
Polarization qubits can be created and measured using polarizers
and waveplates oriented at various angles.

\paragraph{Spatial-mode qubits}

Another possibility to realize a qubit is shown in
Fig.~\ref{modeqbit}. Using a variable coupler and a phase shifter,
Alice can create any desired superposition of a photon being in
mode $\ket{0}$ and in mode $\ket{1}$. A similar set-up serves to
analyze the qubit. If Bob uses for instance a symmetrical coupler
and a phase $\varphi$=0, a click in detector 0 collapses the
photon state to
$\ket{\psi_0}=\frac{1}{\sqrt{2}}\big(\ket{0}+i\ket{1}\big)$ and in
detector 1 to the orthogonal state
$\ket{\psi_1}=\frac{1}{\sqrt{2}}\big(\ket{0}-i\ket{1}\big)$. Using
a completely asymmetrical coupler, the photon state is projected
onto the basis spanned by the states $\ket{\phi_0}=\ket{0}$ and
$\ket{\phi_1}=\ket{1}$. Spatial-mode qubits would not be very
practical for transmitting quantum information since the phase
between $\0$ and $\1$ is easily randomized by different
environments acting on the different modes.

\fig{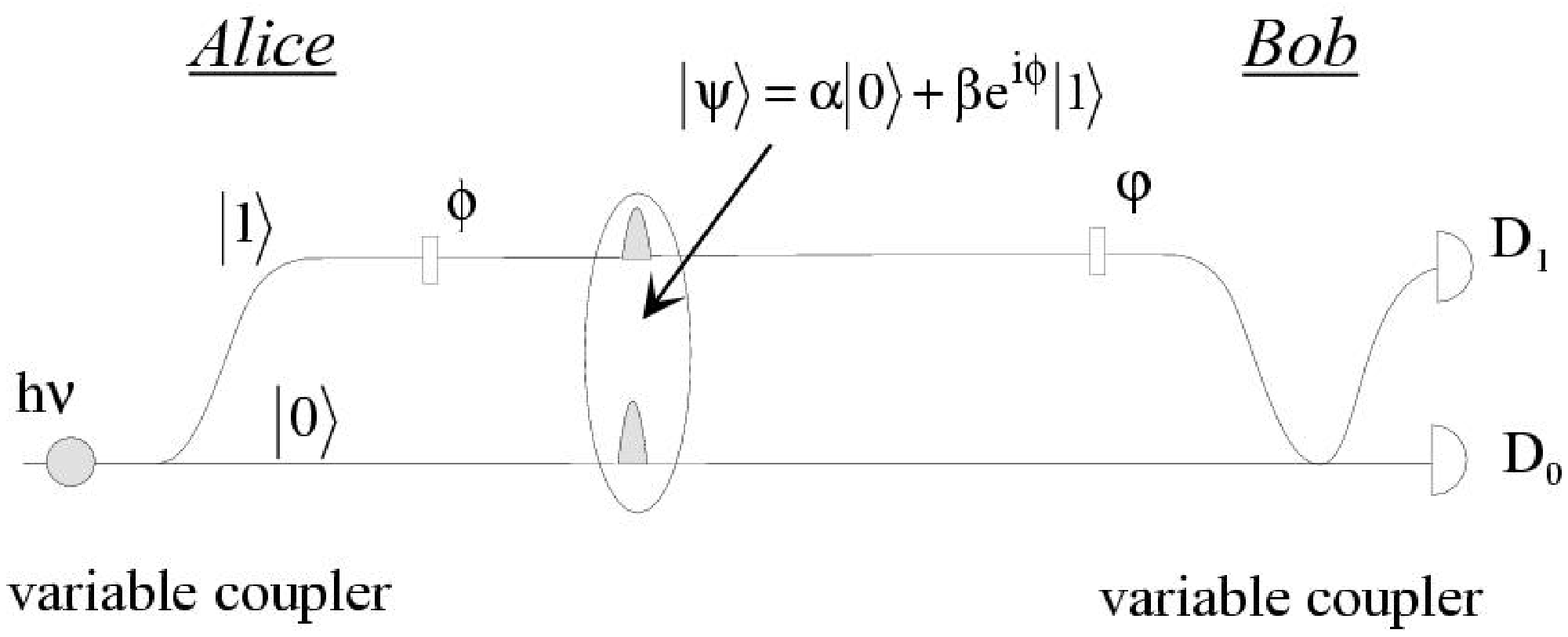}{0.7}{Creation and measurement of a spatial-mode
qubit.}

\paragraph{Time-bin qubits}

A much more robust realization of $\0$ and $\1$ in so-called
time-bin qubits
is shown in Fig.~\ref{timeqbit}. The switch at Alice's is used to
transfer the amplitudes of both spatial modes --- arriving with
time-difference $\Delta$t large compared to the photon's coherence
time (localization) --- without losses into one mode.  The net
effect is to create a superposition of amplitudes describing a
photon in two different time-bins. To undo this transformation,
Bob uses a second switch, delaying now the amplitude of the first
time-bin with respect to the amplitude of the second one so that
both arrive simultaneously at the variable coupler ---
identically to the measurement of the mode qubit. This set-up
corresponds to systems developed for faint laser-pulse based
quantum cryptography (see Section~\ref{cryptography}) by British
Telecom\cite{Townsend98a}, Los Alamos National
Laboratory\cite{Hughes00a} and, in a modified and even more robust
``plug\&play" form, by one of our groups
(GAP).\cite{Zbinden97a,Ribordy00a} The good performance of these
systems underlines the robustness of time-bin qubits with respect
to decoherence effects as encountered while transiting down an
optical fiber.

\fig{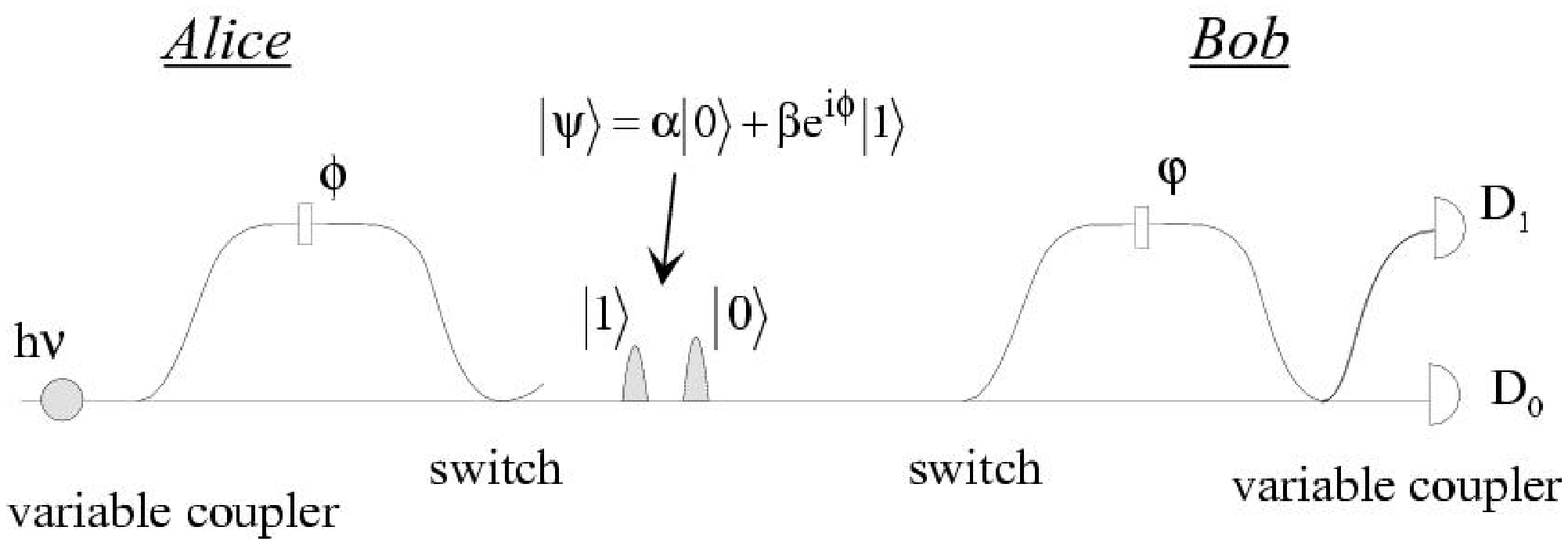}{0.7}{Creation and measurement of a time-bin qubit.}

\paragraph{Frequency qubits}

Finally, qubits can in principle be realized using a superposition
of basic states at frequencies $\ket{\omega_1}$ and
$\ket{\omega_2}$. This resembles much schemes in atomic physics
where different energy levels are used to realize a qubit.
However, the superposition of the two basic states is probably
difficult to achieve with photons, and to our knowledge, no
experiment has been reported to date. Note nevertheless that there
is related work concerning cryptography with frequency
states\cite{Molotkov98a} or based on phase-modulated light
\cite{Merolla99a}.

\paragraph{Superposition in higher dimensions: qu-nits}
\label{qunits}

All we said so far was based on superposition of two orthogonal
states. Although this is general for polarization, two dimensions
are only one possibility for superpositions of different modes,
emission times, frequencies, or orbital angular
momenta\cite{Mair01a} which are not restricted to two dimensional
Hilbert space. Fig.~\ref{qu-quart} shows the straight-forward
generalization of a time-bin qubit to a 4-dimensional
qu-quart.\cite{Bechmann00a}

\fig{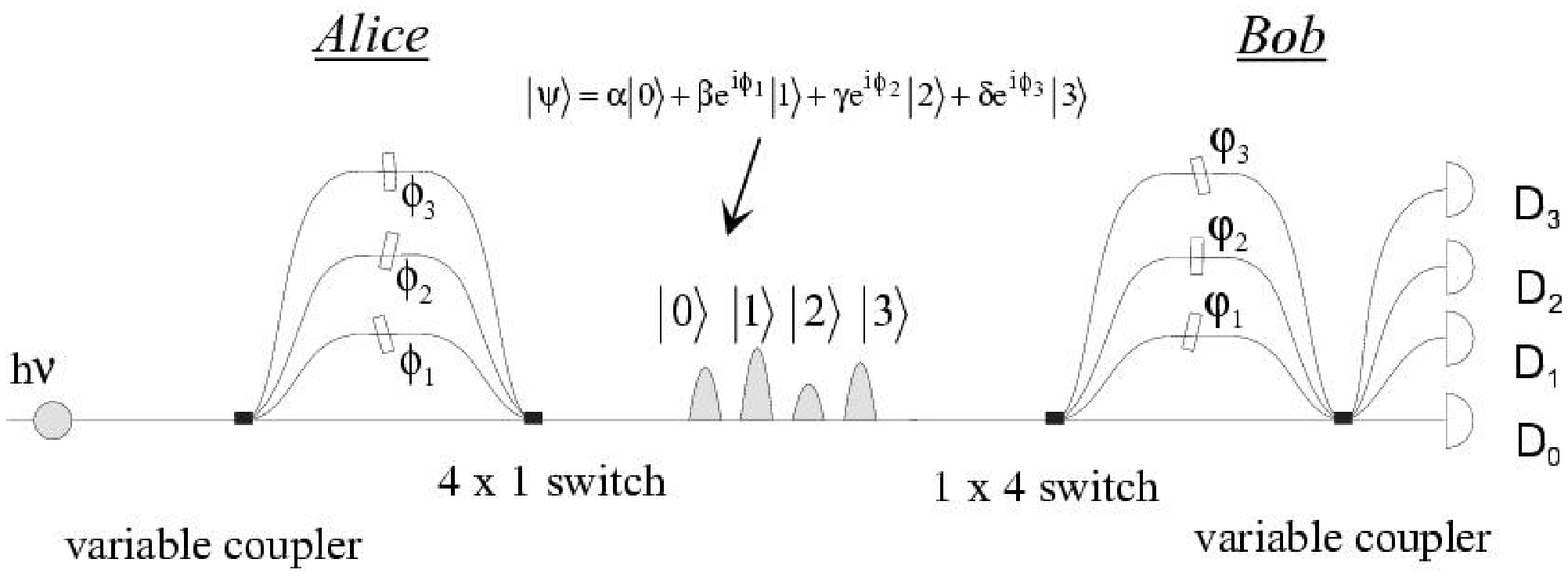}{0.8}{Generation and detection of qu-quarts.
Depending on the coupling ratios and the phases $\phi_1$ to
$\phi_3$, Alice can create any four dimensional time-bin state.
The analyzing device at Bob's is identical to Alice's preparation
device. A click in one of his detectors corresponds to the
projection on one of the four eigenstates.}


\subsection{Two-particle entanglement}
\label{entanglement}

Entanglement can be seen as a generalization of the superposition
principle to multi-particle systems, a principle which is already
at the heart of the qubit. Entangled qubits can be described by

\begin {equation}
\ket{\psi}=\alpha\ket{0}_A\ket{0}_B+\beta
e^{i\phi}\ket{1}_A\ket{1}_B \label{entanglement1}
\end{equation}

\noindent or
\begin {equation}
\ket{\psi}=\alpha\ket{0}_A\ket{1}_B+\beta
e^{i\phi}\ket{1}_A\ket{0}_B \label{entanglement2}\end{equation}

\noindent and again $\alpha^2+\beta^2=1$. The indices label the
entangled photons. For equal amplitudes $\alpha$ and $\beta$, and
$\phi=0, \pi$, Eqs.~\ref{entanglement1} and \ref{entanglement2}
reduce to the well-known Bell states

\begin {equation}
\ket{\psi^{\pm}}=\frac{1}{\sqrt{2}}\big( \ket{0}_A\ket{0}_B\pm
\ket{1}_A\ket{1}_B \big) \label{psi+-}
\end{equation}
\noindent and
\begin {equation} \ket{\phi^{\pm}}=\frac{1}{\sqrt{2}}\big(
\ket{0}_A\ket{1}_B\pm \ket{1}_A\ket{0}_B \big)  \label{phi+-}
\end{equation}

\noindent Entangled states states are states where each of the
entangled particles has no property of its own but where the
property of the global state is well defined. This becomes clear
when we look at the density matrix representations for the global
and the reduced (one particle) state. Here we give the example of
the $\psi^-$-Bell state:

\begin{equation}
 \rho_{\mbox{\scriptsize global}}=\ket{\psi^-}\bra{\psi^-}=
 \bordermatrix{
 & \ket{00} & \ket{01} & \ket {10} & \ket{11} \cr
 &0 & 0 & 0 & 0 \cr
 &0 & \frac{1}{2} & -\frac{1}{2} &0 \cr
 &0 & -\frac{1}{2} & \frac{1}{2} &0 \cr
 &0 & 0 & 0 & 0
 }
\end{equation}
and
\begin{equation}
\rho_{\mbox{\scriptsize
reduced}}=\mbox{Tr}_A(\rho)=\mbox{Tr}_B(\rho)=\mathbf{1}
\end{equation}

All properties that can be used to realize photonic qubits can be
used to create entangled qubits as well. Before focusing on the
various realizations, we will briefly present the two main types
of photon pair sources --- sources that always create photons in
pairs, however, not necessarily in an entangled state.

\subsubsection{Photon pair sources} \label{pairsources}
\noindent

\paragraph{Atomic cascades}
\label{atomiccascades}

The first sources for entangled photons were constructed using
two-photon transitions in various elements with either very
short-lived or even virtual intermediate
states.\cite{Freedman72a,Fry76a,Aspect81a} The most notable
elements used were Ca and Hg.
All these sources suffered from the common drawback that the
atomic decay with two emitted photons is a three-body process.
Therefore, the relative direction of one emitted photon with
respect to the other is completely uncertain. This reduces the
achievable collection efficiency to an extremely low value leading
to numerous experimental problems.

\paragraph{Parametric down-conversion}
\label{PDC}

When Burnham and Weinberg\cite{Burnham70a} discovered the
production of photon pairs by spontaneous parametric
down-conversion (SPDC) in 1970, they did not foresee the enormous
wealth and precision of experiments that this technique would
allow. Ou and Mandel\cite{Ou87b,Ou88a} then triggered the
extensive work on entanglement from SPDC.

Spontaneous parametric down-conversion is part of a $\chi^{(2)}$
nonlinear optical effect also known as three-wave mixing. In the
spontaneous case only one of the three interacting fields ---
usually called pump --- is initially excited. The two others are
in the vacuum state at first. Classically they would remain
unexcited but quantum mechanically there exists a small chance
that a pump photon decays into two photons which emerge within
their coherence-time, $\approx$ 100 fs, simultaneously from the
crystal.\cite{Hong87a}

The rate of this process scales linearly with the pump intensity
and the magnitude of the nonlinear coefficient. Non-vanishing
$\chi^{(2)}$-nonlinearities can only appear in non-centrosymmetric
materials. KDP (KD$_2$PO$_4$), LiIO$_3$, KNbO$_3$, LiNbO$_3$, and
BBO ($\beta$-BaB$_2$O$_4$) are a few of the most widely used
crystals.

Naturally, the process of SPDC is subject to conservation of
energy and momentum.  The latter one is also called phase matching
condition. They determine the properties of the created photons,
namely their polarization, the wavelength and the direction of
propagation (the mode). Phase matching can be achieved in
birefringent crystals if either both, or one of the down-converted
photons are polarized orthogonally with respect to the pump
photons. The first type is referred to as type I phase matching,
the second type as type II phase matching. If the birefringence of
the crystal can not be exploited, it is possible to achieve
so-called quasi phase matching in crystals where the sign of the
$\chi^{(2)}$ non-linear coefficient is periodically reversed. SPDC
is a very inefficient process; it needs around $10^{10}$ pump
photons to create one photon-pair in a given mode. Lately, two
groups demonstrated down-conversion in periodically poled lithium
niobate waveguides,\cite{Tanzilli01a,Sanaka01a} reporting
unprecedented efficiencies as high as $10^{-6}$ --- four orders of
magnitude more than what has been achieved with bulk crystals
.\cite{Tanzilli01a}

Spontaneous parametric down-conversion is possible using a
continuous as well as a pulsed pump (for problems associated to
very short pump pulses, see Ref.~\onlinecite{Kim01b} and
references therein). The latter is necessary if the creation time
of a photon pair must be determined exactly (see
Sections~\ref{GHZ}, \ref{teleportation} and
\ref{entanglementswapping}).

\subsubsection{Types of entanglement} \label{typesofentanglement}

In order to produce photon pairs in entangled states (here
entangled qubits), there must be two possible ways of creating
such a pair --- for example both photons in state $\ket{0}$ or
both photons in state $\ket{1}$. First sources where based on the
already mentioned cascaded transitions in atoms
(Section~\ref{atomiccascades}), going from a well defined state
of total angular momentum of the atom to another such state of
lower energy. The final state could be reached via two possible
ways producing polarization entangled photon pairs in its net
effect. In the following, we will omit these first realizations
and focus only on recent sources of entangled photons based on
spontaneous parametric down-conversion. However, as already
mentioned, we will neither comment on squeezed states
entanglement\cite{Ou92a,Furusawa98a} nor on entanglement of
external angular momentum states.\cite{Mair01a} Finally, problems
arising from the fact that the number of photon-pairs is (almost)
thermally distributed will be addressed only in
Section~\ref{cryptography} (quantum cryptography). Here we assume
photon-pairs in a n=1 Fock-state.

\paragraph{Polarization entanglement}

Most experiments to date have taken advantage of polarization
entanglement. The first down-conversion sources only generated
pairs of photons in product states and the entanglement was
created using some additional optics. The set-ups were rather
simple, but they suffered from necessary postselection by
coincidence measurements. Nowadays more sophisticated
configurations for down-conversion can produce polarization
entanglement directly.

\begin{itemlist}

\item Entanglement with post-selection

The first sources were so-called type-I sources, which means that
the two down-converted photons carry identical polarization.
Momentum conservation rules the emission directions such that two
photons out of an individual down-conversion process emerge on
cones centered around the pump beam. 
At the degenerate wavelength (twice the pump wavelength) both
photons will always be on the same cone opposite of each other
with respect to the pump beam.

The type-I down-conversion source can be used to produce
polarization entanglement if the polarization of one of the beams
is rotated by 90$^\circ$ before it is superposed with the other
beam on a beam-splitter\cite{Ou88a} (see Fig.~\ref{postsel}).
Post-selecting the cases where the photons exit in different
spatial modes by means of a coincidence measurement yields
polarization entanglement. An appropriate description of the
post-selected state is given by
\begin{equation}
\ket{\psi}=\frac{1}{\sqrt{2}}\left[\ket{0}_A\ket{1}_B + e^{i\phi}
\ket{1}_A\ket{0}_B\right]
\end{equation}
where $\ket{0}$ and $\ket{1}$ stand for horizontal and vertical
polarized photons. This state can easily be transformed into one
of the Bell states (Eqs.~\ref{psi+-} and \ref{phi+-}).

\fig{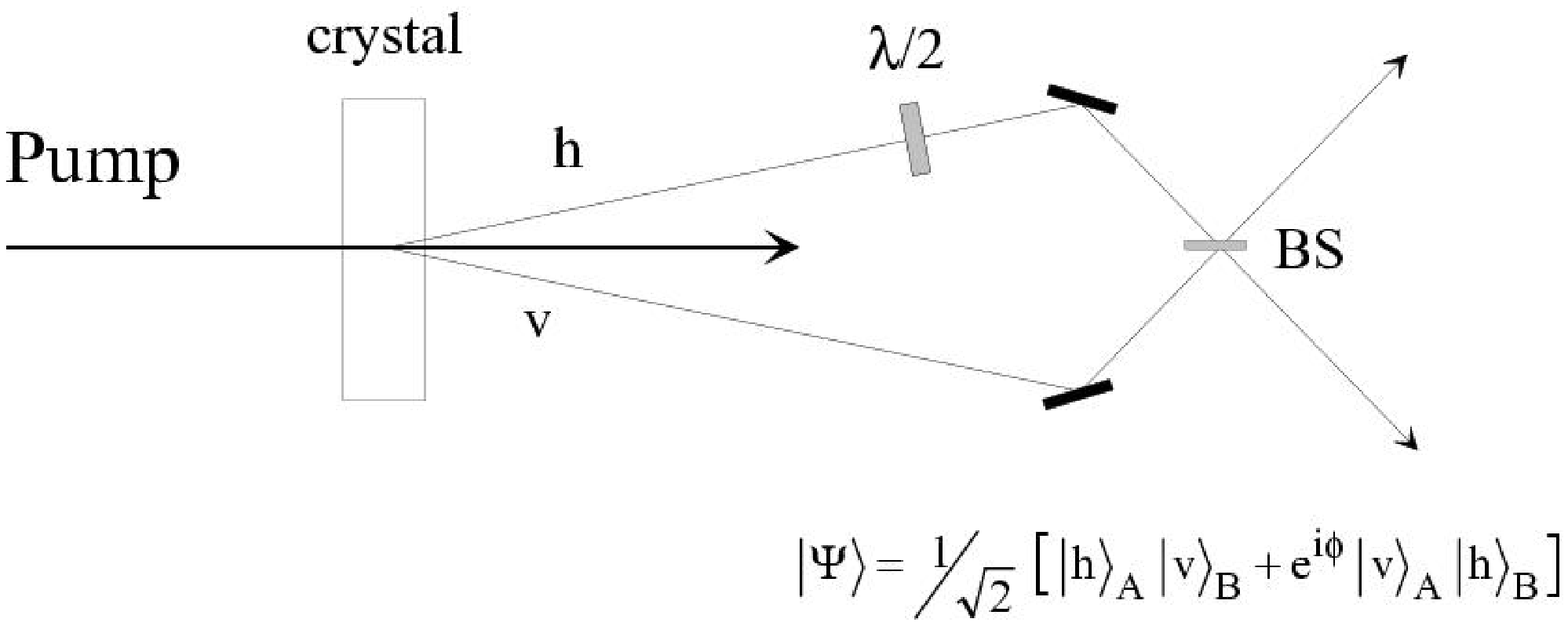}{0.7}{Schematic of a non-collinear type-I SPDC source
creating polarization entanglement after coincidence
post-selection behind a beam-splitter (BS).}

A more direct way to achieve polarization entanglement is type-II
down-conversion where the down-converted photons are polarized
orthogonally with respect to each other. In the simplest way the
crystal is cut such that all beams are collinear. The pump is then
separated by prisms or filters and the down-converted photons
forming a pair are probabilistically separated on a
beam-splitter.\cite{Kiess93a} Similar to the type-I source, this
procedure again necessitates post-selection by coincidence
techniques.

\item Entanglement without post-selection

The non-collinear extension of type-II down-conversion helps to
get rid of post-selection since the two down-converted photons are
emitted into different spatial modes already from the beginning
on. A schematic is shown in Fig.~\ref{type2}. This kind of source
has been realized for the first time by Kwiat {\it et al.} in
1995.\cite{Kwiat95b} The high degree of entanglement and
brightness (up to 400.000 coincidence counts per second  as
counted with Silicon APDs) of this source made quantum
teleportation and other more complicated quantum communication
protocols feasible (see Section~\ref{quantumcommunication}).

\fig{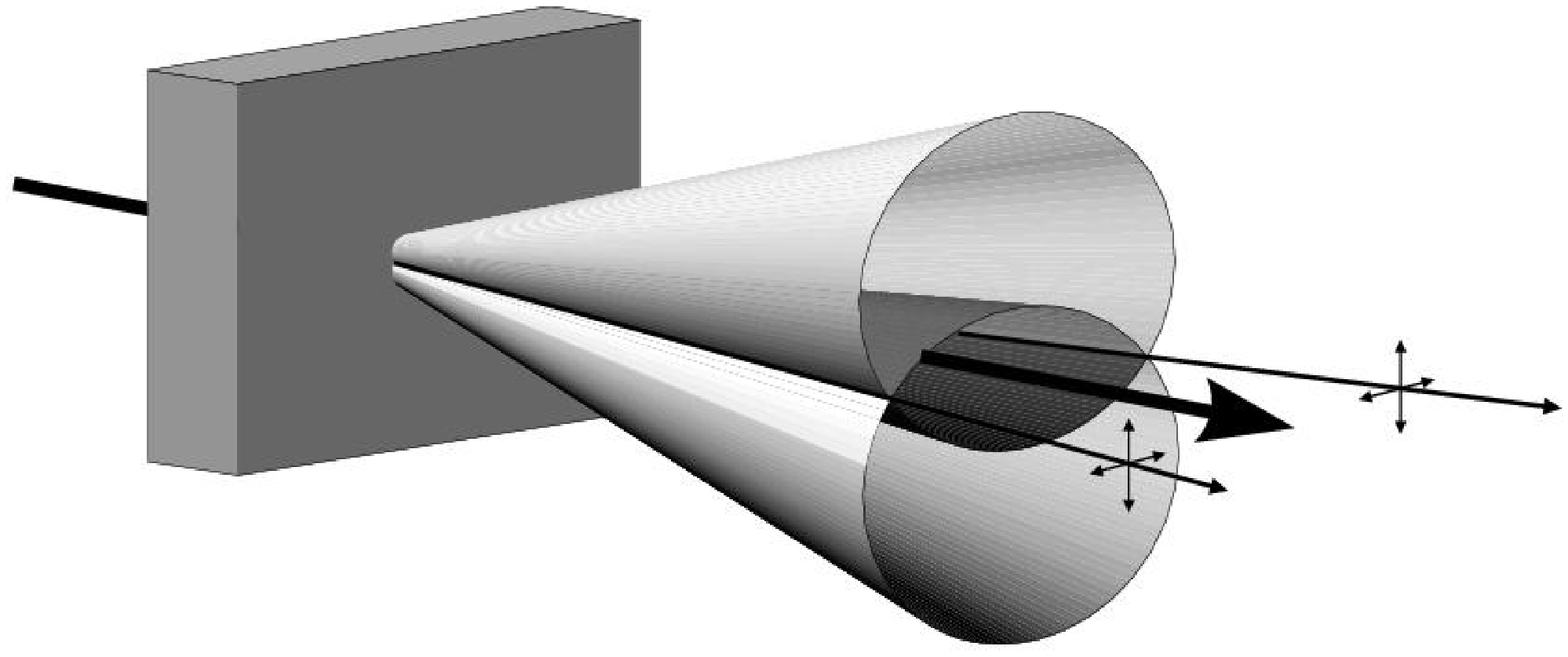}{0.7}{Schematic of non-collinear type-II parametric
down-conversion. Extraordinary (vertical, V) photons of of a
certain wavelength emerge on the upper cone, ordinary (horizontal,
H) on the lower cone. The intersections are unpolarized and
display polarization entanglement after proper compensation of the
birefringent delay incurred in the conversion crystal.}

The quest for better sources and stimulated by the success of
polarization entanglement from type-II sources led Kwiat {\it et
al.}\cite{Kwiat98a} to invent a new scheme in which they stack
two thin type-I crystals with their optic axes at 90$^\circ$ of each
other (Fig.~\ref{2crystal}). A pump photon that has linear
polarization at 45$^\circ$ with respect to the two optic axes will
equally likely down-convert in either crystal. The photons
created in the first crystal will be polarized orthogonally to the
ones created in the second crystal. If the crystals are thin
enough this leads to polarization entanglement. Furthermore, by
varying the pump polarization one can also realize non-maximally
entangled states.

\fig{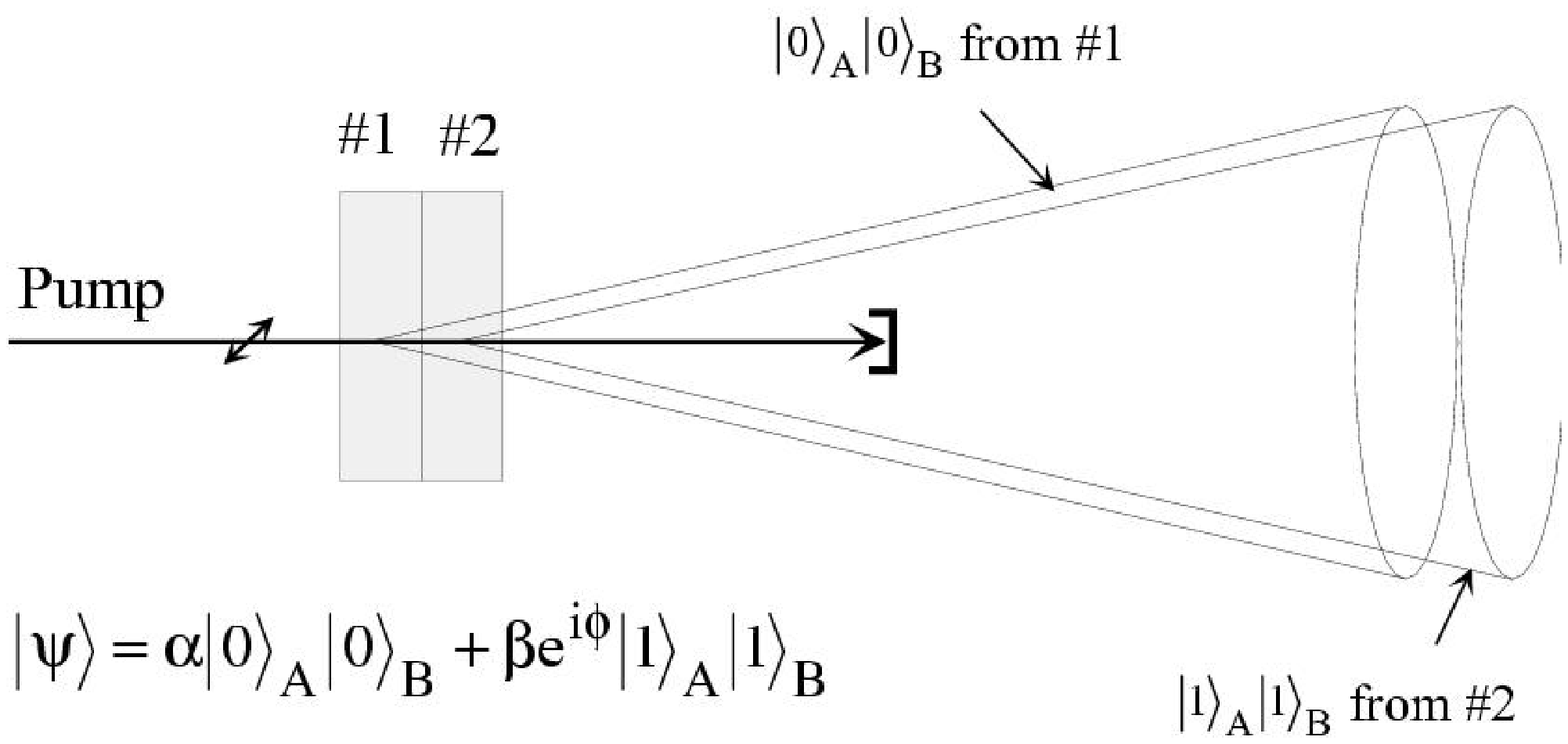}{0.7}{Polarization entanglement using two stacked
type-I down-conversion crystals with optic axis oriented at 90$^\circ$
with respect to each other. Depending on the polarization of the
pump, maximally as well as non-maximally entangled states can be
created.}

\end{itemlist}

\paragraph{Momentum or mode entanglement}

Rarity and Tapster have first realized momentum entanglement in
1990.\cite{Rarity90a} A schematic is shown in Figure
\ref{mom_all}, left-hand picture.

\fig{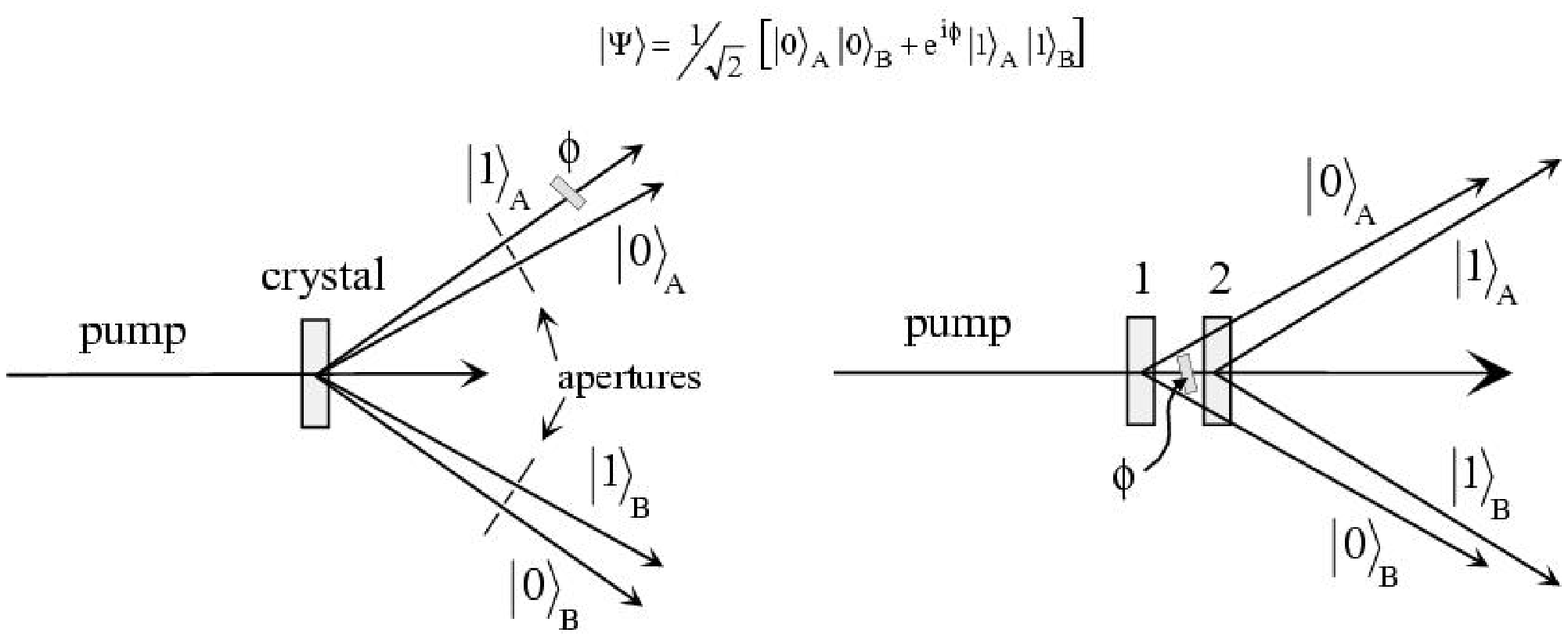}{0.9}{Schematics of two different mode-entangled
source. The phase shifter can act either locally on one of the
modes as shown in the left-hand picture (Rarity {\it et
al.}\cite{Rarity90a}), or on both modes $\1_A$ and $\1_B$ as shown
in the right-hand picture (Ribeiro {\it et
al.}\cite{Ribeiro00a}).}

From the emission of a non-linear crystal, two pairs of spatial
(momentum, direction) modes are extracted by pinholes. Due to the
phase matching conditions, photon pairs are created such that
whenever a photon at frequency
$\frac{1}{2}\omega_{\mbox{\scriptsize pump}}+\delta\omega$ is
emitted into one of the inner two modes its partner at frequency
$\frac{1}{2}\omega_{\mbox{\scriptsize pump}}-\delta\omega$ will be
found in the opposite outer mode. The momentum entanglement very
much resembles the original EPR idea of a state of two particles
whose momenta are correlated in continuous
space.\cite{Einstein35a}

Another realization has recently been published  by Ribeiro {\it
et al.}\cite{Ribeiro00a} (see right-hand side of Fig.
\ref{mom_all}). Two subsequent crystals are pumped by a laser
having a coherence length larger than the distance between the
crystals. The superposition of the two amplitudes describing a
photon pair created either in crystal 1 or in crystal 2 leads to a
mode- or momentum-entangled state.


\paragraph{Time-bin entanglement}

Using a set-up similar to the one shown in Fig.~\ref{timeqbit},
Brendel {\it{et al.}}\cite{Brendel99a} proposed and realized the
first source for time-bin entangled qubits in 2000 (see
Fig.~\ref{timeent}).\footnote{For related work, see also
Refs.~\onlinecite{Keller98a,Kim99a}.} A classical light-pulse is
split into two subsequent pulses by means of an interferometer
with a large path-length difference. Pumping a nonlinear crystal,
a photon pair is created either by pulse 1 (in time-bin 1) or by
pulse 2 (in time-bin 2). Depending on the coupling ratios of the
couplers in the interferometer and the phase $\phi$, any
maximally as well as non-maximally entangled (pure) state can be
realized --- similar to the polarization entangled source
mentioned above. Furthermore, this set-up can easily be extended
to create time-bin entangled qu-nits (see also
Section~\ref{qunits}).

\fig{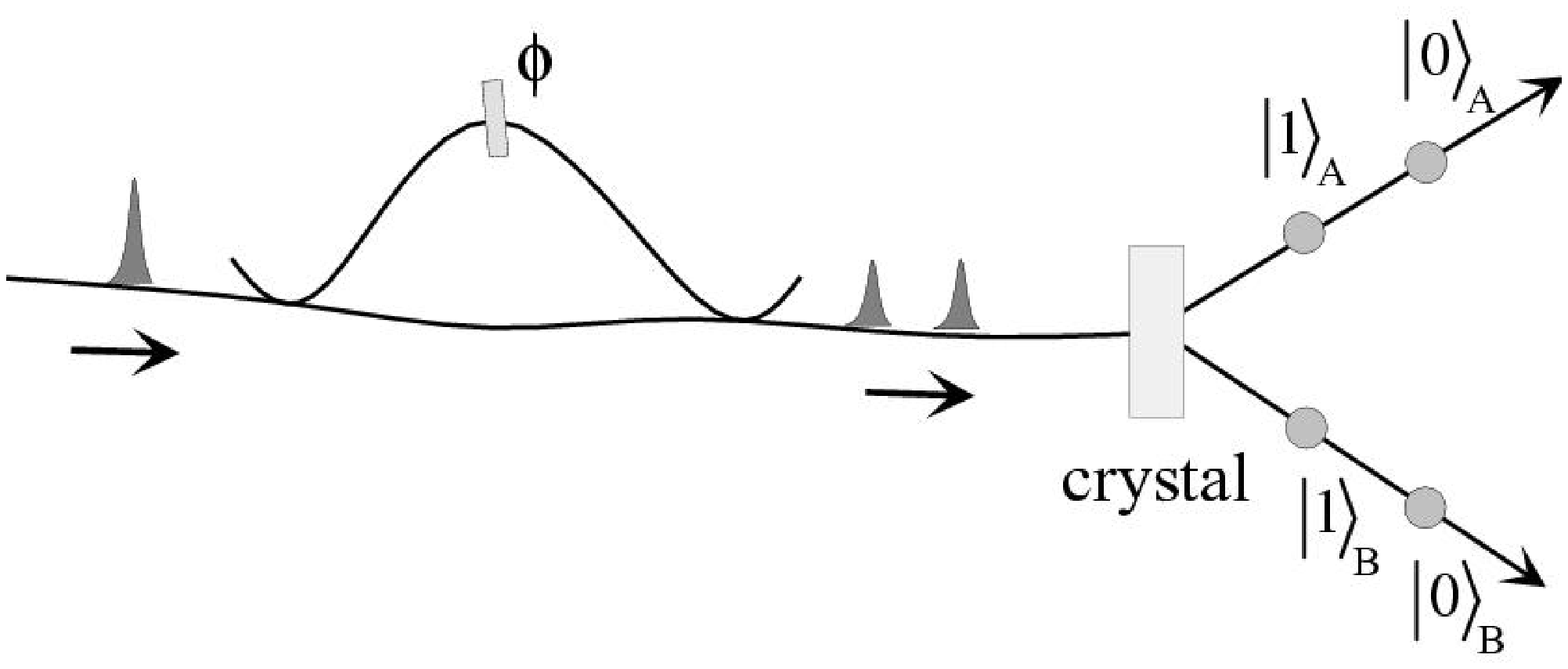}{0.7}{Schematics of a source creating time-bin
entangled qubits. Here we show a fiber-optical realization of the
interferometer.}

\paragraph{Energy-Time entanglement}

Energy-time entanglement can be considered the ``continuous"
version of time-bin entanglement, hence does not belong to the
class of entangled qubits. It has been proposed already in 1989
(long before time-bin entanglement) by Franson\cite{Franson89a}
in connection with a novel test of Bell inequalities (see
Section~\ref{Bell}). Initially, Franson considered a three-level
atomic system with a relatively long lifetime for the initial
state $\psi_1$ and for the ground state $\psi_3$, and a much
shorter lifetime for the intermediate state $\psi_2$ (see
Fig.~\ref{3level}). Therefore, the sum energy of both photons is
very well defined although the energy of each of the two emitted
photons is very uncertain. Or, in the time domain, although the
precise emission time of a pair can be predicted only to within
the long lifetime of the initial atomic state, both photons are
emitted almost simultaneously --- only depending on the short
lifetime of the intermediate state.

The first to realize and employ energy-time entanglement in terms
of Franson-type tests of Bell inequalities were Brendel {\it et
al.}\cite{Brendel92a} in 1992 and, almost simultaneously, Kwiat
{\it et al.}\cite{Kwiat93a} In contrast to the initial proposal
which is based on cascaded atomic transitions, both took
advantage of SPDC in a non-linear crystal pumped by a coherent
laser. The long coherence time of the pump-laser now bounds the
emission time of a photon pair --- equivalent to the lifetime of
the atomic state $\psi_1$ --- and the coherence time of the
down-converted photons, which can be as small as 100~fs,
determines the degree of simultaneity of the emission of the
photons.

\fig{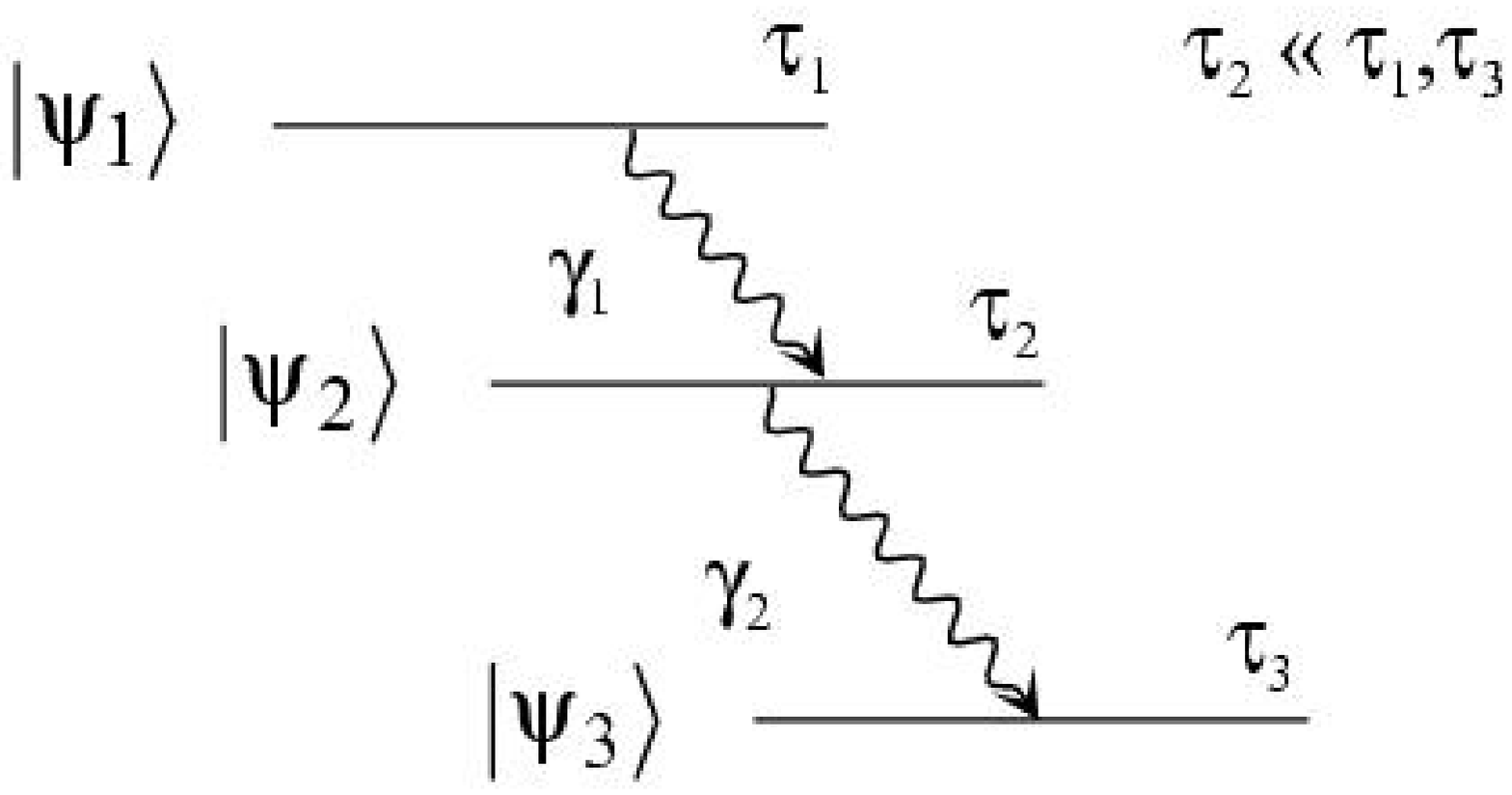}{0.5}{Schematics for the creation of energy-time
entangled photon pairs based on electronic transitions in a
three-level atom. The lifetimes $\tau_1$ and $\tau_3$ of the
initial and the final state are supposed to be large compared to
the lifetime $\tau_2$ of the intermediate state.}

\subsubsection{Measuring entanglement}
\label{measuringentanglement}

\paragraph{Projection on Bell states}
\label{bellmeasurement}

Common to many protocols of quantum communication is the necessity
to determine the state of a two particle system. For qubits, this
means to project on a basis in 4-dimensional Hilbert space,
spanned for instance by the four Bell states (Eq.~\ref{psi+-} and
\ref{phi+-}). Such a measurement is known as a Bell-, or
Bell-state measurement.

Figure~\ref{bellmeas} shows a possible set-up for such a
measurement in the cases of polarization (left hand figure) and
time-bin (right hand figure) qubits: The two particles enter a
beam-splitter (BS) from modes $a$ and $b$. Behind the
beamsplitter, each particle is subjected to a {\it{single qubit}}
projection measurement. In case of polarization qubits, this is
often a projection on the horizontal and vertical axis using a
polarizing beam-splitter (PBS),\cite{Michler96a,Mattle96a} in case
of time-bin qubits, the simplest way is to chose a projection on
time-bins by looking at the detection time.\cite{Brendel99a} If
now both photons exit the BS via different output modes ($c$ and
$d$, respectively) and are found to be projected on different
eigenstates of the analyzers --- one horizontal, one vertical, or
both with one time-bin difference, respectively, --- and there is
no possibility to know which photon entered the BS from which
mode, then the {\it{two-photon}} state is projected on the
$\psi^-$-state. If both photons exit the BS in the same mode but
are still detected in orthogonal states, then the
{\it{two-photon}} state is projected on to the $\psi^+$ state.

\fig{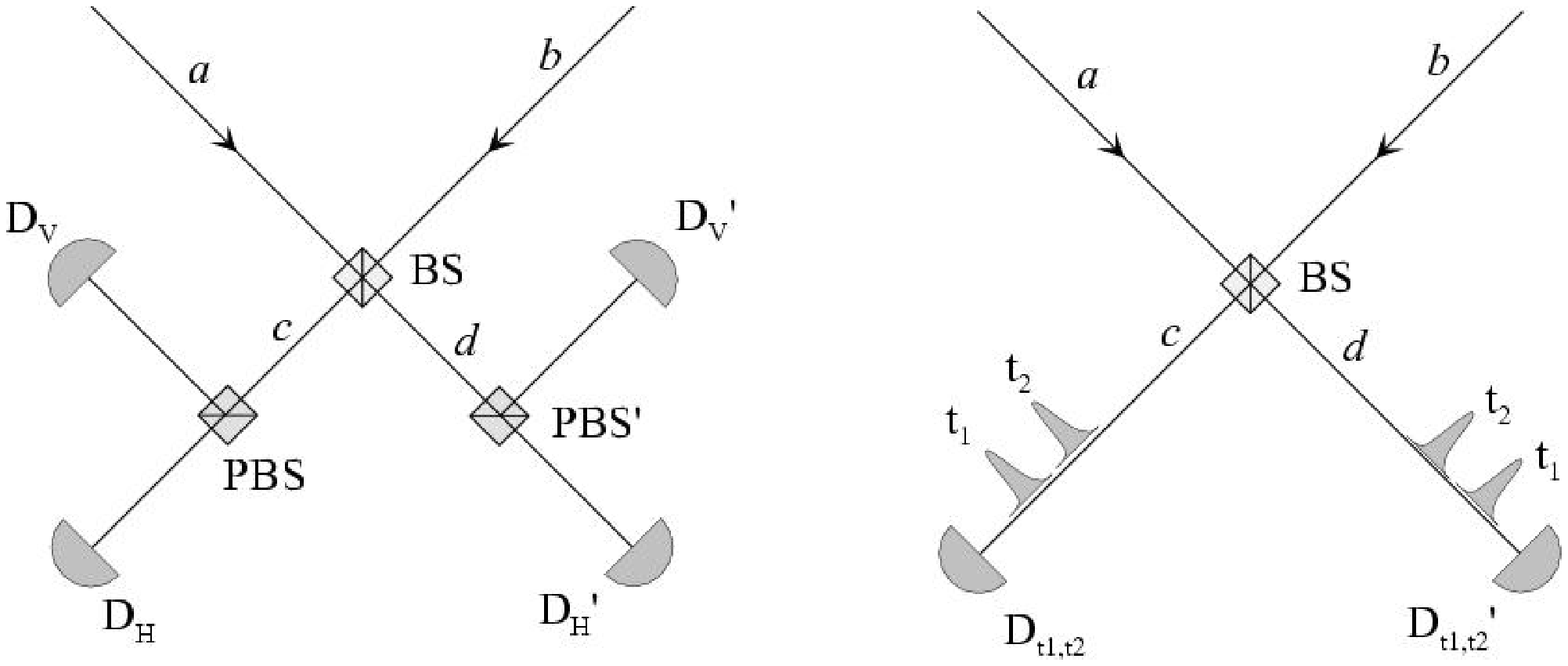}{0.9}{Set-up using linear optics for projecting two
polarization (right-hand picture) and two time-bin (left-hand
picture) qubits on a basis spanned by the four Bell states. The
two photons enter a beam-splitter (BS) via modes $a$ and $b$ and
are then subjected to a polarization-, or arrival-time
measurement, respectively. For polarization qubits, a coincidence
between detectors $D_V$ and $D_H'$ (or $D_H$ and $D_V'$)
corresponds to a projection on the $\psi^-$ state, a coincidence
between $D_V$ and $D_H$ (or $D_H'$ and $D_V'$) on $\psi^+$. For
time-bin qubits, detection of both photons in different time bins
yields a projection on a $\psi$ state: if they are found in
different spatial modes on $\psi^-$, if they leave the
beam-splitter in same spatial mode on the $\psi^+$ state ---
similar to the polarization case. Note that in both schemes only
two of the four Bell states can be distinguished.}

Surprisingly, as shown by L\"{u}tkenhaus {\em et
al.},\cite{Lutkenhaus99a,Calsamiglia01a} there is no experimental
possibility to differentiate between all of the states, at least
as far as linear optics is concerned. The best one can do is to
identify two of the four Bell states with the other two states
leading to the same, third, result. However, as theoretically
shown by Kwiat and Weinfurter in 1998, complete Bell-state
analysis is possible even with linear optics if the two particles
are entangled in another degree of freedom as
well.\cite{Kwiat98b} However, this condition can not be fulfilled
by photons that come from independent sources as required e.g. for
entanglement swapping (Section~\ref{entanglementswapping}).

Recently, Kim {\it et al.}\cite{Kim01a} performed an experiment
within the frame of quantum teleportation (see Section
\ref{teleportation}) which could lead the way to a complete Bell
measurement. The authors were taking advantage of non-linear
interactions, however, even though they used a classical input,
the efficiency was extremely small. Whereas in principle it would
be possible to extend this method to a single photon input, which
could carry any qubit realization, it does not seem feasible with
current technology.

A major problem for every Bell state measurement is to erase the
``identity" of the originally incoherent wave-functions. It must
not be possible to infer by any degree of freedom whether a
detected photon originates from a specific mode. One condition
that emerges from this criterion is that the coincidence window
should be significantly smaller than the coherence time of the
photons (``ultracoincident''). It turns out that this situation is
rather tricky to achieve in an experiment, especially if both
photons come from different sources. As can be seen in
Ref.~\onlinecite{Zukowski95a}, it involves an elaborate
application of a quantum erasure technique using mode-locked
pulsed lasers to make the coincidence window independent of any
electronic limitations while still maintaining reasonable count
rates.

\paragraph{Quantum state reconstruction}
\label{tomography}

The most general approach to measure a quantum state is the
reconstruction of its density matrix (also called quantum
tomography). In the case of two-qubit states, this has first been
investigated by White {\it{et al.}} in 1999\cite{White99a} for
polarization qubits. Obviously, this method can be generalized to
any realization of a qubit. The measuring scheme is depicted in
Fig.~\ref{tomogr}. In contrast to the Bell-state measurements
mentioned before, not only one projection has to be measured, but
the density matrix is reconstructed from the statistical outcomes
of different joint projection measurements.

\fig{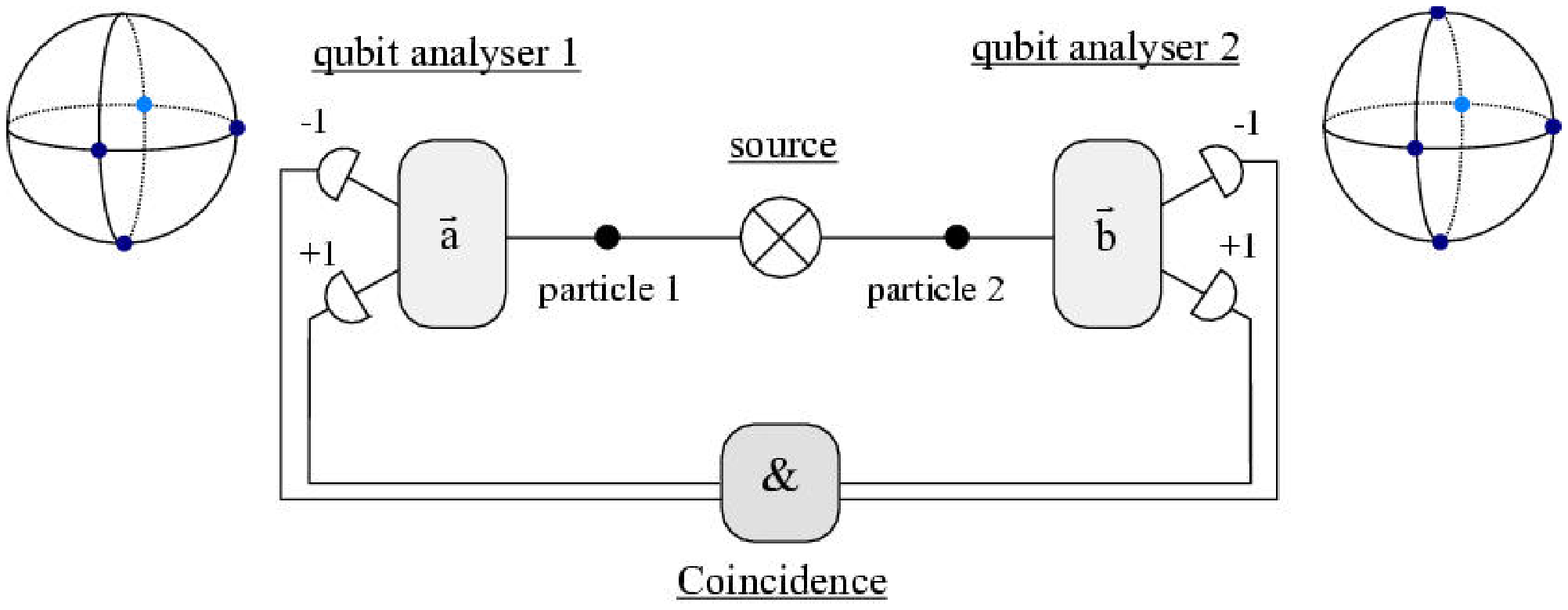}{0.9}{Schematics for reconstruction of the density
matrix of a two-qubit state. Each of the two particles is analyzed
using a qubit-analyzer. The vectors $\vec{a}$ and $\vec{b}$
specify the bases to be projected on. Note that the eigenvalues of
the different bases lie on a two-dimensional subspace within the
shell of the respective qubit-spheres.}

In general there is some freedom in the choice of the measurements
that will ultimately be performed. A density matrix for a
partially mixed state in a $d$-dimensional Hilbert space contains
$d^2-1$ independent parameters. Therefore one needs at least as
many independent measurements to be able to reconstruct such a
matrix. For a two-qubit state this would be 15 measurements.

Still, in general it is difficult to calculate the matrix elements
given only some marginal probabilities. Of course, it is possible
to just directly invert the set of measurement results but taking
into account the ubiquitous measurement errors it turns out that
direct inversion is not always suitable because it can sometimes
even lead to unphysical density matrices.\cite{Rehacek01a} A
different approach --- this was also applied to experimental data
in Ref.~\onlinecite{Kwiat01a} --- is the maximum likelihood
method, in which a suitably chosen likelihood functional is
maximized by the physical state that is most likely to having
produced the given measurement results.

\subsection{N-particle entanglement}
\label{nparticles}

With the pioneering work by Greenberger {\it et
al.}~\cite{Greenberger89a} it became clear that entangled states
of three or more particles are very important for fundamental
tests of quantum theory. Also, applications in quantum
communication like quantum secret sharing have been found that
require multi-particle entanglement.\cite{Zukowski98a,Hillery99a}

Although the interest in GHZ states is quite high, no efficient
practically usable sources have been discovered to
date.\footnote{One might speculate about a generalization of
parametric down-conversion to processes, where more than two
photons are generated. However, the corresponding non-linear
coefficients are generally many orders of magnitude lower than in
the parametric down-conversion case. As we restricted ourselves to
photonic entanglement, it would be beyond the scope of this
article to discuss many-particle entanglements that have been
produced with Rydberg atoms and ions in a
trap.\cite{Rauschenbeutel00a,Sackett00a}} As long as
deterministic quantum gates are not available we depend on
combining two-particle entangled states and projective
measurements to construct entangled states of three or more
particles by postselection.\footnote{As always, postselection
means that a state is not actually prepared at any stage in the
experiment. Still, it is possible to observe the correlations that
quantum theory predicts for the specific state.} Examples of this
technique will be presented in Section~\ref{GHZ}.

Just as in the two particle case these higher entanglements could
in principle be prepared for any degree of freedom (for time-bin
entanglement see Ref.~\onlinecite{Brendel99a}), and methods for
GHZ state analysis have been suggested.\cite{Pan98b} Still, as
long as there is no efficient way to produce these states, the
corresponding applications will probably remain in the academic
interest only. An exception is the quantum secret sharing scheme
explained in Section~\ref{paircrypto}, where GHZ state
correlations are mimicked by the correlations observed between
the pump photon and the two downconverted photons.\cite{Tittel01a}

\section{Fundamental tests of nonlocality} \label{fundamentals}
\noindent

\subsection{Bell inequalities}\label{Bell}
\noindent

Since the early days of quantum physics the intrinsic randomness
of measurement outcomes puzzled many physicists. Even very
prominent proponents believed that there should be a more
complete, ``realistic" theory complementing quantum physics with
some ``extra'' information in order to be able to describe the
observed randomness in classical terms of statistical ensembles.
This ``extra'' information has become widely known as ``hidden
variables''.\cite{Bohm52a,Bohm52b} A state of a system with a
certain set of hidden values would then be called dispersion-free
or super-pure.

The debate whether or not hidden variable theories correctly
describe nature was triggered in 1935 by the now famous paper by
Einstein, Podolsky and Rosen (EPR).\cite{Einstein35a} However,
the debate remained a purely philosophical one until 1964, when
Bell derived a statement which is in principle experimentally
testable.\cite{Bell64a,Bell87a} He started from EPR's example in
the version given 1957 by Bohm.\cite{Bohm57a} The latter
considered a gedankenexperiment where a source emits pairs of
spin-1/2 particles --- we would now call them qubits --- into
opposite directions (see Fig.~\ref{bell}). The particles are
analyzed by two independent ``qubit-analyzers": Stern-Gerlach
apparata in separate regions of space. From the idea of spatial
separation, the assumption of locality, namely that the setting of
the analyzer on one side can not influence the properties of the
particle on the other side, is inferred as a very natural
restriction for any otherwise most general hidden-variable model.
Bell found that the correlation between the two measurements as
predicted by any such model must necessarily comply with a set of
inequalities nowadays known as Bell inequalities. The most widely
used form reads
\begin{equation}
S(\mathbf{a},\mathbf{b},\mathbf{a'},\mathbf{b'}) :=
|E(\mathbf{a},\mathbf{b}) - E(\mathbf{a},\mathbf{b'})| +
|E(\mathbf{a'},\mathbf{b'}) + E(\mathbf{a'},\mathbf{b})| \leq 2,
\label{CHSH}
\end{equation}
where $E(\mathbf{a},\mathbf{b})$ is the correlation coefficient of
measurements along $\mathbf{a}, \mathbf{a'}$ and $\mathbf{b},
\mathbf{b'}$. S is sometimes called Bell-parameter and has the
meaning of a second-order correlation.

It is easily seen that a singlet state of two spin-1/2 particles
(the $\psi^-$ Bell state) will violate this inequality with
$S=2\sqrt{2}$ for a specific set of analyzer directions. We
conclude that a system governed by local hidden variables (LHV)
--- a system that can be described by a local theory --- cannot
mimic the behaviour of entangled states and hence that quantum
theory must be a non-local theory. It is interesting to note that
the set-up for testing Bell inequalities is identical to the one
needed to measure the density matrix of a two-qubit state as
explained in Section~\ref{tomography}. Only, in order to get the
full information about a quantum system, there are more analyzer
directions to measure then when testing whether the two-particle
system is correctly described by a local or by a non-local
theory.\footnote{Another approach to see whether hidden variable
theories can mimic the quantum behaviour is based on the concept
of non-contextuality. This means that the measured value of an
observable should not depend on the context, i. e. other commuting
observables, that are measured simultaneously. Kochen and
Specker\cite{Kochen67a} and independently Bell\cite{Bell66a}
showed that given an at least three dimensional Hilbert space and
starting from the assumption of non-contextuality, it is indeed
possible to prove a theorem against hidden variables. For many
years it was an open question how to experimentally test the
Bell-Kochen-Specker theorem. Last year, Simon {\it et
al.}\cite{Simon00b} proposed an experiment based on measurements
on a single particle that is in an entangled state of two
different degrees of freedom.}

Shortly after Bell's discovery it became clear that no existing
experimental data could be used to find out whether or not
non-local behaviour can indeed be observed. Furthermore it turned
out that there were still some problems in applying the inequality
to real experiments because of shortcomings of sources, analyzers,
and detectors. These shortcomings are usually called loopholes
(see Section~\ref{loopholes}) because they leave an escape route
open for local realistic theories. To arrive at a logically
correct argument against LHV one has to supplement the ``strong"
original Bell inequalities with additional assumptions which allow
to calculate the necessary correlation coefficients from the
measured coincidence count rates (see paragraph~\ref{loopholes}).
Such modified ``weak" inequalities as derived e.g. 1969 by
Clauser, Horne, Shimony and Holt (CHSH)\cite{Clauser69a}
(Eq.~\ref{CHSH}), or 1974 by Clauser and Horne,\cite{Clauser74a}
were then used in most experiments to interpret the data.

By today all these shortcomings have been eliminated, though not
all of them simultaneously in one experiment. The overwhelming
evidence suggests that indeed quantum physics accurately describes
nature and that LHV theories have been ruled out.

\fig{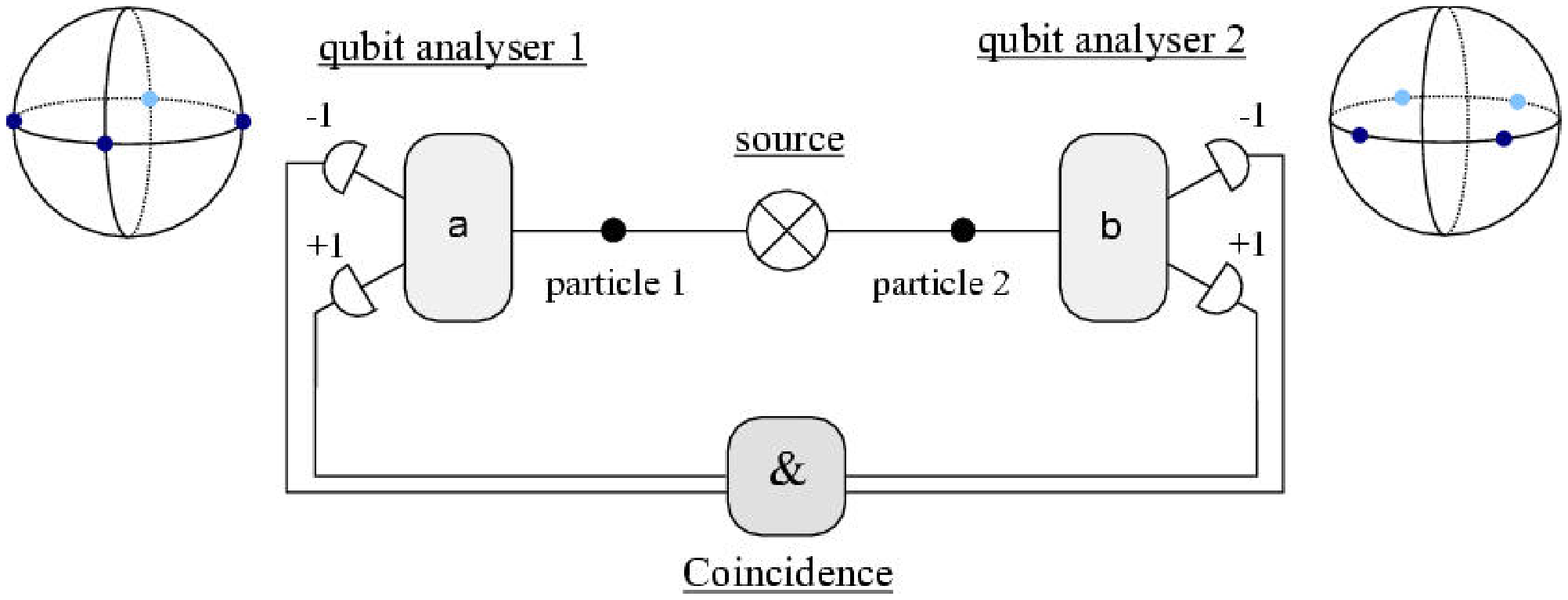}{0.9}{General set-up for Bell experiments. A source
emits correlated particles, each of which can be described in a
two-dimensional Hilbert space. The qubits then fly back to back
towards two qubit analyzers making projection measurements in
bases  defined by parameters $a$ and $b$, respectively. The
outcomes of the measurements are correlated, enabling to test via
Bell inequalities whether the two-particle system can be described
by a local or by a non-local theory.  Note that it is sufficient
to project on two different bases with eigenvalues located on a
one-dimensional subspace within the shell of the respective
qubit-spheres, in contrast to quantum tomography (Section
\ref{tomography}).}

In the following we will present a brief history of 
Bell experiments, give an account of the current state-of-the-art
and discuss the experiments that were instrumental in closing the
loopholes.

\paragraph{History}

The following table is a necessarily incomplete account of events
in the history of tests of Bell inequalities.
 More experiments can be found in
Refs.~\onlinecite{Clauser76a,Fry76a,Aspect81a,Perrie85a,Kiess93a,Pittman95a,Kwiat95b,Kwiat99a,Tapster94a,Strekalov96a,Apostolakis98a}.

\noindent
\begin{minipage}{\columnwidth}
\begin{tabular}{@{}lp{0.8\columnwidth}}

1972  &    Freedman and Clauser perform the first experimental
test based on polarization entangled photons generated via
cascaded atomic transitions, demonstrating that indeed a Bell
inequality is violated for an entangled
system and thus ruling out a local realistic description.\cite{Freedman72a} \\

1982  &    Aspect {\it et al.} measure polarization correlations
with two-channel analyzers.\cite{Aspect82b} Later they carry out
an experiment in which they vary the analyzers
during the flight of the particles under test.\cite{Aspect82a} \\

1988  &    Ou and Mandel and independently Shih and Alley do the
first non-locality experiments using parametric
down-conversion sources to create polarization entanglement.\cite{Ou88a,Shih88a} \\

1990  &    Rarity and Tapster observe momentum entanglement
in their experiment.\cite{Rarity90b} \\

1992  &    Brendel {\it et al.}\cite{Brendel92a} and a bit later
Kwiat {\it et al.}\cite{Kwiat93a} realize Franson's idea of a test
based on energy-time entanglement.\cite{Franson89a} Although the
source itself does not produce entangled qubits directly (see
Section~\ref{typesofentanglement}), the use of qubit-analyzers
post-projects on   such states. \\


1997  &    Tittel {\it et al.} show that the quantum correlations
between energy-time entangled photons are preserved
even over distances of more than 10 km.\cite{Tittel98a,Tittel98b} \\

1998  &    Weihs {\it et al.} close the spacelike separation
(Einstein locality) loophole using randomly switched analyzers.
This experiment was based on polarization entanglement.\cite{Weihs98a} \\

%

%

2001   &    Rowe {\it et al.} perform the first test violating a
strong Bell inequality. In contrast to all other tests that where
based on photons, this experiment took advantage of entangled
ions.\cite{Rowe01a}
\end{tabular}
\end{minipage}

\paragraph{Current status}

As can be seen from this table, various types of entanglement have
been used for tests of Bell inequalities. In addition, Tittel {\it
et al.}  recently  employed time-bin entangled photons for quantum
cryptography (see Section~\ref{cryptography}), an experiment that
can be interpreted as a test of Bell inequalities as
well.\cite{Tittel00a} This supports the notion of a completely
abstract formulation of Bell's gedankenexperiment in terms of two
apparata each with a variable parameter that produce output
correlated results. The generalization of the original formulation
based on Bohm's example of two spin 1/2 particles and Stern
Gerlach apparata can for instance be found in a paper by
Mermin.\cite{Mermin85a}

Presently parametric down-conversion sources (s.
Section~\ref{pairsources}) in various configurations deliver the
highest quality entangled states. The entanglement contrast can be
as high as 99.5\% while maintaining reasonable count rates. In
terms of standard deviations and coincidence count rates, some of
the most impressive violations of Bell inequalities were published
in Refs.~\onlinecite{Kwiat95b,Kurtsiefer01a}.

\subsubsection{Closing loopholes} \noindent
\label{loopholes}

As already stated above, there  are  certain loopholes in most of
the realized experiments. Here we discuss their current status
from the experimentalist's perspective

\paragraph{Detector efficiency}

Pearle\cite{Pearle70a} first noticed already in 1970 that in real
tests of Bell's inequality, which fall short of detecting all
particles that are emitted by the source, one can still construct
a hidden variable model that accurately predicts the observed data
(see also recent work by Santos\cite{Santos96a} and by
Gisin\cite{Gisin99a}). Afterwards this argument has been named
efficiency loophole because of the fact that various
inefficiencies reduce the ratio of counted to emitted particles
sometimes to less than a few percent. These inefficiencies include
incomplete collection of particles from the source, imperfect
transmission of optical elements and analyzer devices, and most
important, the far-from ideal detectors.

Violation of a Bell inequality without any supplementary
assumption requires an overall efficiency of more than 82.8\%. For
photon based experiments this is very difficult to achieve with
present photon counting technology. The threshold can be reduced
to 67\% by the help of non-maximally entangled
states.\cite{Eberhard93a} In both cases the experimental
visibility must be perfect. Although in principle there exist
detectors which have high enough efficiency, it is practically
very unlikely that a photonic Bell experiment could achieve these
efficiency levels. To date, all these experiments invoked
so-called ``fair sampling''\cite{Clauser69a} and
``no-enhancement'' assumptions\cite{Clauser74a} which allow to
derive inequalities into which only coincidence count rates
enter. However, very recently a beautiful experiment based on two
entangled ions in a microscopic trap succeeded in violating a
strong Bell inequality at nearly 100\% efficiency.\cite{Rowe01a}
Many physicists consider this a closure of the discussed
loophole.\cite{Grangier01a}

\paragraph{Separation}

Spatial separation is an important ingredient in the derivation of
Bell's inequality. The lack of large spatial separation does not
by itself constitute a loophole but it was said that with
increasing distance the quantum correlations could
diminish.\cite{Furry36a} Therefore it is natural to try to extend
the range of proven quantum phenomena to as large a distance as
possible. In addition, the fact that, and the question how
entanglement can be maintained over large distances is very
important for quantum communication protocols like quantum
cryptography (see also Section~\ref{distillation}).

\fig{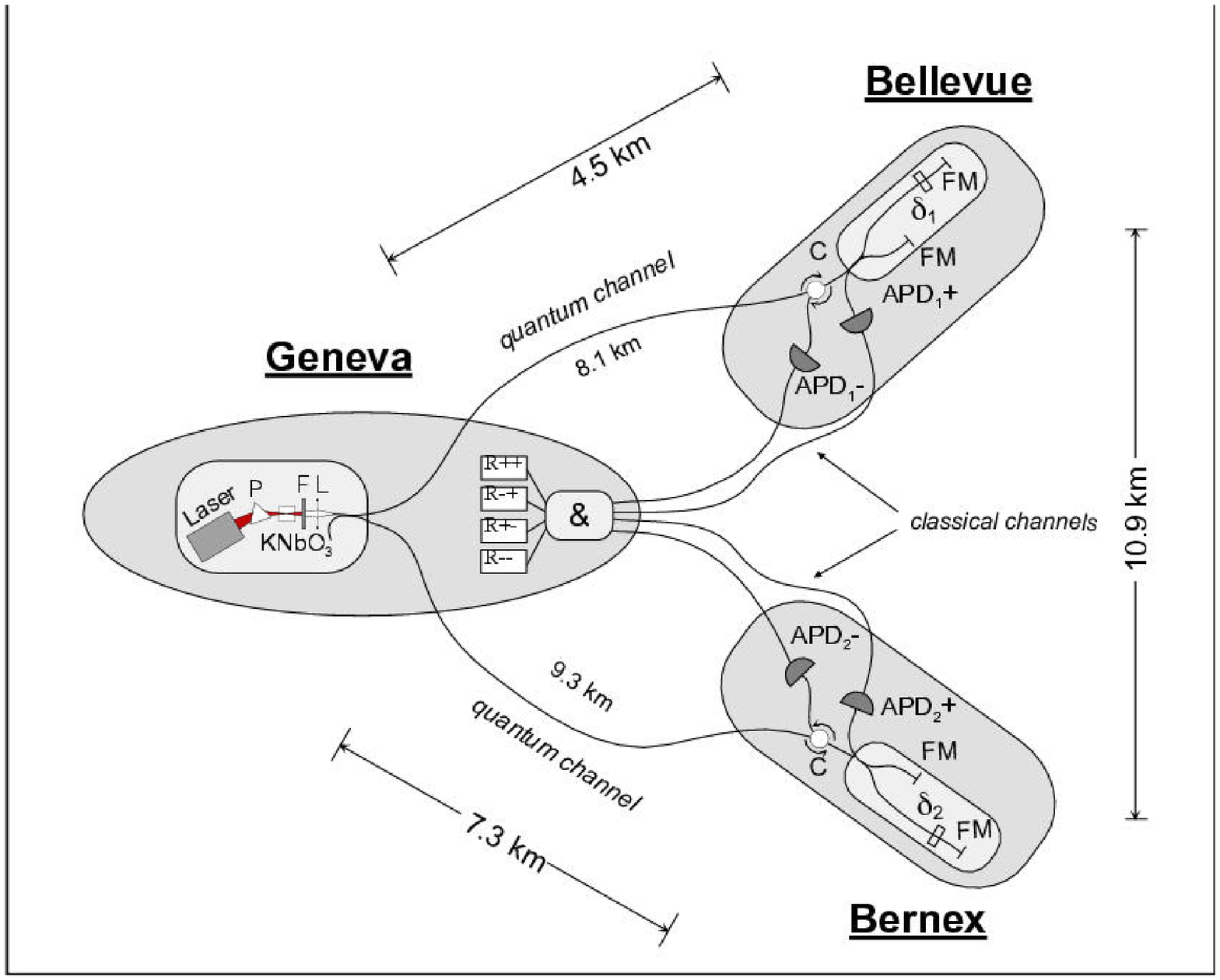}{0.8}{Experimental arrangement for a test of
Bell's inequality with measurements made more than 10~km apart.
Source (in Geneva) and observer stations (interferometers in
Bellevue and Bernex, respectively) were connected by a fiber optic
telecommunications network.}

The largest spatial separation has been achieved in tests that
were carried out by one of our groups (GAP) in  1997  and
1998.\cite{Tittel98a,Tittel98b,Tittel99a} The experiments utilized
energy-time entanglement created by parametric down-conversion at
a wavelength of 1310 nm, suitable for transmission in standard
telecommunication optical fibers. Clear violations of Bell
inequalities of up to 16 standard deviation were achieved for
measurements that were more than 10~km apart (see
Figs.~\ref{10kmBell_setup} and~\ref{10kmBell_results}).

\fig{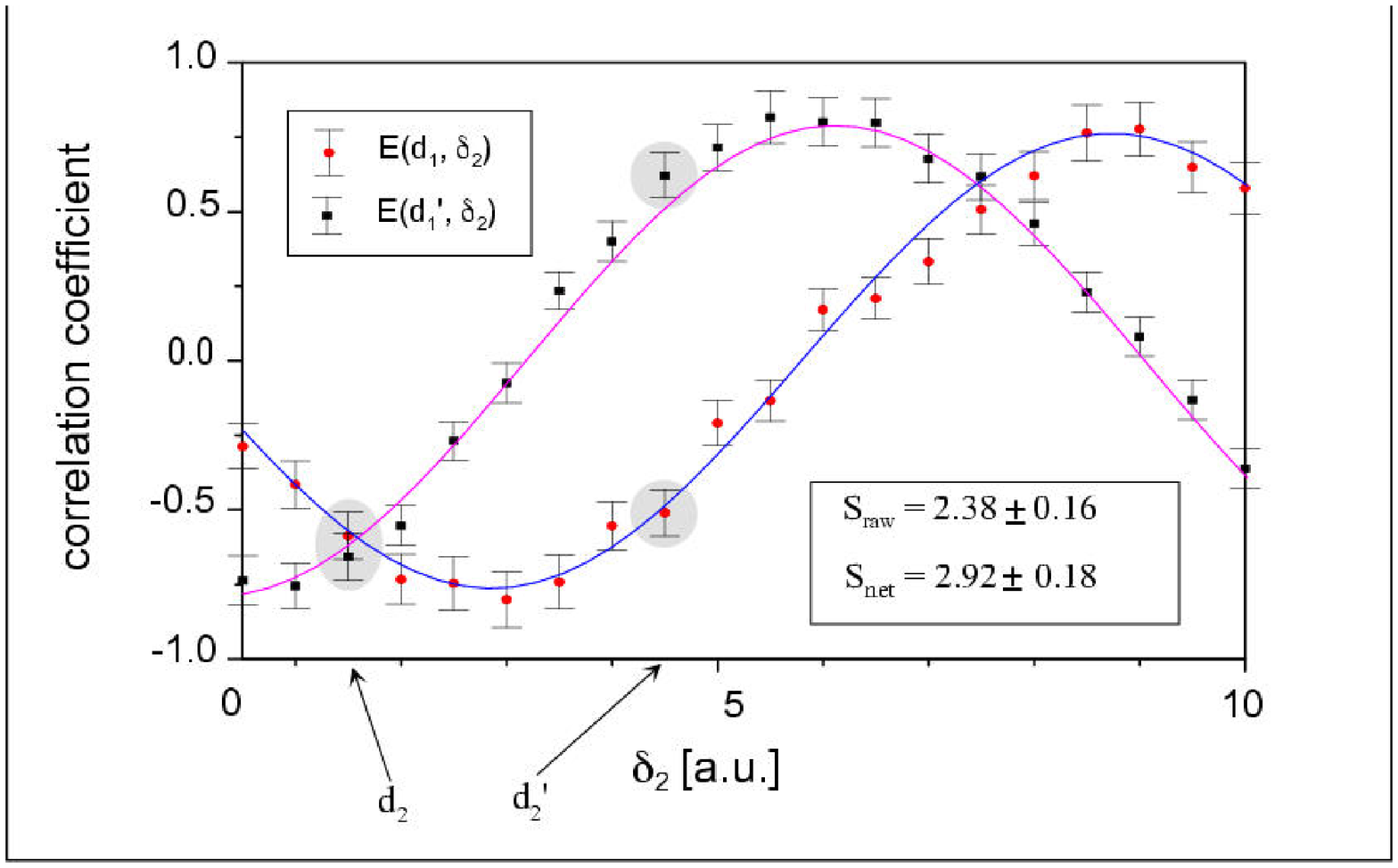}{0.7}{Data from the long distance Bell test.
 The two curves show the correlation coefficients for two
different analyzer settings at Alice's while varying the setting
at Bob's. A clear violation of the CHSH inequality is observed.}

\paragraph{Einstein locality}

The other prominent loophole in experiments on Bell's inequality
 is  called the ``spacelike separation'' or ``Einstein
locality'' loophole.\cite{Santos95a} It is constituted by the fact
that for most experiments it was possible to explain the observed
correlations by a hypothetic subluminal (slower than or equal to
the vacuum speed of light) link between the two particles or
apparata and particles. Given such a link it would in principle be
possible that the analyzer direction or even the measurement
outcome is communicated to the other side.

\fig{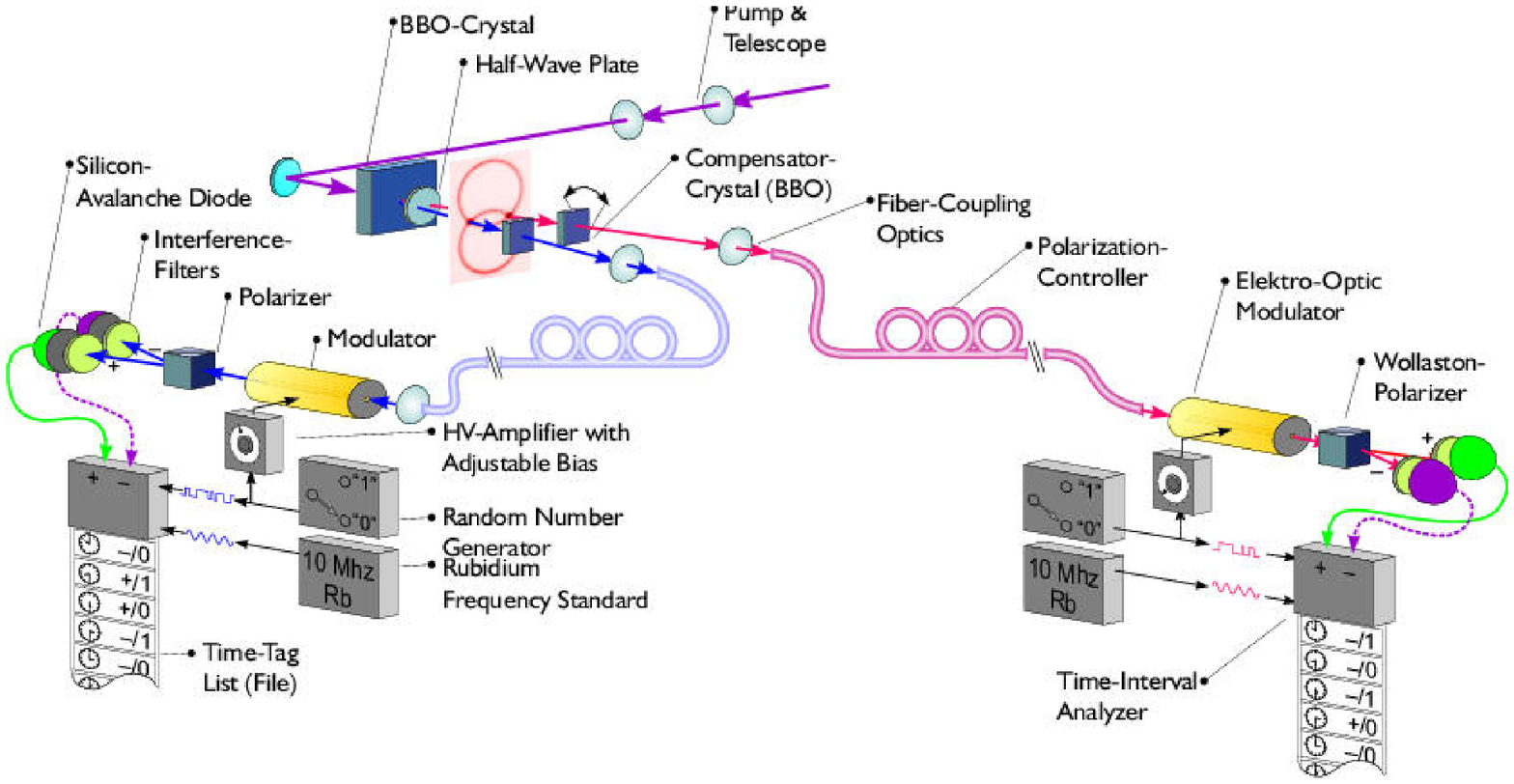}{1}{Experimental set-up for a test of Bell's
inequality with independent observers. Data are collected locally
at the observer stations only and can be compared after the
measurement is completed.}

Bell himself considered this fact as being important, calling it
``the vital time factor''.\cite{Bell81a} However, operationally
and theoretically it is hard to define this ``time factor''. There
is no criterion that is generally agreed upon. Bohm talked already
in 1957 about variation of the analyzers while the particles are
in flight.\cite{Bohm57a} Over the years the idea emerged that it
would be necessary to vary the analyzers in a random way, where
the randomness would have to be drawn from local sources or from
distant stars in opposite directions of the universe. Including
the generation of randomness and other delays the measurements
should then be completed within a time that is short compared to
the time it takes to signal to the other observer station. This
prescription amounts to performing the measurement in spacelike
separated regions of spacetime.

Obviously these are extremely vague concepts and therefore it is
not astonishing that only two experiments have tried to pin down
and answer these questions. The first one was performed by Aspect
{\it et al.}\cite{Aspect82a}  in 1982  and included a periodic
variation of the analyzers. Because periodic functions are in
principle predictable it has been said that this experiment was
not definitive in closing the spacelike separation loophole.

 In 1998, 16 years later, one of our groups (UNIVIE) was able
to include the randomness factor and, at the same time, to extend
the spatial separation to 360~m yielding a large safety margin for
the spacelike separation issue.\cite{Weihs98a} The experimental
set-up is shown in Fig.~\ref{bellsetup}. It yielded a violation of
Bell's inequality by 30 standard deviations.

Another approach to attack the locality loophole has been
demonstrated by Tittel {\it et al.} in 1999.\cite{Tittel99a} In
this experiment, two analyzers with different parameter settings
were attached to each side of the source, and the random choice
was done by a passive optical coupler, sending the photons to one
or the other analyzer. Hence, in contrast to the experiment
mentioned before, the randomness does not come from an external
random number generator but is engendered using the photons
themselves and might therefore seem ``less
good".\footnote{However, note that the borderline between ``good"
and ``bad" randomness is very vague.\cite{Gisin99b}}

\fig{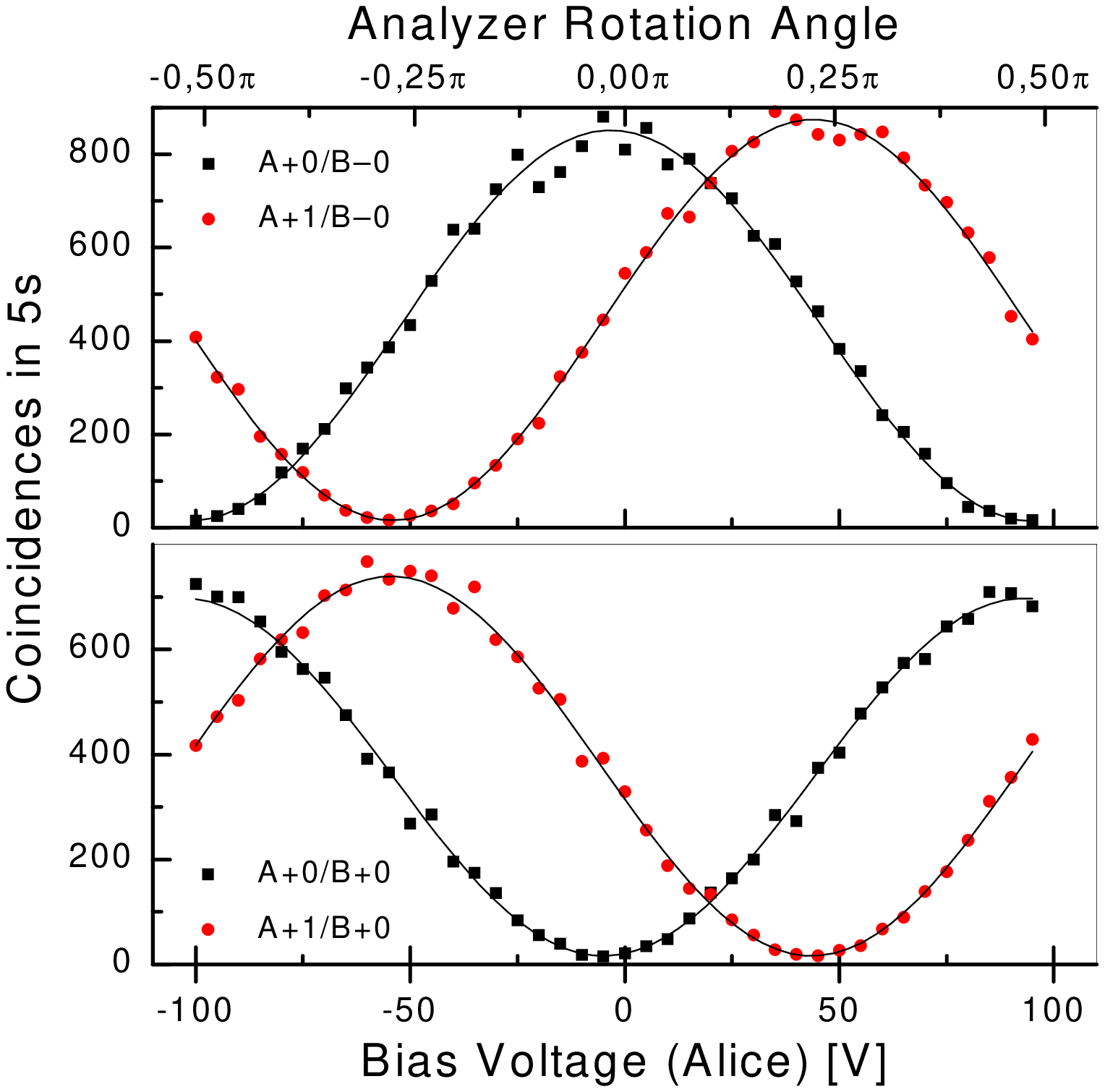}{0.7}{Correlation curves taken for spacelike
separated measurements on a polarization entangled photon pair at
a distance of 360~m. The measurements yield a violation of Bell's
inequality by 30 standard deviations.}

\subsubsection{Relativistic configurations}
\label{relativistic}

To find the quantum mechanical predictions for the results of
Bell-type measurements, one can think of the first measurement
(traditionally at ``Alice's") as a non-local state preparation for
the photon traveling to the second analyzer (at ``Bob's") --- like
in entanglement based quantum cryptography (Section
\ref{paircrypto}): In a first step one calculates the
probabilities for the different outcomes of Alice's measurement
that depend only on the setting of her analyzer and the
{\it{local}} quantum state.\footnote{The local state is obtained
by tracing over the distant system. In case of a maximally
entangled global state, it is completely mixed and the outcome of
Alice's measurement is completely random.} Knowing the
{\it{global}} two-particle state enables in a second step to
calculate Bob's {\it{local}} state. In a third step, equivalent
to the first one, one calculates the probabilities of the
different outcomes of the measurement at Bob's (which are now
joint probabilities), determined by his local state --- hence by
the setting of the first analyzer and the specific two-particle
(global) state used
--- and by the setting of the second analyzer.

Many people believe that the state vector is not endowed with
reality but that it is only a mathematical tool that helps to
calculate the statistical outcomes of experiments. Consequently,
the reduction of the state of Bob's subsystem by Alice's
measurement must be understood as an instantaneous modification
of the {\it{knowledge}} of an observer of the first measurement
concerning the quantum state of the second particle. Indeed,
possessing only the second system, it is impossible to see any
change as a result of the first measurement: The density matrix
describing the local state remains unchanged and thus there is no
possibility for superluminal signaling and hence no contradiction
with special relativity.\cite{Ghirardi80a} However, it has never
been  proven experimentally whether the state vector does indeed
not represent reality and that the collapse only changes the
{\it{knowledge}} of the observer, or whether the state vector is
real and its collapse has to be considered a {\it{real physical
phenomenon}} as assumed e.g. by Ghirardi, Rimini and
Weber.\cite{Ghirardi86a}

\paragraph{The speed of quantum information.}

If one assumes the collapse to be real, it is natural to ask how
fast it propagates from Alice's to Bob's subsystem. The lower
bound of this ``speed of quantum information" (or, following
Einstein's words, the speed of the ``spooky action at a distance")
can easily be calculated from the distance between both analyzers
(the distance that has to be traveled) and the time left for the
second particle until reaching its analyzer. However, before doing
so, one still has to define which parts of the analyzers are
considered crucial for the alignment. The most natural choice
seems to be the detectors where the transition from quantum to
classical takes place, although there are other possibilities as
well.\cite{Bohm51a,Suarez97b} It is assumed in the following that
the important parts are indeed the detectors. From standard Bell
experiments, one already knows that quantum information propagates
faster than with the speed of light $c$. However, these
experiments have not been devised in order to investigate the
lower bound, and the precision of the alignment is not discussed
in most works. In 2001, Zbinden {\it et al.}\cite{Zbinden01a}
reported on an experiment performed again with analyzers separated
by more than 10 km (similar to Fig.~\ref{10kmBell_setup}) where
the fibers connecting the source with the detectors, this time
each of 10 km length, were aligned such that the arrival time
difference was smaller than 5 ps. This allowed to set a lower
bound of the speed of quantum information of $2/3\times 10^7 c$
as seen from the laboratory (Geneva) reference frame. Obviously,
this is not the only possible choice of a reference frame but one
can argue that this is a very natural one.

The speed of quantum information is very important for a class of
problems that can be labeled ``the search for a covariant
description of the measurement process" with the example of the
impossibility of a causal description of the instantaneous
collapse in an EPR experiment that would be valid in all frames.
The introduction of a preferred frame PF that would allow a
realistic (obviously non-local) description of the quantum
measurement is a way out of these problems.\cite{Hardy92a}
However, the introduction of a PF is still an intellectual tool
(or trick)  and there is no experimental evidence to support this
hypothesis. A good candidate for such a preferred frame is the
frame from which the cosmic microwave background radiation (CMB)
is seen to be isotropic. Analyzing the same data and taking into
account the relative motion between the Geneva and the CMB frame,
Scarani found a lower bound of the speed of quantum information
of $2\times 10^4 c$ as seen from the CMB frame.\cite{Scarani00a}
 Repeating this experiment with different alignments, each
corresponding to simultaneous measurements in a different frame,
it would be possible to test whether there is one (the preferred)
frame from which the speed of quantum information is seen to be
limited.\cite{Eberhard01a}

\paragraph{Moving detectors in different reference frames.}

The experiment establishing the lower bound of the speed of
quantum information has been extended\cite{Zbinden01a} to perform
a first test of relativistic non-locality (RNL)(or
multisimultaneity) --- a theory unifying non-locality and
relativity of simultaneity that was proposed in 1997 by Suarez and
Scarani.\cite{Suarez97b} RNL predicts that the quantum
correlation should disappear in a setting where both analyzers
are in relative motion such that each one in its own inertial
frame is considered to cause the collapse. In the initial
proposal, the crucial part of an analyzer was assumed to be the
last beamsplitter. In the experiment, it was supposed again that
the alignment has to take into account the positions of the
detectors. Although many assumptions concerning the nature of a
detector had to be made in order to make this experiment
feasible\footnote{For example, it was assumed that a black
absorbing surface is sufficient to engender the collapse of the
wave function: A detected photon can {\it{in principle}} be
observed, e.g. via increase of the temperature of the black
surface. This is similar to the fact that it is sufficient to
observe distinguishability {\it{in principle}} to make the
fringes in an interference experiment disappear.}, it was
possible for the first time to test an interpretation of quantum
mechanics. The data always reproduced the quantum correlations
regardless the motion of the detectors, yielding thus
{\it{experimental}} evidence that the tested version of RNL does
not correctly describe nature, and making it more difficult to
consider the collapse of the wave function as a real phenomenon.

\subsection{GHZ states and nonlocality} \label{GHZ}
\noindent In 1989 (G)reenberger, (H)orne, and
(Z)eilinger\cite{Greenberger89a} found that entangled states of at
least three quantum systems can exhibit contradictions with local
realistic models that are much more striking than those found for
two particles in violations of Bell inequalities: Not even the
perfect correlation predicted by quantum mechanics for coincidence
measurements between more than two particles can be described by
realistic models. The basic set-up for such a GHZ-type test is
shown in Fig.~\ref{GHZsetup}. While it became clear later, that,
from a fundamental point of view, there is no difference between
GHZ- and Bell-type tests of non-locality, at the conceptual level
the so-called GHZ correlations remain a remarkable feature of
quantum physics.

\fig{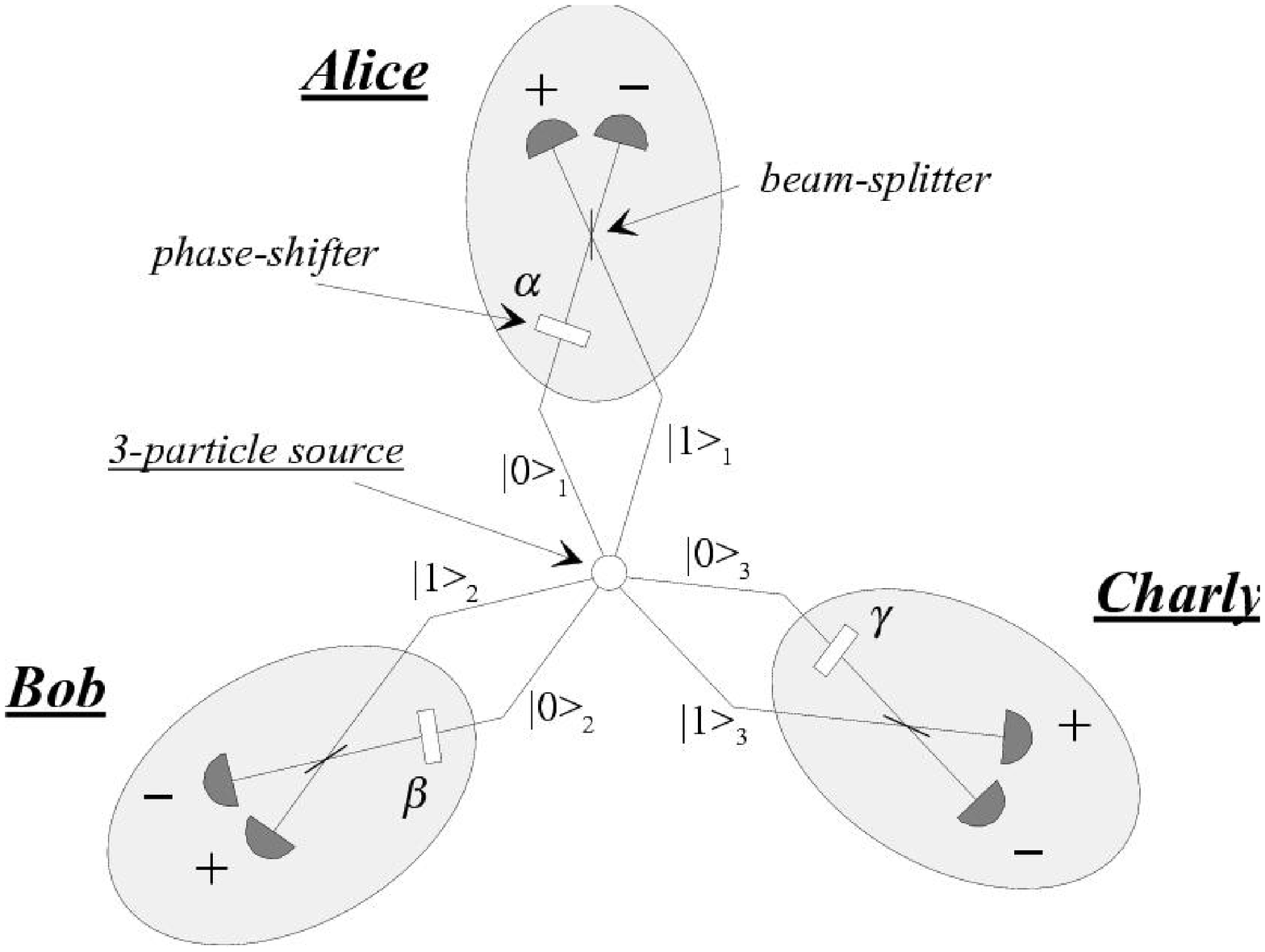}{0.7}{Schematics for a test of 3-particle GHZ-type
non-locality using momentum entangled states. The three qubits are
each send to an analyzer. Coincidence measurements in identical or
orthogonal bases enable to test whether the three-particle state
is described by a local, or by a non-local model. Note that, in
contrast to Bell-type tests of non-locality, only settings
leading to deterministic outcomes --- either perfect coincidences
or perfect anti-coincidences --- are required.}

Naturally, people have tried to construct such states and
investigate their properties but as there are no efficient and
controllable natural sources for three or more photon states the
researchers had to resort to techniques allowing to construct
higher dimensional entanglement from two-particle entangled states
(see Section \ref{nparticles}).

\paragraph{Generation of three-photon entangled states.}

In 1999 Bouwmeester {\it et al.}\cite{Bouwmeester99a} reported the
first observation of three-particle entanglement. The team
employed two photon pairs produced in non-colinear parametric
downconversion from a pulsed UV pump beam and combined them via
beam-splitters to a conditional three particle-entangled state.
The schematic (Fig.~\ref{3ghz}) shows that whenever a trigger
photon is received and simultaneuosly photons are observed in
modes 1--3, in subsequent correlation measurements (not shown) one
will observe results which can only be described by the state
\begin{equation}
 \frac{1}{\sqrt{2}}\ket{H}_T\left[\ket{HHV}_{123}+\ket{VVH}_{123}\right].
\end{equation}
The fact that such a state exists must be considered a
manifestation of quantum non-locality.

\fig{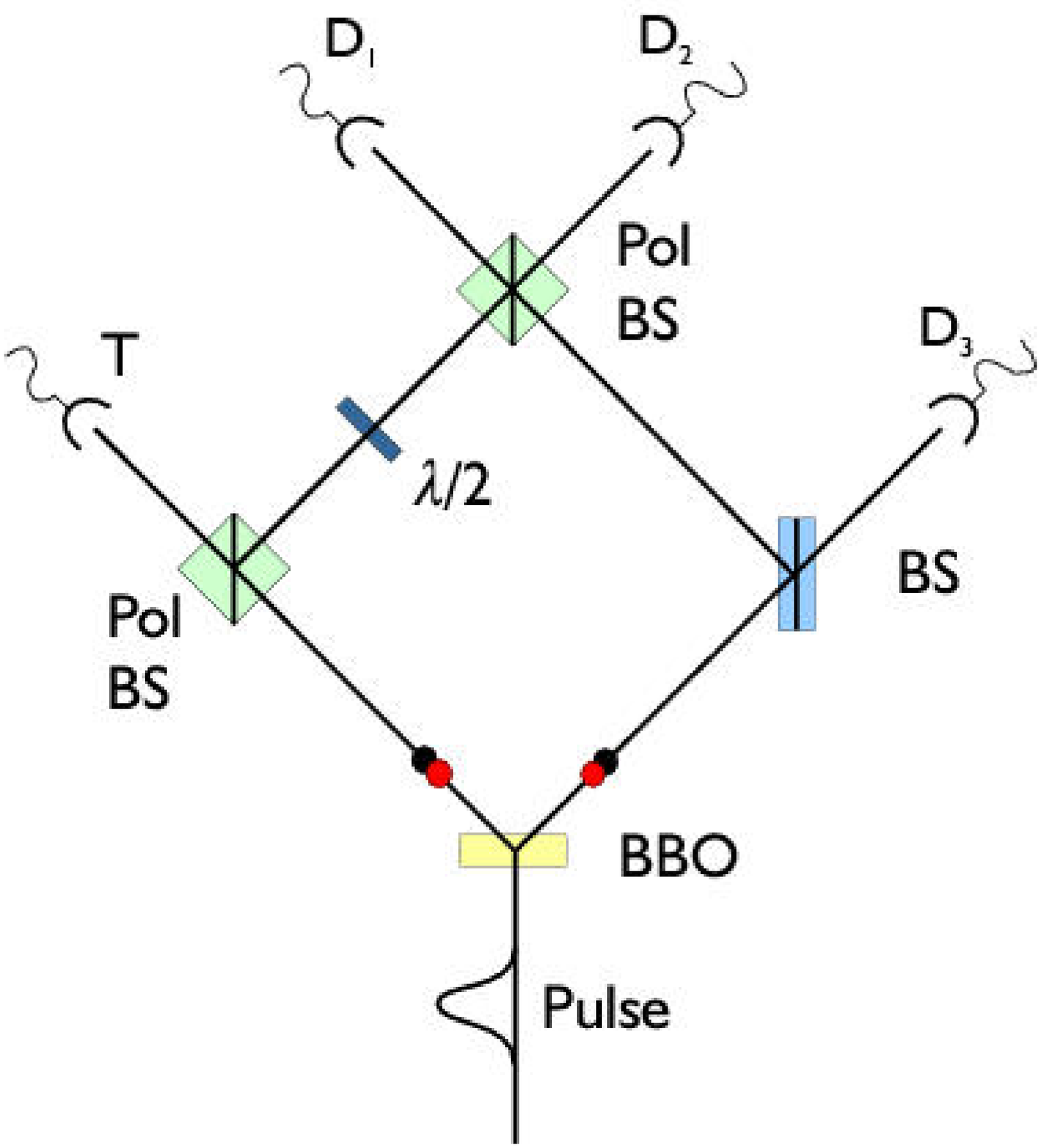}{0.5}{Schematic of the set-up used by Bouwmeester
{\it{et al.}} to produce three-photon GHZ correlations. Two
independent pairs are generated simultaneously and one photon from
each pair is send ``right", one ``left". By action of the left
polarizing beam-splitter (PBS), only horizontally polarized
particles can reach the trigger detector T. To register three-fold
coincidences between detectors 1, 2, and 3, the other photon must
be reflected from that PBS, i.e. it must be vertically polarized.
This photon is subsequently rotated by 45$^\circ$ and can end up as V
in detector 1 or as H in detector 2. The only two possible ways
that a triple coincidence event arises in detectors 1, 2, and 3
are therefore when the right two photons split at the right
beam-splitter and when the two photons meeting at the upper PBS
have are found to have identical polarization --- either both
horizontal or both vertical.}

\fig{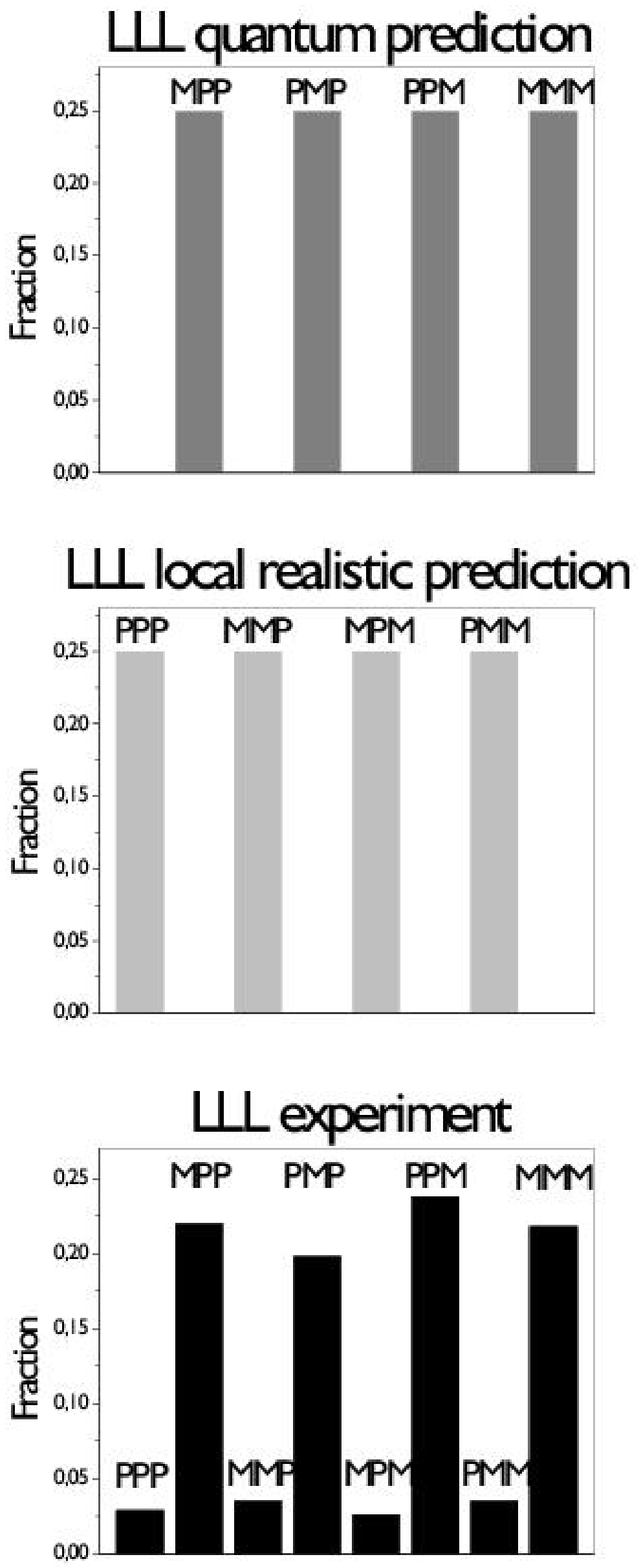}{0.5}{The graphs show the relative frequencies of
three-fold coincidence events between outputs D$_{1}$, D$_{2}$,
and D$_{3}$ as predicted and as measured (bottom). All analyzers
project onto a L=linear $45^\circ$ basis. ``P'' and ``M'' refer to
plus and minus $45^\circ$ linear polarization, respectively. MPM for
example marks the probability to measure a three-fold coincident
event if polarizer settings are M for particle 1, P for particle
2, and M for particle 3. The data violate the local realistic
prediction by 10 standard deviations.}

While Bouwmeester demonstrated the existence of a three-photon GHZ
state, Pan {\it{et al.}}\cite{Pan00a} reported last year the first
and still only experiment in which the test against local realism
was carried out in full. Although GHZ-type tests theoretically
permit a clear cut contradiction to local realism, in a real
experiment one has to resort to an inequality just like in Bell's
theorem. For this task, Pan measured in total 32 (4 groups of 8)
different combinations of polarizer settings, and the reasoning
followed Mermin's argument for an entangled system of three
spin-1/2 particles,\cite{Mermin90c} adapted to an imperfect
(noisy) experimental set-up. Fig.~\ref{3ghzdata} shows the results
of the measurements of one of the four groups and compares them to
the local realistic as well as to the quantum predictions. The
data clearly violates the local realistic prediction, and agrees
within experimental errors with quantum physics.

\paragraph{Four-photon GHZ}

\fig{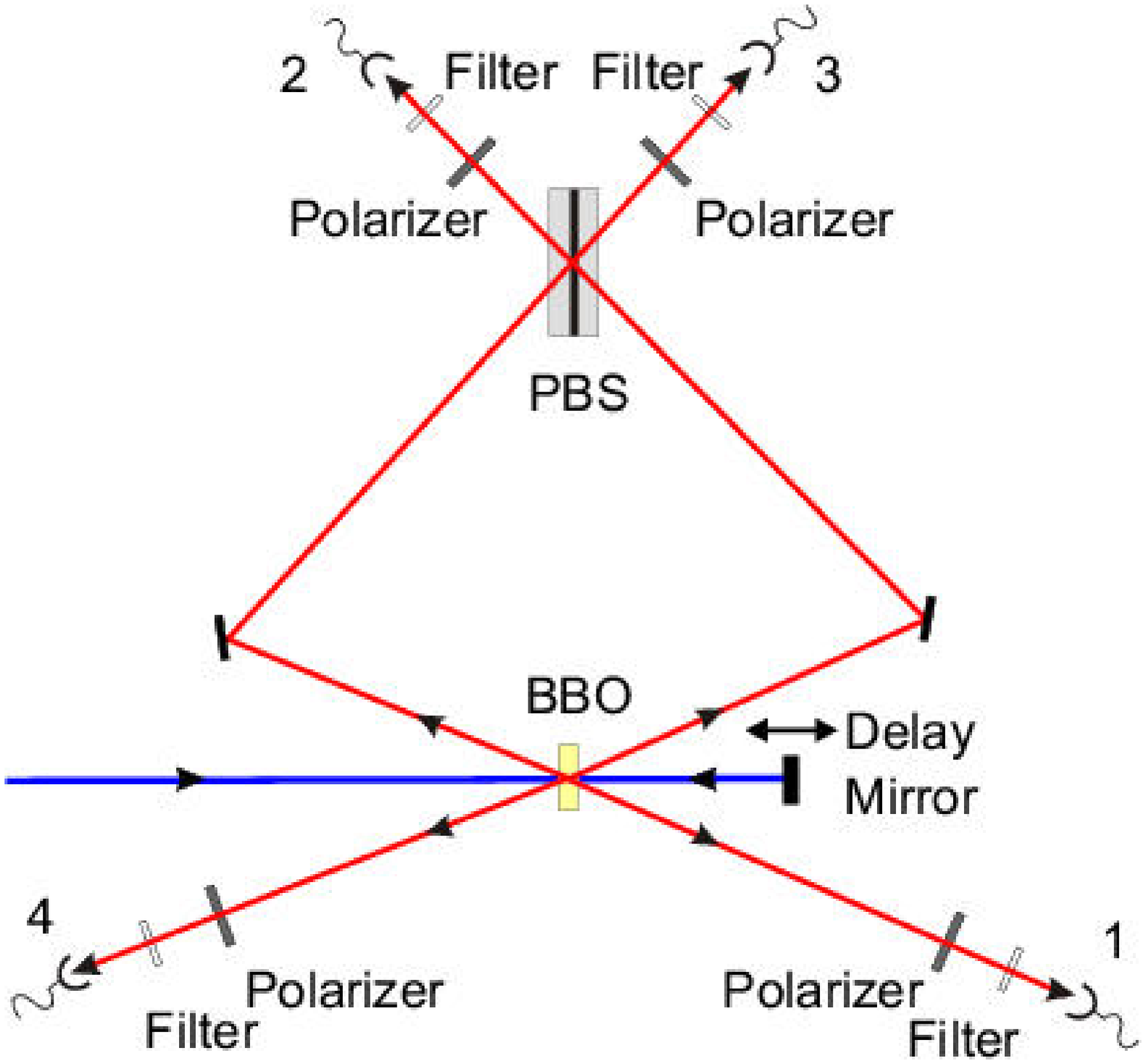}{0.5}{Experimental set-up to create and confirm the
existence of a four-photon GHZ states.}

Extending the teleportation and entanglement swapping experiments
discussed in sections \ref{teleportation} and
\ref{entanglementswapping}, one of our groups (UNIVIE) recently
succeeded in demonstrating four-photon entanglement.\cite{Pan01b}
In this experiment again two down-conversion pairs are used as an
initial resource, but this time from processes occurring in
different spatial modes. One photon of each pair is directed
towards a polarizing beam splitter. (s. Fig.~\ref{4ghz})
Coincidence detection after this beam-splitter projects the
incoming product state of two entangled photon pairs onto a
four-photon GHZ state.\footnote{Note that the state is never
realized as a freely propagating system.}

\begin{equation}
    \ket{\Psi}_{1234} = \frac{1}{\sqrt{2}} \left( \ket{H}_{1}
    \ket{V}_{2} \ket{V}_{3} \ket{H}_{4} + \ket{V}_{1} \ket{H}_{2}
    \ket{H}_{3} \ket{V}_{4} \right).
\end{equation}

A comparison of the measured 4-photon coincidence probabilities
for various combinations of H and V projections confirms that
indeed only the desired
$\ket{H}_{1}\ket{V}_{2}\ket{V}_{3}\ket{H}_{4}$ and
$\ket{V}_{1}\ket{H}_{2}\ket{H}_{3}\ket{V}_{4}$ components have
been created. The contrast in this measurement was of more than
100:1. In addition, measurements in the conjugate 45$^\circ$ basis (s.
Fig.~\ref{ghz_45}) demonstrate the coherent superposition of the
two components, hence show the existence of a four photon GHZ
state and confirm the existence of non-locality.

\fig{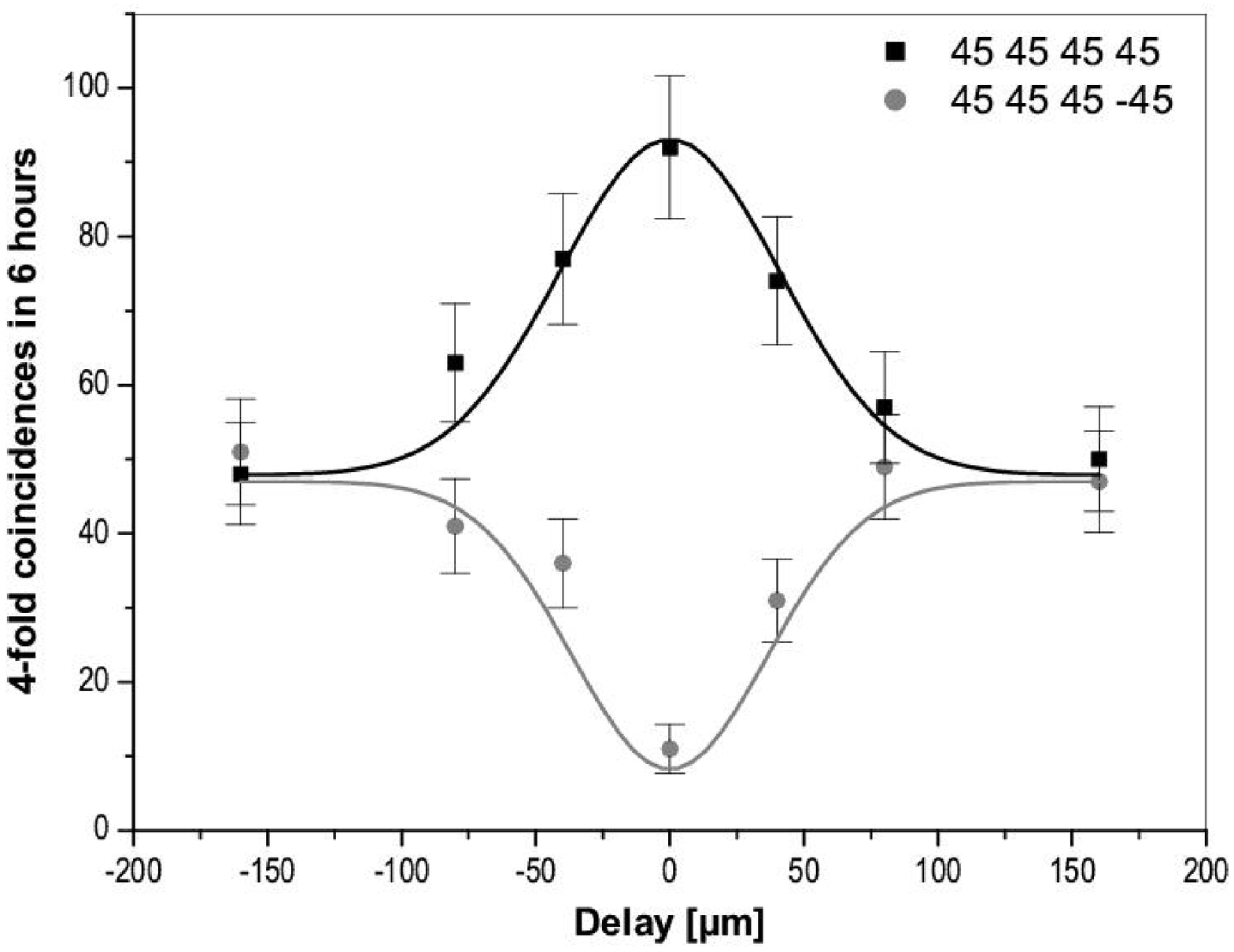}{0.7}{Experimental data showing four-photon
polarization correlation in the 45$^\circ$ linear basis. At zero delay
between the two photons that are superposed on the polarizing
beam-splitter, indistinguishability is granted and interference
occurs, demonstrating the coherent superposition of $\ket{H}_{1}
\ket{V}_{2} \ket{V}_{3} \ket{H}_{4}$ and $\ket{V}_{1} \ket{H}_{2}
\ket{H}_{3} \ket{V}_{4}$.}

\section{Quantum communication} \label{quantumcommunication}
\noindent

\subsection{Quantum cryptography}\label{cryptography}
\noindent

Quantum cryptography (QC) is certainly the most mature application
of quantum communication. Based on the non-classical feature of
the quantum world, it provides two parties, a sender Alice and a
receiver Bob, with a means to distribute a secret key in a way
that guarantees the detection of any eavesdropping (Eve): Any
information obtained by an unauthorized third party about the
exchanged key goes along with an increase of the quantum bit error
rate (QBER) of the transmitted data which can be checked using a
suitable subset of the data. It has been shown that, as long as
the QBER of the sifted key (the key after bases reconciliation) is
below a certain threshold (11 or 15 \%, respectively, depending on
the eavesdropping strategy assumed\cite{Fuchs97a,Mayers98a}),
Alice and Bob can still distill a secure key by means of classical
error correction and privacy amplification
protocols.\cite{Bennett92d} This secret key can then be used
together with the one-time pad to exchange a confidential message
in complete privacy.

There is vast literature covering theoretical as well as
experimental aspects of quantum cryptography. Since there is an
extensive and very recent article\cite{Gisin01a} reviewing both
sides, we will keep this section short and give only the necessary
information to understand {\it{entanglement-based}} quantum
cryptography.

\subsubsection{Quantum cryptography based on faint laser pulses}


The first QC protocol has been published by Bennett and Brassard
in 1984\footnote{The work was inspired by an unpublished article
by Wiesner from 1970 (published only in 1983\cite{Wiesner83a})};
it is known today as BB84 or four-state
protocol.\cite{Bennett84a} Fig.~\ref{BB84} illustrates the
protocol with the example of four polarization qubits of linear
polarization. However, any property can in principle be used to
realize a qubit. Furthermore, any four states that fulfill the
requirement that they form two bases in the qubit space and that
any two states belonging to a different basis have an overlap of
$\frac{1}{2}$ will do as well.

\fig{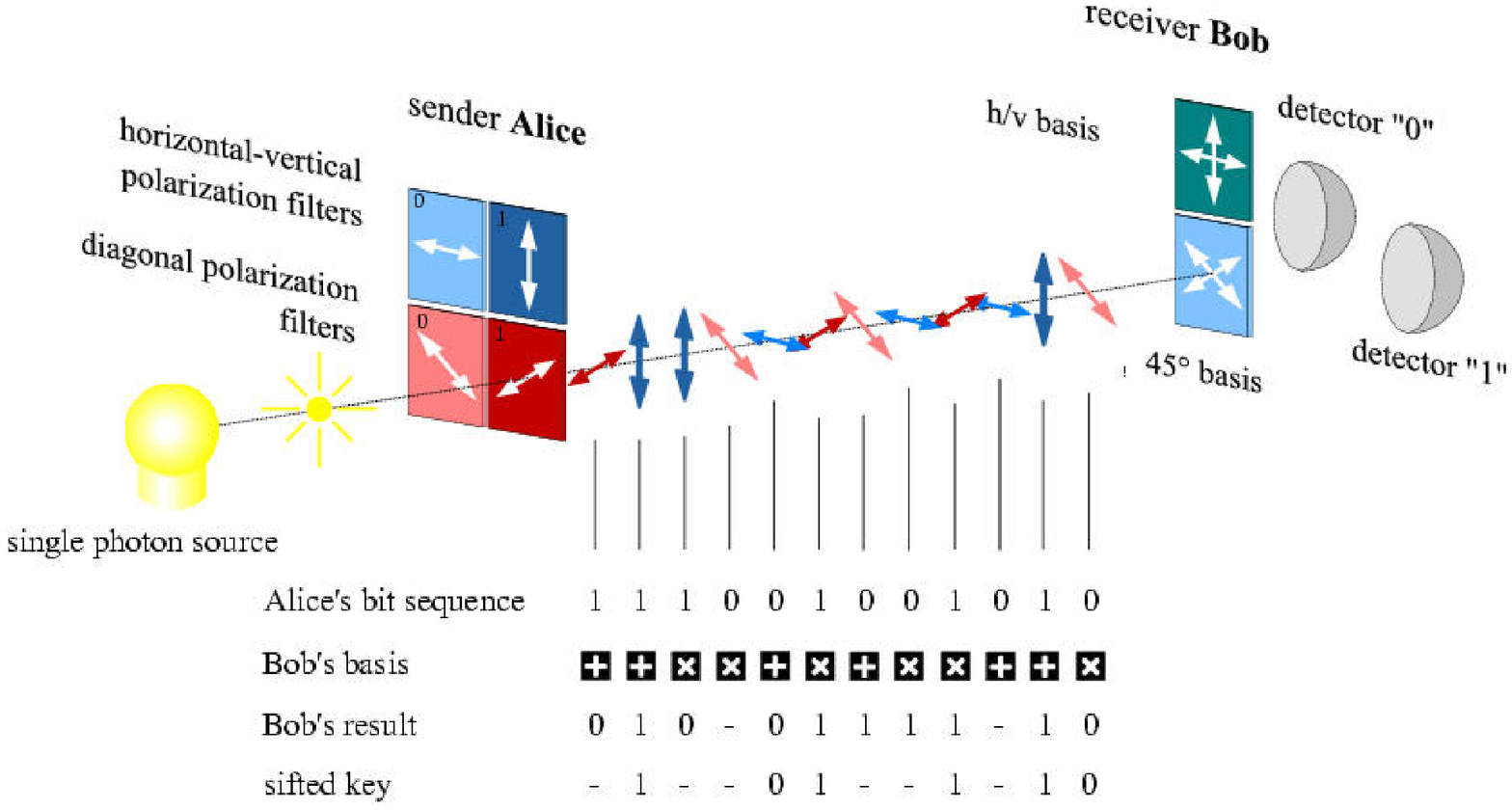}{1}{For each photon she sends to Bob, Alice chooses
randomly a bit value (row 1) and a basis (h/v, or $\pm 45^o$), and
prepares the photon in the corresponding state. Every time Bob
expects a photon to arrive, he activates his detectors and chooses
randomly to analyze in the h/v basis, or in the $\pm 45^o$ basis.
He records which basis he used (row 2) and, in case of a
successful detection, which result (in terms of bits) he got (row
3). After exchange of a sufficient large number of photons, he
publicly announces the cases where he detected a photon and the
basis used for the measurements. However, he does not reveal which
results he got. Alice compares event by event whether or not Bob's
analyzer was compatible to her choice of bases. If they are
incompatible or if Bob failed to detect the photon, the bit is
discarded. For the remaining bits (row 4), Alice and Bob know for
sure that they have the same value. These bits form the so-called
sifted key. The security of the key distribution is, roughly
speaking, based on the fact that a measurement of an unknown
quantum system will, in most cases, disturb the system: If Alice's
and Bob's sifted keys are perfectly correlated, no eavesdropper
tried to eavesdrop the transmission and the key can be used for
encoding a confidential message using the one-time pad. If the
sifted keys are not 100\% correlated, then, depending on the QBER,
Alice and Bob can either distill a secret key via error correction
and privacy amplification, or the key is discarded and a new
distribution has to be started.}

The first experimental demonstration of quantum cryptography took
place in 1989 at IBM when Bennett {\it{et al.}} realized the
so-called B92 (or two-state) protocol\cite{Bennett92a} based on
polarization coding with ``single photons" over a distance of 30
cm in air.\cite{Bennett92d} Since then, a lot of experimental
progress has been made, and from 1995 on, several groups
demonstrated that quantum cryptography is possible outside the
laboratory over distances of tens of kilometers as
well.\cite{Gisin01a} All experiments relied on faint laser pulses
-- strongly attenuated pulses that contain less than 1 photon on
average --- to mimic single photons. These realizations could,
even today, provide secure communication in case cryptographic
protocols that are based on mathematical complexity turn out to be
unsafe. However, they still suffer from low bit-rates\footnote{To
give an example, the secret key rate (after error correction and
privacy amplification) reported by Ribordy {\it{et
al.}}\cite{Ribordy98a} over a distance of 23 km was of 210 Hz.},
and the maximum span with today's technology is only of around 100
km.

Let us briefly elaborate on the maximum distance. In theory, i.e.
using perfect experimental equipment, the sifted key rate
decreases exponentially with increasing transmission losses but
never drops to zero. It is given by the product of Alice's pulse
rate $f_{\mbox{\scriptsize rep}}$, the number of photons per pulse
$\mu$, the probability $t_{\mbox{\scriptsize link}}$ that a photon
arrives at Bob's, and the quantum efficiency $\eta$ that it is
detected.
\begin{equation}
R_{\mbox{\scriptsize sifted}}=\frac{1}{2}\cdot
f_{\mbox{\scriptsize rep}}\cdot\mu \cdot t_{\mbox{\scriptsize
link}}\cdot\eta \label{siftedkey}
\end{equation}
\noindent The factor 1/2 is due to key sifting --- the fact that
Alice and Bob use compatible bases in only 50 \% of the cases. In
practice, there are always experimental imperfections, and there
will always be some errors in the sifted key --- even in the
absence of an eavesdropper. For the sake of the presentation, we
assume here that the errors are only due to detector dark counts
(a signal generated by a detector without the presence of a
photon), arising with probability $p_{\mbox{\scriptsize dark}}$:
\begin{equation}
R_{\mbox{\scriptsize wrong}} \approx R_{\mbox{\scriptsize
dark}}=\frac{1}{2}\cdot\frac{1}{2}\cdot f_{\mbox{\scriptsize
rep}}\cdot p_{\mbox{\scriptsize dark}}\cdot n \label{R}
\end{equation}
\noindent The first factor 1/2 is again due to key sifting, the
second one to the fact that a detector dark count leads only in
half of the cases to a wrong result, and $n$ is the number of
detectors. Finally, the QBER is given by
\begin{equation}
QBER=\frac{R_{\mbox{\scriptsize wrong}}}{R_{\mbox{\scriptsize
sifted}}}\approx\frac{n\cdot p_{\mbox{\scriptsize dark}}}{2\cdot
t_{\mbox{\scriptsize link}}\cdot\eta\cdot\mu} \label{QBER}
\end{equation}

\fig{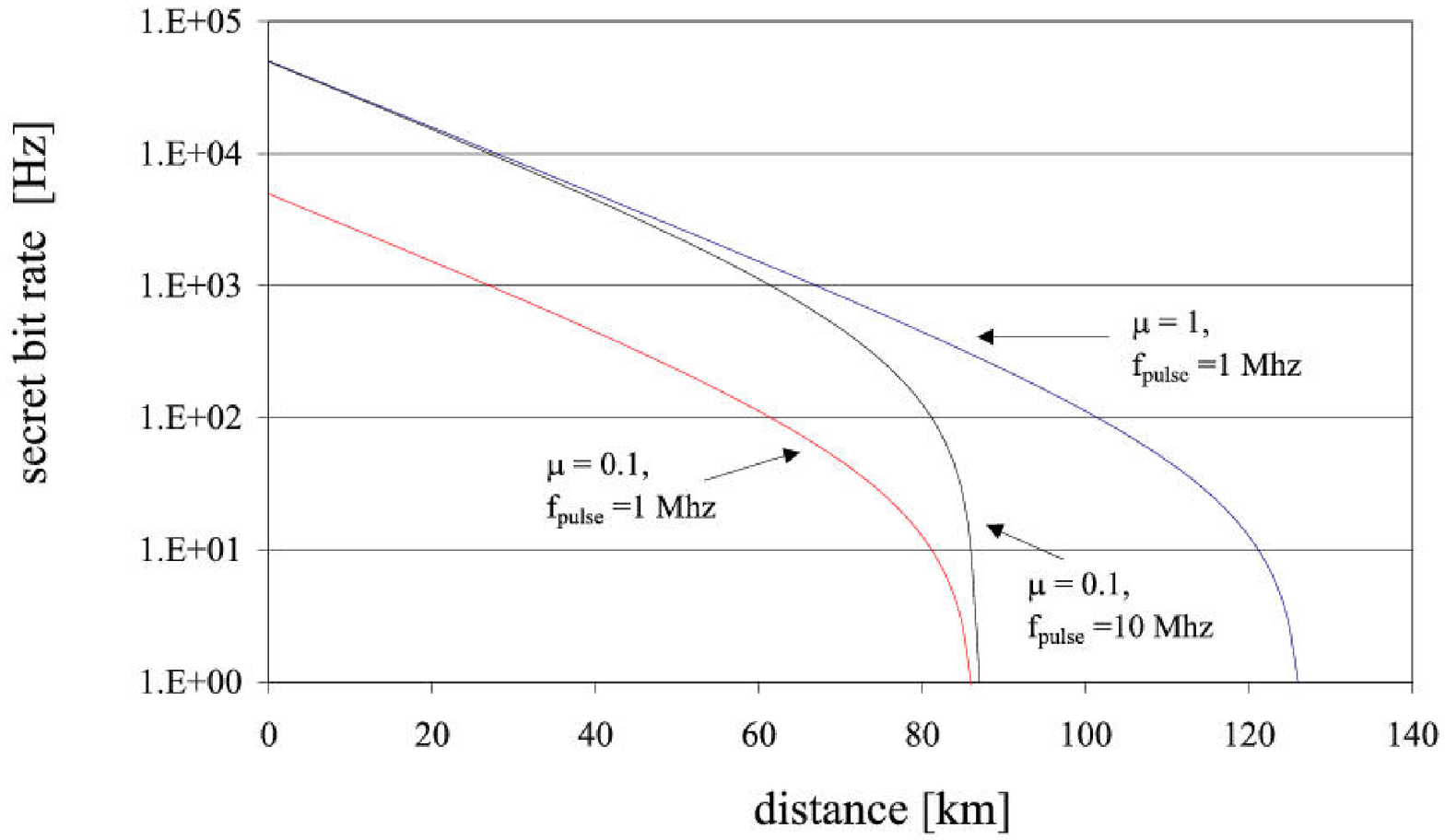}{0.9}{Secret bit rate after error correction and
privacy amplification. The maximum transmission span is given by
the distance where the QBER equals 15\% (assuming symmetric
individual eavesdropping attacks\cite{Fuchs97a}). Here we assume a
initial pulse rate $f_{\mbox{\scriptsize pulse}}$ of 1 and 10 MHz,
respectively, losses of 0.2 dB/km of optical fibers at 1550 nm
wavelength (close to the fundamental limit), a quantum efficiency
of 10\%, dark count probability of $10^{-5}$ and a mean photon
number per pulse $\mu$ of 0.1 and 1, respectively. The use of a
higher mean photon number leads to a higher secret bit rate for a
given distance and pulse rate, as well as to a larger maximum
transmission span, i.e. 130 instead of 90 km. A higher pulse rate
engenders a higher secret key rate, however, does not change the
maximum span. In practice, if we take into account non-ideal
error correction and privacy amplification algorithm,
multi-photon pulses and other optical losses not considered here,
the maximum distance is likely to be reduced by a factor of
around two.}

Since Alice and Bob can never know for sure whether the observed
QBER is due to the imperfections of their equipment, or whether it
is engendered by the presence of an eavesdropper, they always have
to assume to worst case, i.e. that there is an eavesdropper that
has the maximum information compatible with the observed QBER.
Therefore, they have to apply classical error correction and
privacy amplification protocols to the sifted key in order to
distill a secret key.

Fig.~\ref{secrate} shows the secret bit rate after error
correction and privacy amplification as a function of distance.
Here we assume that the photons are transmitted using optical
fibers with losses of 0.2 dB/km. The bit rate decreases
exponentially for small distances. Then, with larger distance
(i.e. with decreasing transmission probability
$t_{\mbox{\scriptsize link}}$, hence increasing QBER), the bit
reduction due to error correction and privacy amplification gets
more and more important, and at a QBER (hence distance) where the
Alice-Bob mutual Shannon information is equal to Eve's Shannon
Information, there are no bits left and the curve representing
the bit rate drops vertically to zero. Note that the maximum
distance does not depend on Alice's pulse rate.

To achieve a better performance of a cryptographic system
concerning bit rate or maximum transmission distance, there are
basically two things to improve: the detectors, and the sources.

\begin{itemlist}
\item Today's {\it{single photon detectors}}, capable of counting photons at
telecommunication wavelength of 1.3 and 1.5 $\mu$m where fiber
losses are low, feature quantum efficiencies of only around 10\%,
paired with a high dark count probability of $\approx 10^{-5}$ per
1 ns time window. These figures determine the secret key rate for
a given distance and pulse rate, and the maximum transmission span
as shown in Fig.~\ref{secrate}. In addition, the detectors'
performance limits the pulse rate via the effect of so-called
afterpulses: These are avalanches that are not caused by the
detection of a photon but by the release of charges from trapping
levels populated while a current transits through the diode. Since
the probability for observing an afterpulse after a detection of a
photon decreases exponentially with time, they can be suppressed
using suitable dead times --- with the drawback of limiting the
maximum pulse rate. It is thus obvious that the use of better
detectors would have an important impact on experimental quantum
cryptography.

\item Mimicking single photons by {\it{faint pulses}} has a very important
advantage: it is extremely simple. Unfortunately this advantage is
paired with two drawbacks. First, a mean photon number smaller
than 1 (the upper limit for quantum cryptography) leads to a
reduction of the bit rate (see Eq.~\ref{siftedkey} and Fig.
\ref{secrate}). Second, since the photon-number statistics for
faint pulses is given by a Poissonian distribution, there is
always a possibility to find more than one photon in a weak pulse.
This opens the possibility of an eavesdropper attack based on
multi-photon splitting.\cite{Lutkenhaus00a,Brassard00a} The
smaller the mean number of photons per pulse, the smaller this
threat, however, the smaller the bit rate as well.
\end{itemlist}

\subsubsection{Quantum cryptography based on photon-pairs}
\label{paircrypto}

\paragraph{``Single-photon" based realizations}

In order to get around the problem of faint pulses where the
probability of having zero photons in a pulse is rather high, a
good idea is to replace the faint pulse source by a photon-pair
source (see Section~\ref{pairsources} and
Fig.~\ref{cryptosources}b) where one photon serves as a trigger to
indicate the presence of the other one.\cite{Hong86a} In this
case, Alice can remove the vacuum component of her source, and
Bob's detectors are only triggered whenever she sends at least one
photon.\footnote{Here we assume that the collection efficiency for
the photon traveling towards Bob is 1. In practice, a more
realistic value is $\approx$ 0.70.} This leads to a higher sifted
key rate (assuming the same trigger rate than in the faint-pulse
case) and a lower QBER for a given distance (for given losses) and
therefore to a larger maximum span (see Fig.~\ref{secrate}). It is
important to note that photon-pairs can not be created in Fock
states, similar to single photons mimicked by faint pulses.
Therefore, depending on the probability to create more than one
photon pair, the danger of multi-photon splitting eavesdropping
attacks exists as well.

\paragraph{Entanglement based realizations}

Finally, the potential of a source creating entangled pairs is not
restricted to create two photons at the same time --- one serving
as a trigger for the other one. It is possible to use the full
quantum correlation to generate identical keys at Alice's and
Bobs, and to test the presence of an eavesdropper via a test of a
Bell inequality (see Fig.~\ref{cryptosources}c). This beautiful
application of tests of Bell inequalities has been pointed out by
A. Ekert in 1991\cite{Ekert91a} --- without knowing about the
``discovery" of quantum cryptography by Bennett and Brassard 7
years earlier. The set-up is similar to the one used to test Bell
inequalities with the exception that Alice and Bob each have to
chose from three different bases. Depending on the bases chosen
for each specific photon pair, the measured data is either used to
establish the sifted key, to test a Bell inequality, or it is
discarded.

The security of the Ekert protocol is very intuitive to
understand: If an eavesdropper gets some knowledge about the state
of the photon traveling to Bob, she adds some hidden variables
(hidden in the sense that only the eavesdropper knows about their
value). If she gets full knowledge about all states, i.e. the
whole set of photons analyzed by Bob can be described by hidden
variables, a Bell inequality can not be violated any more. If Eve
has only partial knowledge, the violation is less than maximal,
and if no information has leaked out at all, Alice and Bob observe
a maximal violation.

However, the Ekert protocol is not very efficient concerning the
ratio of transmitted bits to the sifted key length. As pointed out
in 1992 by Bennett {\it{et al.}}\cite{Bennett92e} as well as by
Ekert {\it{et al.}},\cite{Ekert92a} protocols originally devised
for single photon schemes can also be used for entanglement based
realizations. This is not surprising if one considers Alice's
action as a non-local state preparation for the photon traveling
to Bob (see also Section~\ref{relativistic}). Interestingly, it
turns out that, if the perturbation of the quantum channel (the
QBER) is such that the Alice-Bob mutual Shannon information equals
Eve's maximum Shannon information, then the CHSH Bell inequality
(Eq.~\ref{CHSH}) can not be violated any
more.\cite{Gisin97a,Fuchs97a} Although this seems very natural in
this case and a similar connection has recently been found for
n-party quantum cryptography and some n-particle Bell
inequalities\cite{Scarani01a}, it is not clear yet to what extend
the connection between security of quantum cryptography and the
violation of a Bell inequality can be generalized.

Compared to the faint pulse schemes, entanglement based QC
features two advantages. First, similarly to photon-pair based
realizations, Alice removes the vacuum component of her source.
Actually, the entanglement based case is even more efficient since
even the optical losses in Alice's preparation device are now
eliminated as can be seen from Fig.~\ref{cryptosources}c.

Second, even if two pairs are created within the same detection
window --- hence two photons travel towards Bob within the same
pulse --- they do not carry the same bit, although they are
prepared in states belonging to the same basis. Beyond this
passive state preparation, it is even possible to achieve a
passive preparation of bases using a set-up similar to the one
depicted in Fig.~\ref{cryptosources}d. There is no external switch
that forces all photons in a pulse to be measured in the same
basis but each photon independently chooses its basis and bit
value. Therefore, eavesdropping attacks based on multi-photon
pulses do not apply in entanglement based QC. However multi-photon
pulses lead to errors at Bob's who detects from time to time a
photon that is not correlated to Alice's.

\fig{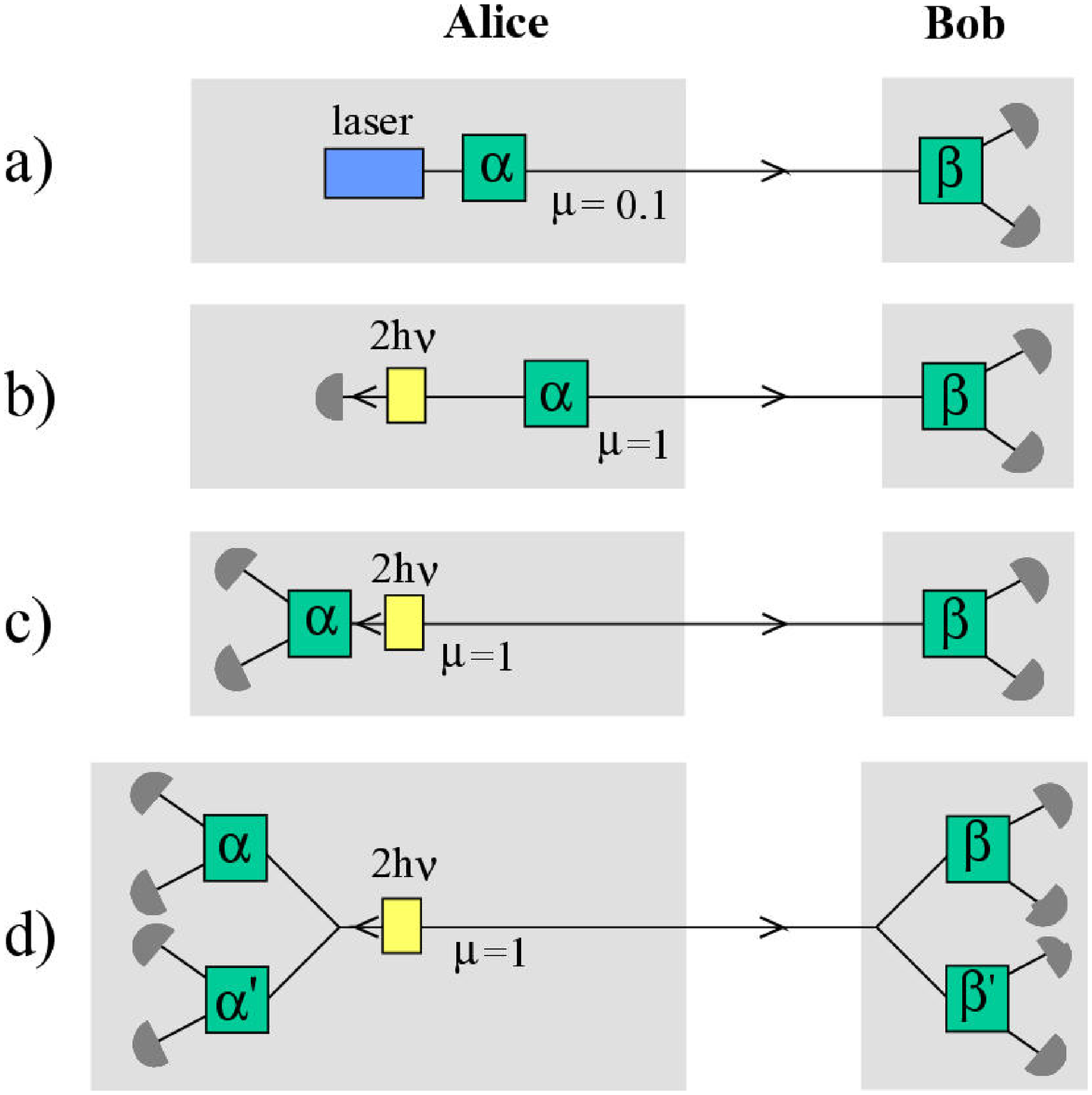}{0.7}{Single photon based quantum cryptography
using a) a faint-pulse source, b) a two-photon source, c)
entanglement based quantum cryptography with active, and d) with
passive choice of bases. ``2h$\nu$" denotes the photon-pair
source, and the parameters $\alpha$ and $\beta$ characterize the
settings of the qubit-analyzers.}

Although all Bell experiments intrinsically contain the
possibility for entanglement based QC, we list here only
experiments that have been devised in order to allow a fast change
of measurement bases.

\begin{minipage}{\columnwidth}
\begin{tabular}{@{}lp{0.8\columnwidth}}

1982    &   Interestingly enough, the first experiment that
            fulfills the above definition is
            the test of Bell inequalities using time-varying
            analyzers, performed by Aspect {\it{et al.}}\cite{Aspect82a} with
            polarization entangled qubits
            in order to close the locality loophole (see also
            Section~\ref{loopholes}) ---
            at a time where quantum cryptography was not
            yet known, not even the single photon based version.\\

1998    &   Weihs {\it{et al.}} demonstrate a violation of Bell
            inequalities with polarization entangled qubits at 700 nm wavelength
            and randomly switched analyzers, separated by 360 km of
            optical fiber.\cite{Weihs98a} This experiment has been devised
            to close the locality loophole.\\

1999    &   Tittel {\it{et al.}} perform a Bell experiment, again
            to be seen in the context of the locality loophole, incorporating a
            passive choice of bases.\cite{Tittel99a} Two fiber-optical
            interferometers are attached to each side of a source creating
            energy-time entangled photons at 1.3 $\mu$m wavelength. However, similar to
            both before-mentioned experiments, the bases chosen for the measurements
            are chosen in order to allow a test of Bell
            inequalities and not to establish a secret key.\\

2000    &   Three publications on entanglement based cryptography
            appear in the same issue of Phys. Rev. Lett.:\\

        &   1.) Using a set-up similar to the one mentioned already in the second
            entry of this table, Jennewein {\it{et al.}} realize a quantum
            cryptography system including error correction over a distance of
            360 m.\cite{Jennewein00a} Two different protocols are implemented,
            one based on Wigner's inequality (a special form of Bell inequality),
            the other one following BB84. Sifted key rates of around 400 and 800 bits/s,
            respectively, are obtained, and QBERs of around 3\% observed. Using the
            same assumptions that lead to Fig.~\ref{secrate}, this
            amounts to a secret key rate of 300 and 600 bits/s, respectively.\\

        &   2.) Naik {\it{et al.}} demonstrate the Ekert protocol in a free space
            experiment over a short (laboratory) distance.\cite{Naik00a} The experiment takes
            advantage of polarization entangled
            qubits at a wavelength of around 800 nm. Sifted key rates of around 10
            bits/s paired with a QBER of 3\% are reported, leading to a secret key rate of 6
            bits/s after implementation of error correction and privacy amplification.
            In addition to the key exchange, the authors simulate different eavesdropping
            strategies and find an increase of the QBER with increasing information
            of the eavesdropper, according to theory. The experiment has recently been extended\cite{Enzer01a}
            to realize the so-called six state protocol.\cite{Bruss98a,Bechmann99a}\\

        &   3.) Tittel {\it{et al.}}\cite{Tittel00a} report on a fiber-optical realization of quantum cryptography
            in a laboratory experiment using the BB84 protocol.
            This experiment is based on time-bin entangled qubits at telecommunication wavelength of 1.3 $\mu$m
            and takes advantage of phase-time coding and a passive choice of bases.
            Sifted key rates of 33 Hz and a QBER of 4\% are obtained, leading to a calculated
            secret key rate of 21 bits/s.\\

2001    &   Ribordy {\it{et al.}}\cite{Ribordy01a} realize a QC
            system based on energy-time entanglement. In contrast to the schemes
            mentioned before, this realization takes advantage of an asymmetric
            set-up, optimized for QC, instead of a set-up designed for tests of
            Bell inequalities where the source is generally located roughly in
            the middle between Alice and Bob. Here, one photon (at 810 nm
            wavelength) is send to a bulk-optical interferometer, located
            directly next to the source, the other one (at 1550 nm wavelength)
            is transmitted through 8.5 km of fiber
            on a spool to a fiber optical interferometer. Implementing the BB84 protocol
            and a passive choice of bases, a sifted key rate of 134 bits/s and
            a mean QBER of 8.6\% (over 1 hour) is observed. From these values, one
            can calculate a secret key rate of 45 bits/s. \\
\end{tabular}
\end{minipage}

\paragraph{Three party quantum cryptography}

In addition to the mentioned two-party QC schemes, Tittel {\it{et
al.}} reported in 2001 a proof-of principle demonstration of
quantum secret sharing (three party quantum cryptography) in a
laboratory experiment.\cite{Tittel01a} This rather new protocol
enables Alice to send key material to Bob and Charlie in a way
that neither Bob nor Charlie alone have any information about
Alice's key, however, when comparing their data, they have full
information. The goal of this protocol is to force both of them to
collaborate.

In contrast to proposed implementations using three-particle GHZ
states,\cite{Zukowski98a,Hillery99a} pairs of time-bin entangled
qubits were used to mimic the necessary quantum correlation of
three entangled qubits, albeit only two photons exist at the same
time (see Fig.~\ref{ssharing}). This is possible thanks to the
symmetry between the preparation interferometer acting on the pump
pulse and the interferometers analyzing the down-converted
photons. Indeed, the data describing the emission of a bright pump
pulse at Alice's is equivalent to the data characterizing the
detection of a photon at Bob's and Charlie's: all specify a phase
value and an output, or input port, respectively. Therefore, the
emission of a pump pulse can be considered as a detection of a
photon with 100\% efficiency, and the scheme features a much
higher coincidence count rate compared to the initially proposed
GHZ-state type schemes.

\fig{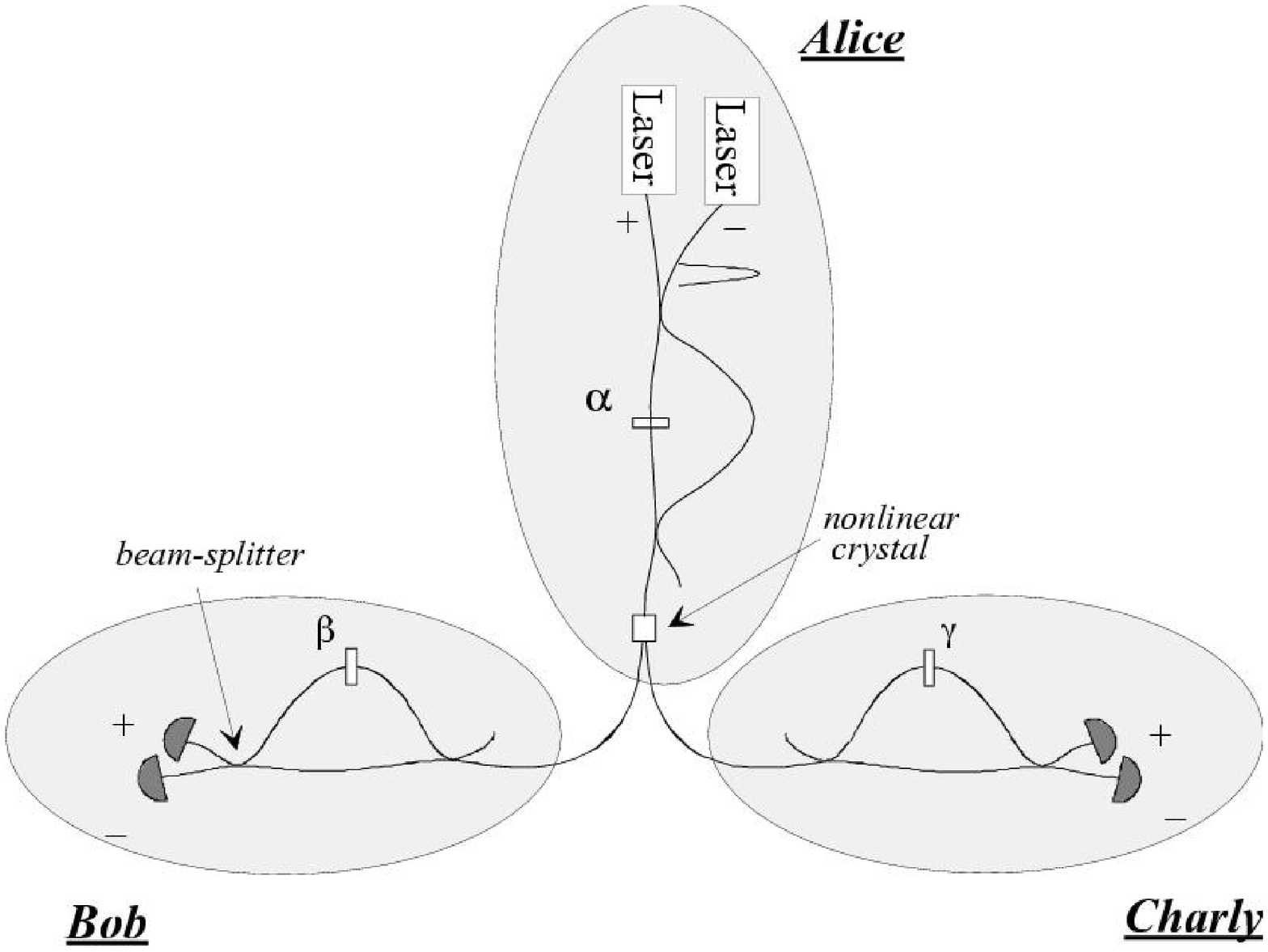}{0.7}{Basic set-up for three-party quantum secret
sharing using ``pseudo-GHZ states". Compare with ``true" GHZ
states as shown in Fig.~\ref{GHZsetup}.}

\subsection{Quantum dense coding}\label{densecoding} \noindent

Whenever two parties A (Alice) and B (Bob) wish to communicate,
they first have to agree on a coding procedure, that is, they
have to associate symbols with physical states. In classical
communication, one usually uses a two letter alphabet where the
different symbols (bit-values) are represented by (classical)
optical pulses with different individual properties. In quantum
physics we can encode information in a novel way into joint
properties of elementary systems in entangled states, leading in
principle to the possibility to transmit two bits of information
by sending only one qubit. This striking application of quantum
communication is known as quantum dense coding.

\fig{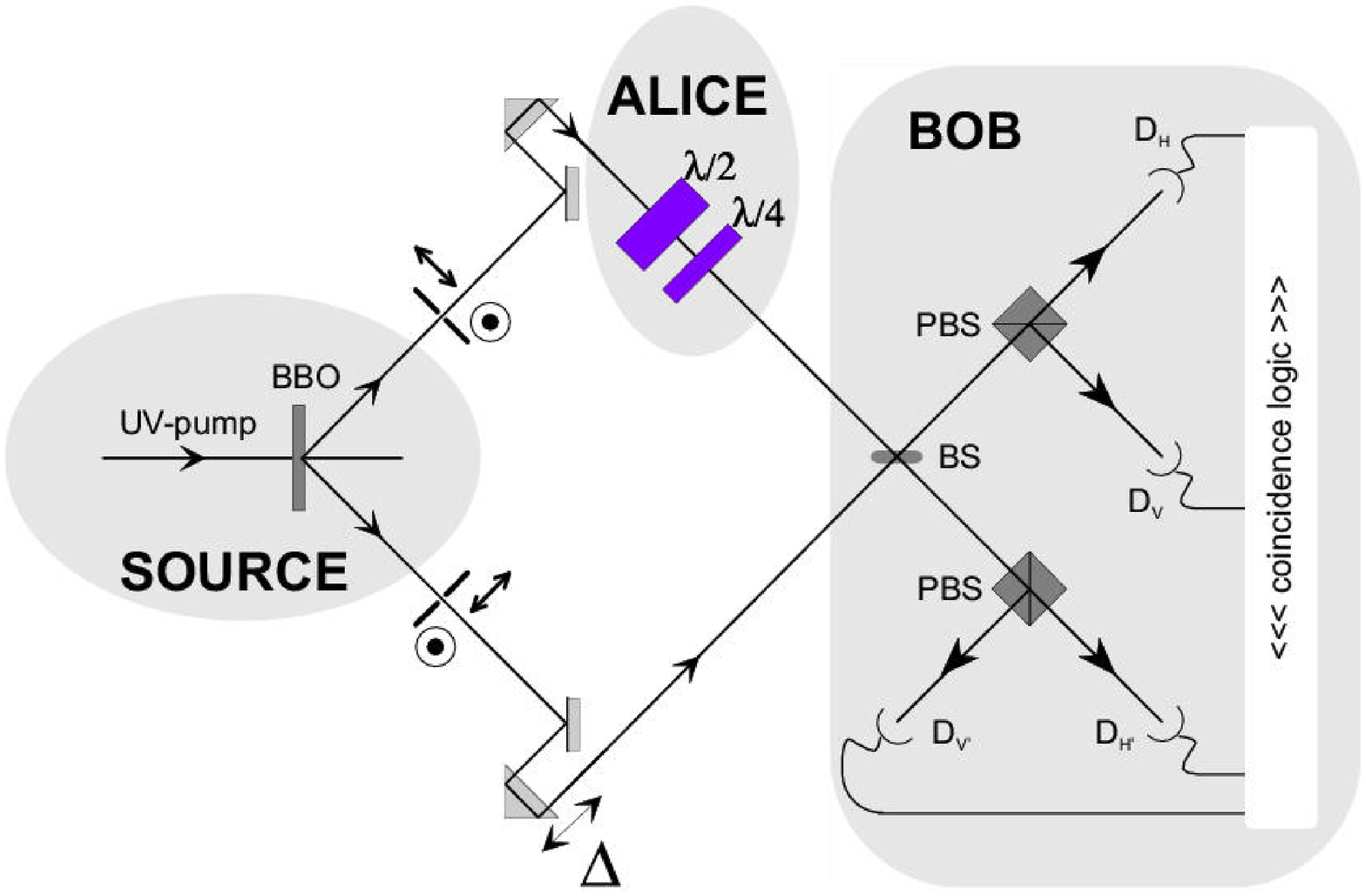}{0.7}{Experimental set-up to demonstrate quantum
dense coding based on Bell state analysis of entangled
polarization qubits.\cite{Mattle96a} After locally preparing the
joint state of the entangled particles by means of wave-plates,
Alice sends her particle to Bob. Performing a Bell measurement on
the two-particle state, Bob can distinguish between two different
Bell states ($\ket{\Psi^\pm}$), with the two other ones
($\ket{\Phi^\pm}$) leading to the same, third, result. Therefore,
Alice can encode 1.58 bit of information sending only one qubit.}

The maximally entangled Bell basis (Eqs.~\ref{psi+-} and
\ref{phi+-}) has a very important and interesting property which
was exploited by Bennett and Wiesner\cite{Bennett92a} in their
proposal for quantum dense coding: In order to switch from any one
of the four Bell states to all others, it is sufficient to
manipulate only one of the two qubits locally. Thus, the sender,
Alice, can actually encode two bits of information into the whole
entangled system by just acting on one of the two qubits.

In order to read out this information, the receiver, Bob, needs to
be able to identify the four Bell states, that is, he needs to
perform a Bell measurement as explained in
Section~\ref{measuringentanglement}. However, using linear optics,
only two out of the four Bell states can be distinguished
unambiguously\cite{Lutkenhaus99a,Calsamiglia01a} whereas the other
two states lead to identical signatures. Still, as has been
experimentally demonstrated in 1996 by Mattle {\it{et
al.}},\cite{Mattle96a} this is enough to encode three-valued
information (corresponding to 1.58 bit of information) into each
transmission event (see Fig.~\ref{dcsetup}).

\subsection{Quantum teleportation}\label{teleportation} \noindent

\fig{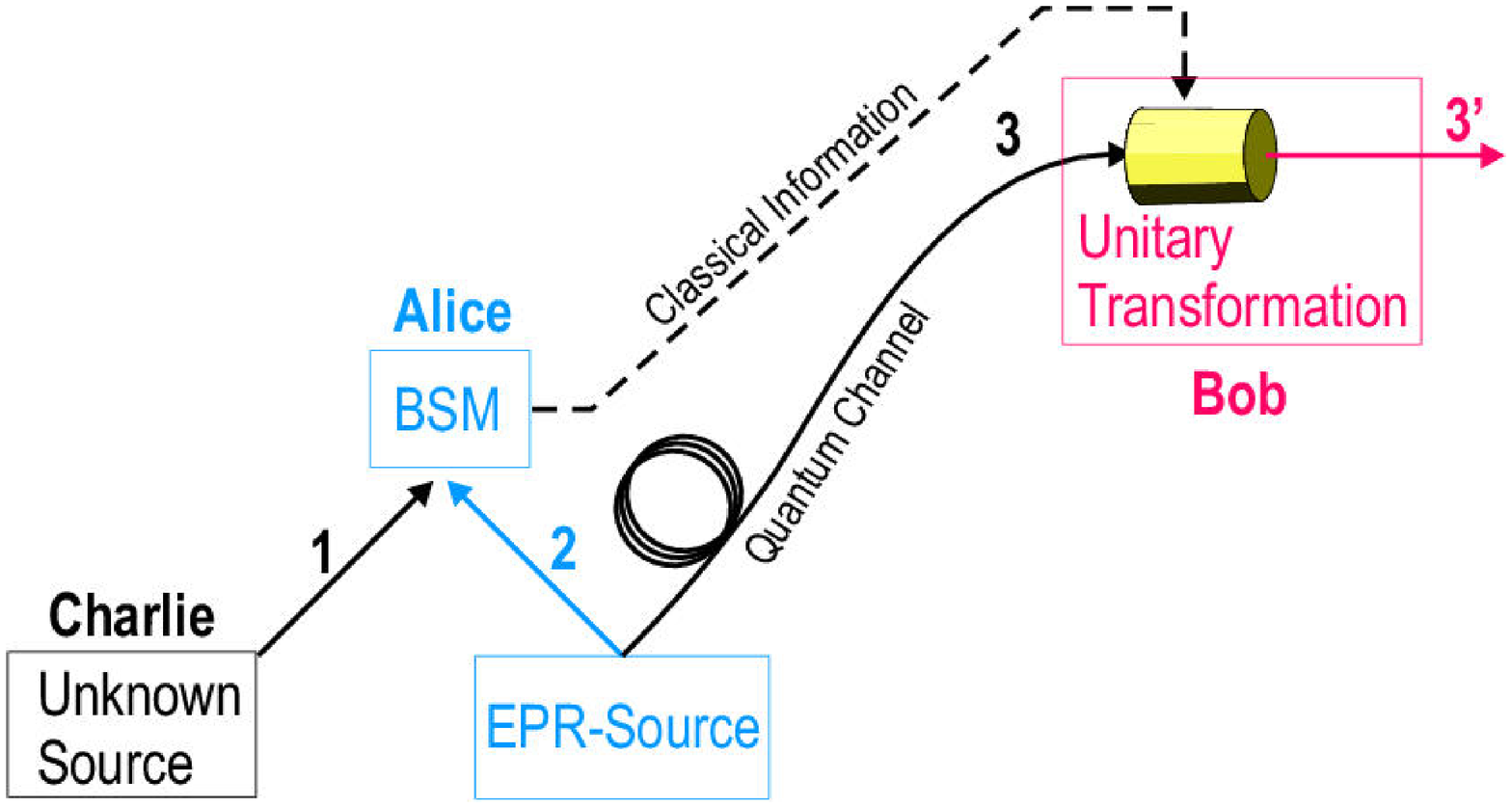}{0.8}{Schematic of quantum teleportation of an
unknown state. Particle 1 is given to Alice who subjects it to a
Bell-state measurement (BSM) together with particle 2, the latter
one being entangled to particle 3 (at Bob's). Depending on the
result of this measurement, Bob applies a unitary transformation
to particle 3 which then ends up in precisely the same state in
which particle 1 was originally.}

If, in some sense, the aim of quantum key distribution is the
communication of classical bits, quantum teleportation, discovered
in 1993 by Bennett {\it et al.},\cite{Bennett93a} can be thought
of as being the exchange of quantum bits. One might define quantum
teleportation as the art of transferring the state of an unknown
qubit located at Alice's to a second quantum system, located at
Bob's, the motivation being that it might be impossible to send
the physical system itself. In this application of quantum
communication, the qubit to be sent is unknown to the parties
involved in the transfer, but it might be known to a third party,
Charlie.

In a world of classical physics, teleportation is nothing
remarkable. It suffices to measure the properties of the
(classical) bit and then communicate the information about its
composition to Bob, who then reconstructs the bit. This strategy
must fail in the quantum case where the measurement of an unknown
qubit without disturbing it is impossible and cloning is
forbidden.\cite{Wootters82a,Dieks82a} Surprisingly, quantum
communication provides a way out of this problem (see Fig.
\ref{telescheme}). Before Charlie hands over the particle to Alice
who then teleports it to Bob, the latter have to share a pair of
entangled particles. Alice now makes a Bell measurement on
Charlie's particle and her part of the entangled pair (see Section
\ref{measuringentanglement}). She thus projects the two-particle
state randomly onto one of the four Bell states. Note that this
measurement only reveals the joint state of both particles, but
not the individual states. The outcome of this measurement
projects Bob's particle onto one of four different states as well.
Using two classical bits, Alice now tells Bob about the outcome of
her measurement and depending on her message, Bob performs one of
four unitary operations on his particle: the identity operation, a
bit flip, a phase flip, or a bit and a phase flip. This finally
leaves it in the state of the particle Charlie had initially
handed over to Alice, albeit neither Alice nor Bob know about this
state. It is important to note that Charlie's particle is left in
an arbitrary state after the Bell measurement, and that no cloning
has taken place. Moreover, since Alice's classical information is
needed to reconstruct the state of Charlie's particle, faster than
light communication is not possible.

The experimental realization of quantum teleportation has invoked
a strong reaction in the public. Whereas one can clearly say that
quantum teleportation in its current form has no relation to
disembodied transport of objects or even humans, it is also true
that the idea to transmit a quantum state without sending a
particle is an intriguing concept. Three different experiments
based on polarization qubits have been reported.

\begin{minipage}{\columnwidth}
\begin{tabular}{@{}lp{0.8\columnwidth}}

1997    &   Bouwmeester {\it{et al.}}\cite{Bouwmeester97a} are
            the first to demonstrate quantum teleportation based
            on a Bell measurement using linear optics.
            Although this allows in principle to teleport in 50\% of all
            cases, only the projection onto the $\psi^-$ state is used in the experiment.
            The result of the measurement is shown
            in Fig.~\ref{teledata}.\\

1998    &   Boschi {\it{et al.}}\cite{Boschi98a} demonstrate a
            teleportation set-up in which all four Bell states can be identified
            --- even using only linear optics. The entangled state is realized
            using k-vector (mode) entanglement, and the polarization degree of
            freedom of one of the entangled photons is employed to prepare the
            unknown state. However, this scheme can not be implemented for photons that come from
            independent sources as required for instance for
            entanglement swapping (see Section \ref{entanglementswapping}).\\

2001    &   Kim {\it{et al.}}\cite{Kim01a} demonstrated quantum
            teleportation based on a Bell measurement implementing
            non-linear interaction. This enables a projection onto all
            four Bell states, however, with very small efficiency
            of around one out of $10^{10}$.
            In order to compensate for the efficiency, the input state
            (send by Charlie) is a classical pulse from a fs
            laser. Nevertheless, this experiment shows that a
            complete Bell measurement is in principle possible,
            even when using single-photons and without having to take advantage of additional degrees
            of freedom of the entangled pair.

\end{tabular}
\end{minipage}
\noindent In addition, Furusawa {\it{et al.}}\cite{Furusawa98a}
demonstrated 1998 quantum teleportation based on continuous
quantum variables.

\fig{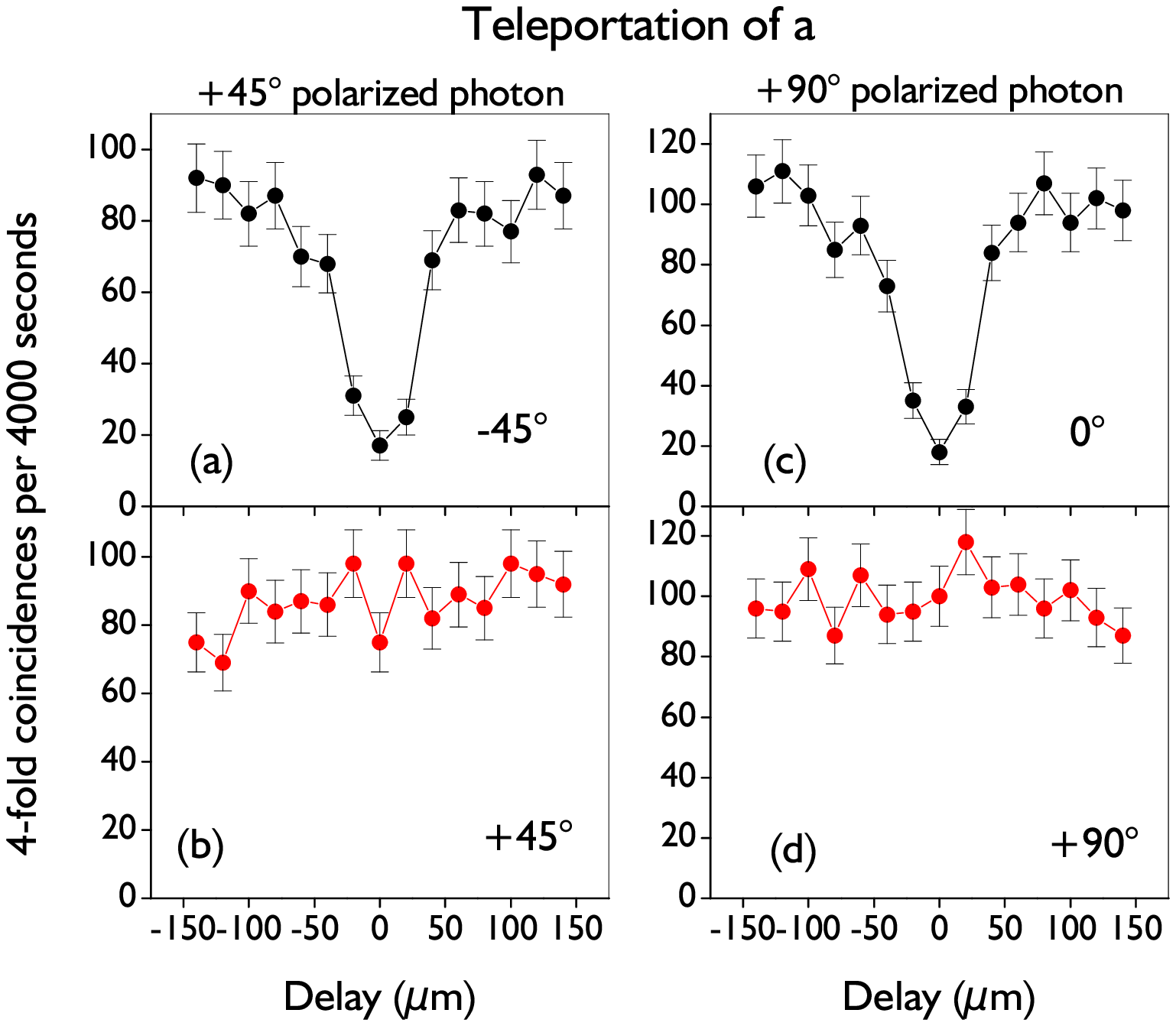}{0.8}{Experimental data showing faithful
teleportation of an independently created polarization qubit. Two
linear polarization states are tested: $45^\circ$ and $90^\circ$. The
fidelity is roughly the same for both cases.}

\subsection{Entanglement swapping}\label{entanglementswapping} \noindent

If we think of quantum teleportation as the transfer of an unknown
(but still well defined) state we usually require that all states
of the space we select are teleported perfectly. A natural
extension is that any relation that the original particle has with
respect to other systems should be transferred as well.
Specifically, if our particle was entangled to another system we
would require from a faithful teleportation machine that this
entanglement is transferred to the particle that ``inherits" the
state at Bob's location.

\fig{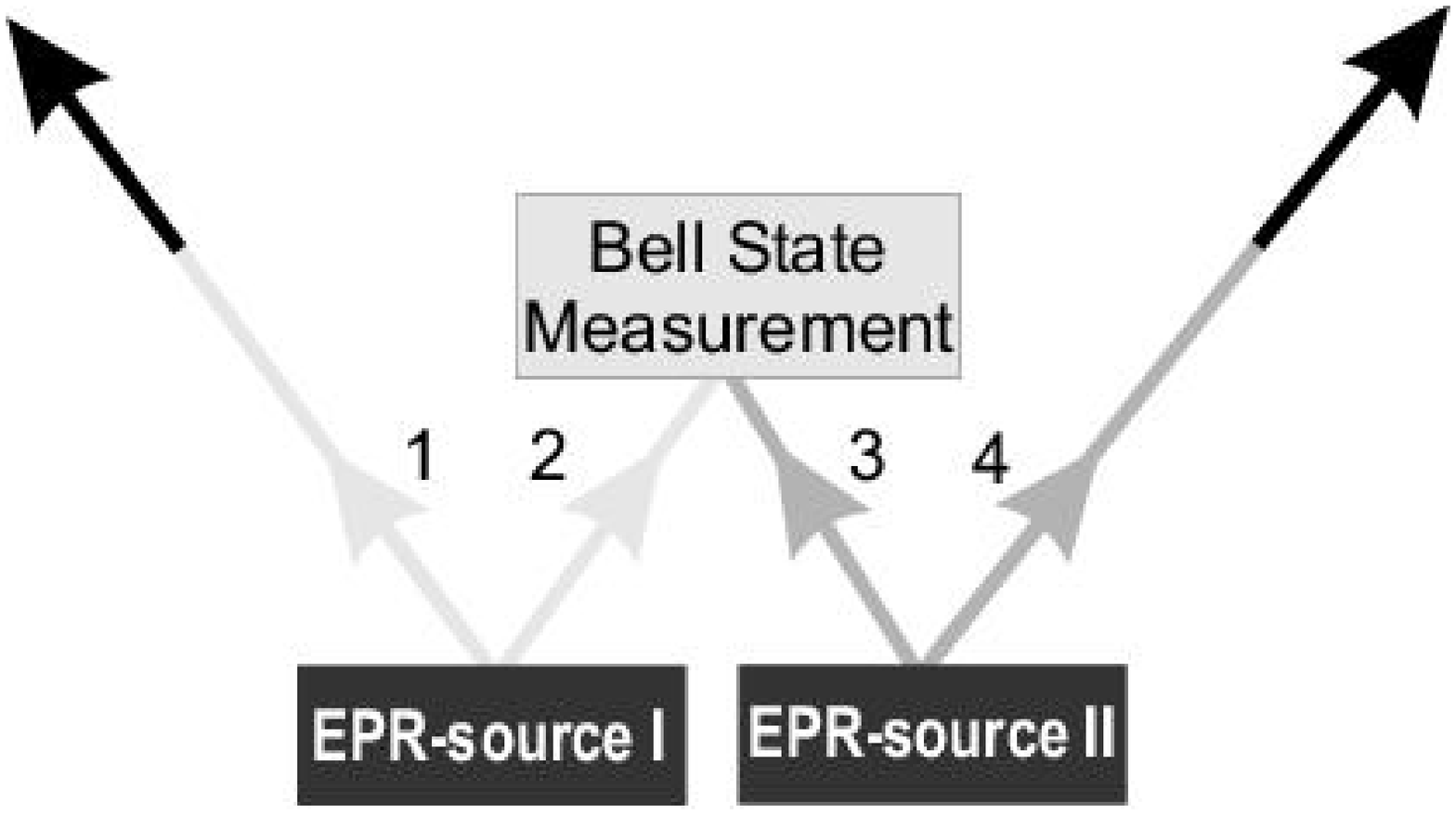}{0.5}{Entanglement swapping between two EPR-pairs by a
Bell-state measurement.}

This generalized concept, mentioned for the first time in 1993 by
\.Zukowski {\it{et al.}}\cite{Zukowski93a}, has become known as
entanglement swapping (or teleportation of entanglement). It
symmetrizes the teleportation scheme to a procedure that can be
applied to two or more entangled systems.\cite{Bose98a} The lowest
order protocol joining two entangled two-qubit systems is depicted
in Fig.~\ref{etswap}. If, say, both entangled systems are produced
in the $\ket{\psi^-}_{12}$ and $\ket{\psi^-}_{34}$ Bell-states,
and we project onto a $\ket{\psi^-}_{23}$ state we find particles
1 and 4 in the state
\begin{eqnarray}
 \bra{\psi^-}_{23} & \left[\ket{\psi^-}_{12} \otimes \ket{\psi^-}_{34}\right] =
 \\
 && =
 \frac{1}{\sqrt{8}}\left[\bra{01}-\bra{10}\right]_{23}
 \left[\ket{01}-\ket{10}\right]_{12} \otimes
 \left[\ket{01}-\ket{10}\right]_{34} \nonumber \\
 && =
 \frac{1}{\sqrt{8}}\left[\bra{01}-\bra{10}\right]_{23}
 \left[\ket{0101}-\ket{1001}-\ket{0110}+\ket{1010}\right]_{1234} \nonumber \\
 && =
 \frac{1}{\sqrt{8}}
 \left[\ket{01}-\ket{10}\right]_{14}=\ket{\psi^-}_{14}.
 \nonumber
\end{eqnarray}
\noindent Therefore, particles 1 and 4 end up in an entangled
state although they never interacted locally.

Any experimental verification of this protocol is a clear
demonstration of the quantum physical projection postulate: the
joint measurement of particles 2 and 3 results in a preparation of
the joint state of particles 1 and 4, regardless of whether one
decides to project on a product, or on an entangled state. In the
specific case considered here, the projection postulate predicts
the change of the joint state of particle 1 and 4 from a product
state to an entangled state. This can easily be verified by
subjecting particle 1 and 4 to a test of a Bell inequality.

A first attempt to demonstrate that the entanglement is indeed
swapped has been reported by Pan {\it{et al.}}\cite{Pan98a} in
1998. The experiment was based on polarization entanglement.
However, although the observed degree of entanglement surpasses
the limit of a classical wave theory, it was not high enough to
manifest itself in a violation of a Bell inequality. Yet, after
some refinements, Jennewein {\it{et al.}} could recently
demonstrate a violation of Bell's inequality with ``swapped"
entanglement.\cite{Jennewein01a} The measurements yielded
$S^{\mathrm{exp}}=2.42 \pm 0.09$ which exceeds the limit of 2 for
local realistic theories by 4 standard deviations.

\subsection{Purification and distillation}\label{distillation} \noindent

In the context of extending quantum information protocols to
larger distances, considerable interest has grown for measures
against decoherence as encountered while transmitting quantum
states through a noisy environment. Various schemes of
entanglement distillation, purification and concentration, and of
quantum error correction have been proposed (see references in
\cite{Lo98a,Bouwmeester00a}), however, to date, only one
experimental demonstration of distillation of photonic
entanglement has been reported.

\fig{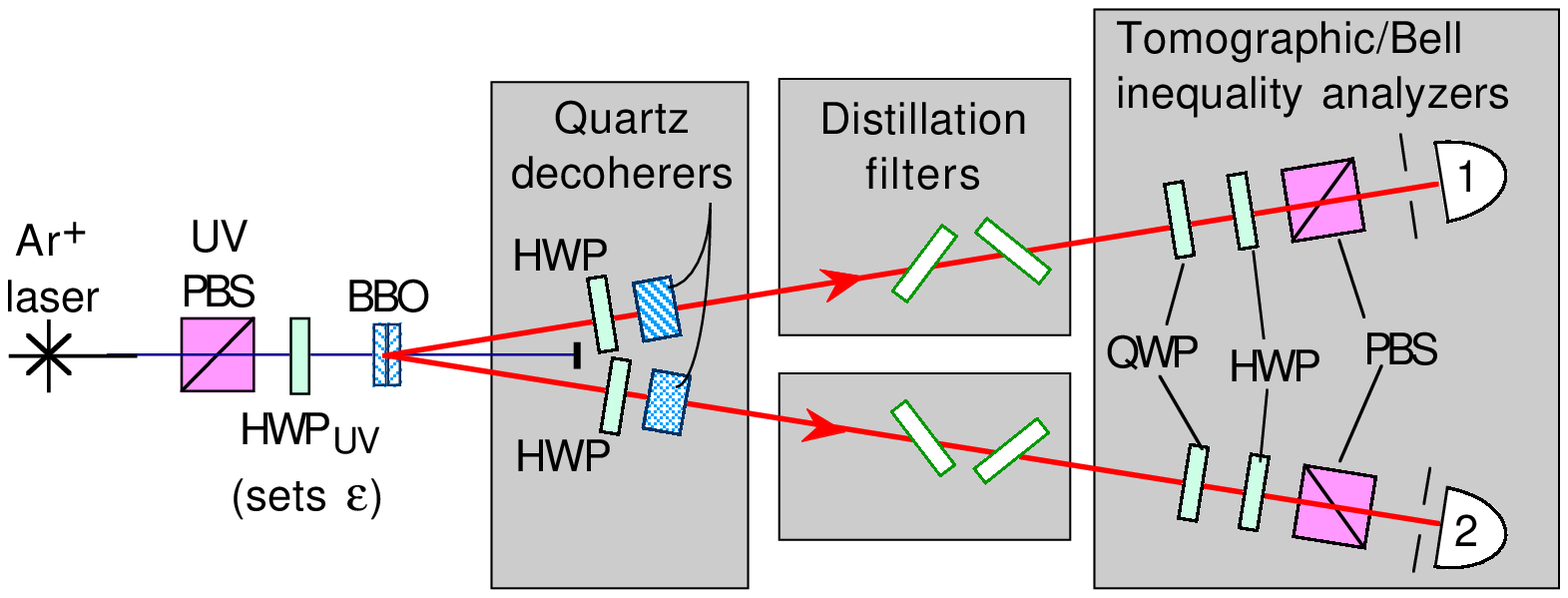}{0.9}{Experimental setup to demonstrate
entanglement distillation and hidden non-locality. Half-wave
plates and two 1cm thick quartz elements (decoherers) allow to
generate partially entangled, partially mixed states. After
filtering they are analyzed by means of two qubit analyzers --
either in order to reconstitute the density matrix via quantum
tomography, or to test the CHSH Bell inequality.}

As predicted by Gisin in 1996,\cite{Gisin96a} entanglement
purification can be achieved by local filtering. This scheme has
been demonstrated experimentally 2001 by Kwiat {\it{et
al.}}\cite{Kwiat01a} (Fig. \ref{setupdistill}). To demonstrate the
phenomenon of ``hidden non-locality", certain partially entangled,
partially mixed states that do not violate Bell inequality were
prepared utilizing the polarization entanglement source based on
two stacked thin type-I crystals with their optic axes at 90$^\circ$
as discussed in Section~\ref{typesofentanglement}. It has been
shown that the resulting states after a suitable local filtering
operation violate Bell-CHSH inequalities (Fig. \ref{datadistill}).

Actually, much more than from decoherence, applications of quantum
communication like quantum cryptography suffer from transmission
losses paired with detector noise.\footnote{This contrasts with
proposals for quantum computing using massive particles where the
particles hardly get lost and good detectors exist, but where
decoherence is the major problem.} As argued in section
\ref{cryptography}, the fact that the QBER increases with losses
(Eq.~\ref{QBER}) limits the maximum transmission distance. A way
out would be to use a quantum non-demolition (QND)
measurement\cite{Grangier98a} at Bob's and to switch on the
detectors only if a photon is known to arrive (hence $\mu$ = 1
and remaining $t_{\mbox{\scriptsize link}}$=1) (see
Fig.~\ref{QND}). Unfortunately, QND measurements for detection of
visible or telecommunication photons do not exist
yet.\footnote{QND measurements for microwave photons have
recently been demonstrated by Nogues {\it et
al.}\cite{Nogues99a}} Another possibility that is in reach with
current technology is to use a concatenation of two-particle
sources and Bell state measurements and to establish entangled
photons at Alice's and Bob's via entanglement swapping. In this
scheme, Bob's detectors are only switched on if Alice detected a
photon {\it{and}} if the last Bell state measurement (at Bob's)
was conclusive. It therefore much resembles a QND measurement.
Interestingly, this idea is closely related to the original
motivation for entanglement swapping:\cite{Zukowski93a} testing
Bell inequalities with ``{E}vent-Ready-Detectors''. However,
although this scheme allows to extend the maximum transmission
span, it will at the same time significantly reduce the bit rate.

\fig{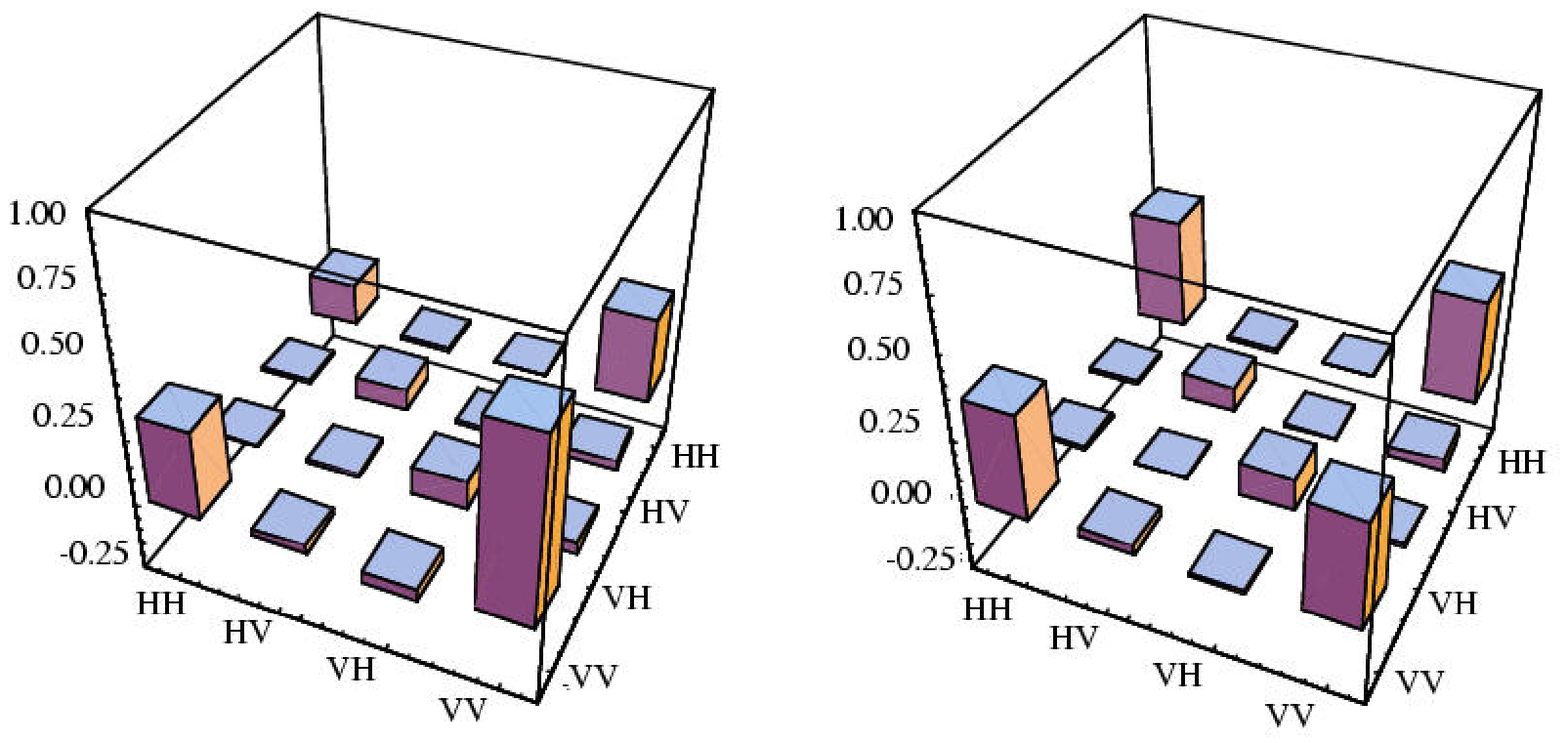}{0.9}{Measured density matrices before (left) and
after (right) distillation. A violation of CHSH Bell inequality is
observed for the state after distillation ($S_{\mbox{\scriptsize
filtered}}$=2.22) while the initial state does not manifest
non-local behavior ($S_{\mbox{\scriptsize initial}}$=1.82)}

\fig{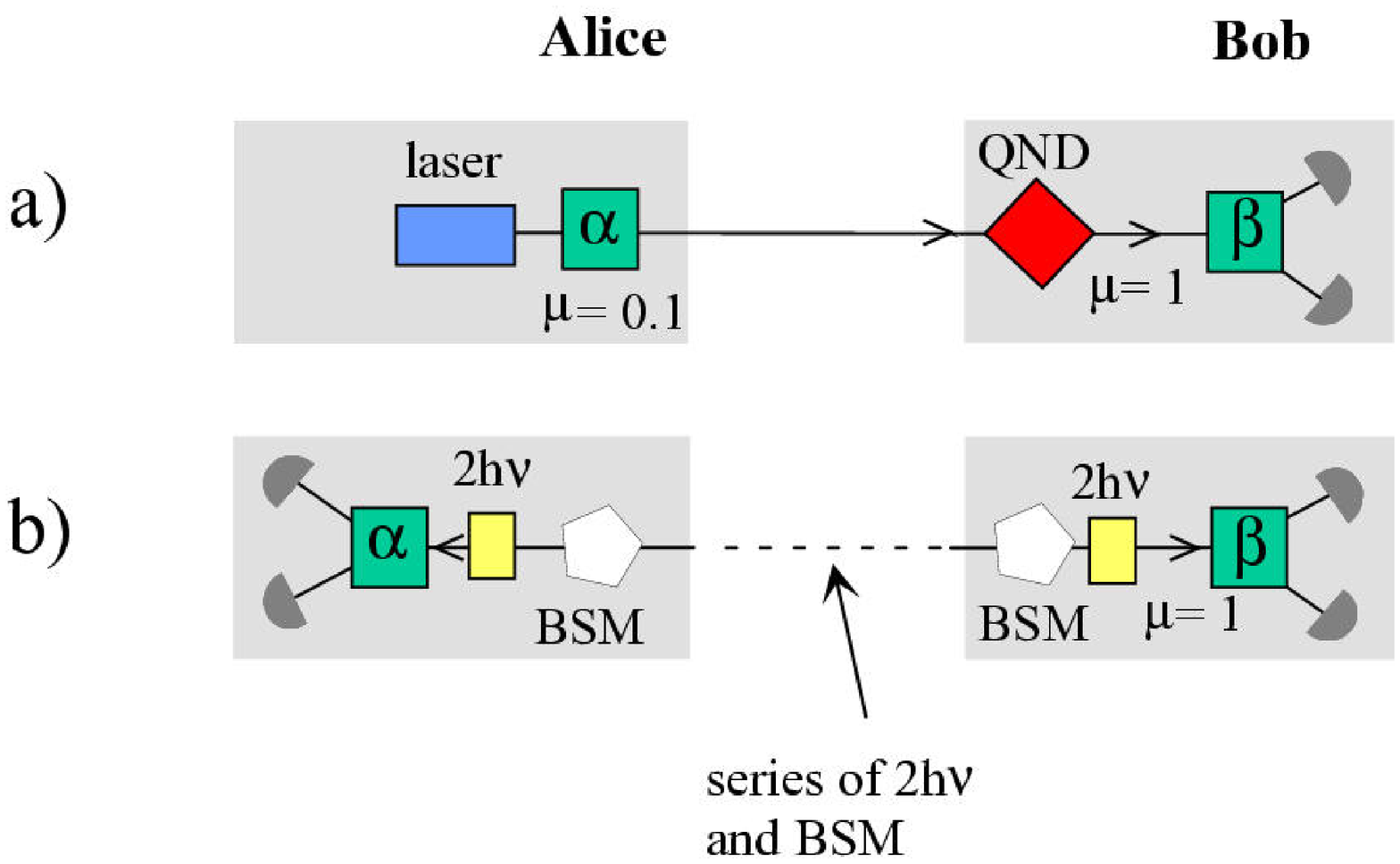}{0.7}{Comparison of QND measurement based (a), and
entanglement swapping based (b) realization of event ready
detectors at Bob's. ``2h$\nu$" denote entangled two-photon
sources, ``BSM" a Bell-state measurement, and $\alpha$ and $\beta$
are the settings at Alice's and Bob's qubit analyzers.}

\section{Conclusion} \label{conclusion}\noindent

In this article, we tried to review the major developments in
experiments based on entanglement of photonic qubits, both in the
traditional domain of fundamental tests of quantum non-locality,
as well as in the new approach of using entanglement as a resource
for quantum communication. In traditional tests of spin 1/2
Bell-inequalities no major surprises are expected any more --- at
least concerning experiments with photons. But we are only at the
very beginning of tests of non-locality using higher dimensional
systems,\cite{Howell01a} systems employing more than two particles
(Section~\ref{GHZ}), or experiments testing other interpretations
of the quantum world (Section~\ref{relativistic}). Furthermore,
the fact that entanglement can be used as a resource for
applications in quantum communication came like a great surprise
and stimulated much interest in the physics as well as in the
general community. Quantum cryptography is a good candidate of
becoming the first industrial application, a development that
would have a major impact on the whole field.

It is very interesting that tests of Bell-inequalities,
traditionally considered part of fundamental research, recently
became an application with the Ekert scheme for quantum
cryptography, and are nowadays routinely employed in the
laboratory to characterize the quality of the entanglement of
photon pairs before passing on to more complicated experiments.
However, not only the change from fundamental to applied aspects
can be observed, the opposite is possible as well: It is for
instance still an open fundamental question if the connection
between security of quantum cryptography and violation of a Bell
inequality can be generalized to all sorts of protocols.

The enormous progress obtained in laser and optical fiber
technology, single photon detectors, and the availability of
non-linear crystals enable nowadays experiments that were only
gedankenexperiments not even long ago. However, new ideas and
technology are needed for further steps. For instance, the secret
bit rate and maximum transmission distance in quantum cryptography
experiments are limited by the performance of detectors and single
photon sources, and there is big need for efficient sources
creating more than two entangled particles. Maybe photon sources
based on individual nitrogen-vacancy color centers in
diamond\cite{Kurtsiefer00a,Brouri00a}, quantum
dots\cite{Gerard98a,Kim99b,Kiraz01a} and parametric
down-conversion in PPLN waveguides\cite{Tanzilli01a,Sanaka01a}
will turn out to enable more refined experiments in the future.

Although quantum non-locality has recently been observed with
atoms, photons still play an outstanding role whenever it comes to
experiments employing entanglement. They are best suited as a
carrier of quantum information since decoherence effects due to
interaction with the environment are very small. But
unfortunately, photons do not interact with other photons, a
major problem when it comes to Bell state measurements or
processing of quantum information in general. A solution might be
to map photonic quantum states onto massive particles like atoms
or ions that are in principle well suited for all applications
where two-particle interaction is needed. A first experiment in
this context has been performed \cite{Julsgaard01a} very recently.

\nonumsection{Acknowledgements} \noindent The authors acknowledge
financial support by the Swiss FNRS and the Austrian Science Fund
(FWF) project no. S1506, as well as by the IST-FET ``QuComm''
project of the European Commission, partly financed by the Swiss
OFES. W. T. would like to thank his colleagues from GAP,
especially N. Gisin and H. Zbinden for theoretical and practical
support during the mentioned experiments. G. W. would like to
thank T. Jennewein and A. Zeilinger for helpful discussions and
continued support.

\nonumsection{References} \noindent
\bibliographystyle{prsty}
\bibliography{main}

\end{document}